\documentstyle[pre,aps]{revtex}

\newcommand{\be}{\begin{equation}}
\newcommand{\ee}{\end{equation}}
\newcommand{\bea}{\begin{eqnarray}}
\newcommand{\eea}{\end{eqnarray}}
\newcommand{\es}{\ell_{ES}}

\begin{document}

\title{Instabilities in crystal growth by atomic or molecular beams}

\author{Paolo Politi~$^{a,b,c,d,}$\footnote{Corresponding author. E-mail: 
{\tt politip@fi.infn.it}}, 
Genevi\`eve Grenet$^{e,}$\footnote{E-mail:
{\tt genevieve.grenet@ec-lyon.fr}}, 
Alain Marty~$^{d,}$\footnote{E-mail:
{\tt amarty@cea.fr}},
Anne Ponchet~$^{f,}$\footnote{E-mail:
{\tt ponchet@cemes.fr}}, 
Jacques Villain~$^{b,c,g,}$\footnote{E-mail:
{\tt villain@drfmc.ceng.cea.fr}}}

\address{$^a$ Fachbereich Physik, Universit\"at GH Essen, 
45117 Essen, Germany}
\address{$^b$ Dipartimento di Fisica dell'Universit\`a degli Studi di
Firenze and Sezione INFM, L.go E.~Fermi 2, 50125 Florence, Italy}
\address{$^c$ Centre de Recherches sur les Tr\`es Basses Temp\'eratures,
CRTBT-CNRS, BP-166 38042 Grenoble Cedex 9, France}
\address{$^d$~CEA,~D\'epartement de Recherche Fondamentale sur la
Mati\`ere 
Condens\'ee, SPMM/NM, 38054 Grenoble Cedex 9, France}
\address{$^e$ LEAME, Ecole Centrale de Lyon, 
36 avenue Guy-de-Collongue. B.P. 163 - F69131 Ecully Cedex, France}
\address{$^f$~CEMES-CNRS
29 rue Jeanne Marvig
BP 4347
31055 Toulouse-Cedex 04, France}
\address{$^g$~CEA,~D\'epartement de Recherche Fondamentale sur la
Mati\`ere 
Condens\'ee, SPSMS, 38054 Grenoble Cedex 9, France}

\date{\today}
\maketitle

\begin{abstract}

When growing a crystal, a planar front is desired for most of the
applications. 
This plane shape is often destroyed by instabilities of various types.
In
the case of growth from a condensed phase, the most frequent
instabilities are 
{\it diffusion instabilities}, which have been 
studied in detail by many authors and 
will be but briefly discussed in simple terms in chapter \ref{snow}. 
The present review is mainly devoted 
to instabilities which arise in ballistic growth, especially 
Molecular Beam Epitaxy (MBE). The reasons of the instabilities can be
geometric, but 
they are mostly {\it kinetic} (when the desired state
cannot be reached because of a lack of time) or {\it thermodynamic}
(when the desired state is unstable).
The kinetic instabilities which will be studied in detail in chapters 
\ref{PP1} and \ref{PP2} result
from the fact that adatoms diffusing on a surface do not 
easily cross steps (Ehrlich-Schwoebel or ES effect). When the growth
front is a high 
symmetry surface, the ES effect produces mounds which often 
coarsen in time according to power laws.  
When the growth front is a stepped surface, the ES effect 
initially produces a meandering of the steps, which eventually may also give
rise to mounds. Kinetic instabilities can usually be avoided by raising
the
temperature, but this favours thermodynamic instabilities of the
thermodynamically 
unstable materials (quantum wells, multilayers~...)
which are usually prepared by MBE or similar techniques. The attention
will
be focussed on   thermodynamic instabilities which result from
slightly different 
lattice constants $a$ and $a+\delta a $ of the substrate and the
adsorbate. 
They can take the following forms.
i) Formation of misfit dislocations, whose geometry, mechanics and
kinetics 
are analyzed in detail in chapter \ref{Marty1}. ii)  Formation of
isolated epitaxial 
clusters which, at least in their 
earliest form, are `coherent' with the substrate, 
i.e. dislocation-free (chapter \ref{dots}). iii) Wavy deformation of the
surface,
which is presumably the incipient stage of (ii) (chapter \ref{tersoff}). 
 The theories and the 
experiments are critically reviewed and their comparison is
qualitatively satisfactory 
although some important questions have  not yet received a
complete answer. Short chapters are devoted to shadowing instabilities,
twinning and stacking faults, as well as the effect of surfactants.

\end{abstract}

\vskip 1cm
\noindent 
PACS codes:
\hfill\newline\noindent
81.15.Hi Molecular, atomic, ion, and chemical beam epitaxy.
\hfill\newline\noindent
81.10.Aj Theory and models of crystal growth.

\vskip 0.5cm


\vfill\newpage

\tableofcontents

\vfill\newpage

\section{Growth, surface roughness and instabilities}
\label{Intro}

\subsection{Growth instabilities}
\label{1.1}

Crystals are known to be the stable form  at
low temperature  of all materials, except helium 
under moderate pressure. 

Nevertheless,
most of the natural materials which surround us are not crystals or
are made of very small crystals. This fact shows that crystals do not
easily grow. The reason is that growing crystals are subject to 
{\it instabilities} which introduce crystal defects and eventually 
result in the formation of polycrystals or even amorphous materials.

On the other hand, crystals are of great technological interest,
because they have reproducible properties. 
Semiconductor technology uses large quantities of big crystals of
silicon
which are artificially pulled from the liquid phase and which are
perfect. This has only been possible because the mechanism of
instabilities 
in this type of growth was well understood. Even so, in the case of 
{\it metals}, it is impossible to 
avoid the formation of a few dislocations in a macroscopic crystal.
The case of semiconducting elements (e.g. Si), which can be prepared as
ideal crystals, is exceptional. These elements, even though they can be 
modified by implantation of appropriately chosen impurities,
are not sufficient for all technological requirements. 
Quantum wells, for instance, are made of a slice of a material $A$,
of well-defined thickness, 
sandwiched between two 
slices of a material $S$. A usual case is $A$=GaAs and 
$S$=GaAlAs. One of the standard  techniques  to do  
that is Molecular Beam Epitaxy (MBE)~\cite{Herman}.

The word {\it epitaxy} designates 
the adsorption of a  crystal  on another crystal, with a well-defined 
relative 
orientation of the two crystals. In most of the examples cited in the
present review,  both crystals are cubic and their principal axes  are
parallel.

In MBE, the atoms are ballistically deposited by a beam
of molecules or atoms. The required thickness is obtained by switching 
the beam of material $S$ during an appropriate time. 
If the atoms do not stick at once, they are sucked out by a vacuum 
pump, so that one can for instance deposit $n$ layers of 
AlAs, then $m$ layers of GaAs and avoid the formation of a GaAlAs
mixture. 

MBE is a very slow technique, which 
has the advantage to permit the production of artificial devices which
could not be produced by  other methods. Other examples are 
{\it multilayers} (Fig. \ref{multi}) or `superlattices' which are
appropriate for
fundamental research because they are periodic and therefore have,
for instance, a simple diffraction spectrum. These materials are generally
not thermodynamically stable but can be metastable during a very long 
time once they have been prepared.

Growth by ballistic deposition, as well as from the melt or a solution, 
is subject to instabilities. These instabilities
generally appear at the place where the dynamics 
is fastest, i.e. the surface.
The initial stage of the instability is a deformation of the growing 
surface (or `front') which ceases to be planar and starts forming bumps 
or holes. Growth instabilities may be classified as 
follows (Fig. \ref{eheh}):

i) Diffusion instabilities (Fig. \ref{eheh}a) are typical of growth from
the melt  or from a solution. 
In a supersaturated solution,
dendrites  can form because diffusing atoms or molecules 
tend to go to the nearest point of the
solid, and therefore to tips which unavoidably form 
at the growth front.

ii) Kinetic instabilities (Fig. \ref{eheh}b) 
which take place because the growth is too
fast and the surface has no time to find its equilibrium shape,
which in the simplest case is a chemically homogeneous plane.

iii) Thermodynamic instabilities (Fig. \ref{eheh}c)  
which take place when one tries to make 
    a thermodynamically 
unstable material. 

iv) Geometric instabilities. A typical example is 
the shadowing instability which occurs in oblique deposition  
(Fig. \ref{eheh}d).

 Diffusion instabilities  have been studied in
many articles and review papers. For this reason, only 
a few words will be said about them in chapter \ref{snow}, and 
the present review will mainly be devoted to the other types of 
instabilities. One might think that diffusion instabilities
do not arise in ballistic deposition and in particular in MBE. 
However, diffusion does occur
{\it on} the surface and is responsible for instabilities as will be
seen 
in chapter \ref{PP2}.

A special type of kinetic instability will be investigated in great
detail
in chapters \ref{PP1}  and \ref{PP2}. It occurs when atoms diffusing on
the surface have difficulties to cross steps. 
Presumably, the reason of its popularity among theorists is that 
this instability is consistent with the so-called S.O.S. model, a
model in which the topology of the solid is fixed, atoms are assumed to 
pile up on preexisting lattice sites. Other instabilities, e.g. stacking
faults, are not so easy to describe by a simple model.

Thermodynamic instabilities often result from a difference in the
lattice 
constants of different materials which one tries to assemble in a 
single crystal. 
If the  lattice constants of the
substrate and the adsorbate are $a$ and $a+\delta a$,
then $\delta a/a$ is called
(relative) `misfit' or `mismatch'.
It will play an important role in 
chapters \ref{Marty1}, \ref{tersoff} and \ref{dots}
where this instability will be studied. 
Another cause of
thermodynamic instability may be of chemical nature. A typical example
is
the solidification of   an eutectic solution, 
which by definition leads to
coexisting clusters of two chemically different alloys. The cluster size
is
determined by atomic motion on and near the growth front, and is related
to the shape of a front, which becomes more or less  complex and typical
of a growth instability~\cite{valance}.

Strictly speaking, thermodynamic instabilities do not appear only
during growth. However, in practice, they mainly appear at high enough 
temperature, and in particular, during growth, while the evolution of
a solid toward equilibrium at room temperature is often geologically
slow.

The shadowing instability will be much more briefly addressed in chapter
\ref{shadow}, as well as kinetic instabilities resulting from
stacking faults and polymorphism, which will be treated in chapter
\ref{poly}.  The very unequal length of the various chapters 
corresponds partly to the unequal competence of the authors,
partly to a different state of the art. In cases which are still
poorly understood, we shall mainly give references, essentially 
experimental ones.

In the case of non-crystalline materials, the bumps and holes which are
formed at the surface are macroscopic. In the case of crystals, which is
of interest in this review, the modulation of the surface may be
microscopic.
It is generally the case, for instance, when dislocations are formed.
	Dislocation formation, which may be in competition with other
types of instabilities, will be addressed in chapter \ref{Marty1}.

Instabilities are often characterized by a sudden burst. 
Before the instability
takes place, 
the surface is smooth. Then, suddenly, an instability develops and its
amplitude increases dramatically for short times, 
for instance exponentially with time. This fast initial increase 
gives no information on the final state of the system, which may be 
a spatially periodic modulation, or a general collapse, or chaotic.
Only in the case of a sharp initial increase of the perturbation  the
word 
`instability' will be used. It will not be used 
for instance if the fluctuation amplitude 
$\delta h(t)$ of the surface increases as 
a power of the time $t$, $\delta h(t)\simeq t^{1/z}$. In the latter
case, 
we will just say that the surface is `rough'.


\subsection{Thermal roughness and roughening transition}
\label{1.2}

A rough surface, with a roughness which increases in time, 
is frequently predicted for a growing
object~\cite{Nozieres,JVAP,book_BS} 
when the growth rate 
undergoes stochastic fluctuations, as is usually the case.
But even at thermal equilibrium, i.e. in the absence of growth, 
the surface of a crystal can be rough. Although the minimization
of the surface (or `capillary') energy corresponds to a plane surface,
thermal fluctuations  can produce some `roughness'. 
At low temperature, this roughness has only a microscopic scale, 
and the surface
is said to be `smooth'. 
The roughness becomes macroscopic, and the surface is then said to be 
`rough', above the so-called roughening temperature 
$T_R$ which depends very much on the surface 
orientation~\cite{Nozieres,JVAP,book_BS}. 
High symmetry surfaces, (001) or (111), of fcc metals are smooth until
very
near the melting temperature, while high index surfaces, e.g. (1,1,11),
have very low roughening temperatures which cannot be observed because
thermodynamic equilibrium cannot be reached~\cite{JVAP}. 
But thermal roughening can be
observed  on (113) or (110) metal surfaces. 
High symmetry surfaces of semiconductors are also believed to 
be smooth at equilibrium until very high temperatures, higher than those
used for MBE.

Elasticity is generally ignored in theoretical treatments of the
roughening 
transitions which have been reviewed for instance by 
Nozi\`eres~\cite{Nozieres}, Pimpinelli \& Villain~\cite{JVAP},
Barab\'asi \& Stanley~\cite{book_BS}. Some properties of  crystal
surfaces above and below $T_R$ are summarized in table 
\ref{rough_table}.
An important property, which will be of interest in chapter
\ref{tersoff},
is that the energy of a plane surface, 
which above $T_R$ is an analytic function 
of its orientation, becomes non analytic below $T_R$. In the case of a
high symmetry surface,
this is a consequence of 
the positive step free energy per unit length $\gamma$: if  the surface
is 
disoriented by an angle $\theta$, the  energy cost is the product of
$\gamma$ by the number of steps, which is proportional 
to $|\theta|$, which is a non analytic 
function for $\theta=0$. Such a surface with a
non-analytic dependence on the
orientation is called {\it singular}. 
Similarly, a height modulation of wavelength
$2\pi/q$ (see  chapter \ref{tersoff})
is an analytic function of $q$ for $T>T_R$ and non-analytic for $T<T_R$.

Another property of a smooth, infinite, high symmetry crystal surface at
equilibrium 
is that its height cannot change continuously. Indeed a continuous
change requires
the creation of an infinitely long step which has an infinite energy. On
the contrary, the height of a rough surface can change continuously
without paying 
any free energy through a motion of the steps. A precise,
renormalisation group
description has been given by   Tang \& Nattermann~\cite{tang}.

The experimental determination of $T_R$ is not always easy~\cite{JVAP}. 
The safest method  uses the equilibrium shape of crystals. 
The presence of facets  of a particular orientation at a temperature $T$
indicates that $T_R>T$ for this orientation. Unfortunately, this method is only
applicable if $T_R$ is pretty close to the melting temperature $T_M$. For instance, 
for Si, where $T_M=1413 ^\circ$C 
Heyraud et al.~\cite{hmb} have very recently found $T_R=1370 ^\circ$C 
for the $\{110\}$ face, $T_R=1340 ^\circ$C 
for the $\{113\}$ face, and $T_R>1400 ^\circ$C 
for the $\{111\}$ face and add: ``The case of  $\{001\}$ is much less clear,
and is left to a future paper~...''. This sentence which reflects the hope of a 
better future also reflects our present ignorance of  thermodynamic
properties of the most common surfaces.

To conclude this paragraph on roughness,  one can say that 
 a `rough' surface, as the
man in the street would call it, can originate
i) either from stochastic fluctuations or ii) from an instability.
In the latter case, the mathematical description is provided by 
deterministic equations of motion. In the first case, the stochastic 
fluctuations can be either thermal fluctuations at equilibrium, or 
for instance the fluctuations of the beam in MBE (the so-called
`shot noise'). The mathematical description of beam fluctuations
requires
the introduction of a stochastic term in the equations, while
thermal fluctuations are just taken into account by the Boltzmann
factor in Gibbs's canonical distribution. We prefer to restrict the use
of the 
word `roughness' to that which arises from stochastic effects, and
to use the expression `instability' when the roughness arises from 
a deterministic mechanism. Some references to stochastic effects
will be made in the next chapters, but most of the present review will
be devoted to instabilities.


\section{Snowflakes,  diffusion instabilities and DLA}
\label{snow}

 The goal of this chapter is to give an intuitive introduction to 
diffusion instabilities. The mathematical treatment will be found 
in  review
papers and books~\cite{wood,pelc,Langer,pom,carol}.

A crystal growing from a dense phase often forms `dendrites'. 
Snowflakes (Fig.~\ref{diff}b) are the most familiar example. 
However,  their formation in clouds is a complicated process~\cite{hobbs},
since  clouds are made of supercooled liquid droplets 
coexisting with  solid particles 
and water molecules which travel from the former to the latter. It is 
simpler to consider growth from a homogeneous phase,
e.g. a solution. As an example, 
Fig.~\ref{DLA}b shows a pattern which 
has a resemblance to a snowflake. It is not a 
crystal, but similar instabilities arise in crystals growth as well.   
Diffusion instabilities arise  (Fig.~\ref{diff}a) because, for some reason 
(e.g. thermal fluctuations) small protruding parts appear. 
If there are protruding parts, the molecules which diffuse through the 
solution to the solid surface go 
preferably to those parts (Fig. \ref{diff}a) because paths leading to
protruding parts are on the average shorter. Thus, protruding parts
become more protruding and form dendrites.

This mechanism can be implemented by simulations. In the simplest
model  (`Diffusion limited aggregation' or DLA~\cite{WS1,WS2})
 molecules are assumed to be independent in the fluid phase. 
They diffuse randomly from a very large distance, 
and stop moving as soon as they meet the solid. This model gives
rise to very irregular shapes (Fig. \ref{DLA}).

The striking difference between the irregular cluster of 
Fig.~\ref{DLA}
and the fairly regular shape of  snowflakes is partly 
due to the hexagonal symmetry of ice crystals. Another factor 
which favours order is the
possibility of atom diffusion along the solid surface, which is ignored
in DLA. It is important in the case of snow, and according 
to Mason \cite{mason} surface diffusion is the reason of
certain shape transitions of snow with temperature.
Finally, interactions between atoms, which are neglected in the DLA model,
make growth more deterministic. An extreme case is growth from the melt. 
Since the fluid is almost incompressible, the above picture,
based on matter diffusion, is
not valid. Actually, the  quantity which diffuses, at least in a pure
melt, is not matter, but energy. The equation which represents energy
diffusion in the liquid, or matter diffusion in a vapour 
(or in a solution) is the  diffusion equation 
\begin{equation}
\partial \rho/\partial t= D \nabla^2 \rho
\label{dif}
\end{equation}
where $\rho(\vec r,t)$ is
the  density of energy in one case, of matter in the other case. 
A third case of interest is solidification from an impure melt. Then
the relevant diffusing quantity is the density of impurities, whose 
diffusion is much slower then energy diffusion. The relevant process
is the slowest one. 

Equation (\ref{dif}) must be coupled with
appropriate boundary conditions at the interface,
which is assumed to be a continuous 
surface~\cite{Langer,pom,carol,mulsek,mulsek'}.
If one wishes to model `directional solidification' (the usual way to
pull a crystal from the melt, e.g. silicon) 
the initial condition is a plane surface. If the pulling velocity
is low enough, the surface remains a plane. For a given temperature
gradient, the instability sets in~\cite{mulsek} if the pulling velocity
$v$ is larger than a critical value $v_c$. It is called the
{\it Mullins-Sekerka instability}. If $v$ is not too large, the 
ultimate state of the surface is a periodic 
modulation~\cite{Langer,pom,carol,mulsek} (Fig. \ref{diff}c). 
The contrast between  
this smooth pattern and  the DLA pattern of Fig. \ref{DLA} is striking.

The DLA cluster is indeed a {\it fractal}~\cite{pietroto}. 
This means that the relation between its
radius $R$ and its mass $M$ is not $M=\hbox{Const} \times R^d$, where
$d=3$ is the dimension of the space, but (for large $R$) 
\begin{equation}
M=\hbox{Const} \times R^{d_f}
\label{frac}
\end{equation}
where $d_f$ is called an
`effective fractal dimension'~\cite{meakin}. Usually, 
it is not an integer and is smaller than the space dimension $d$, which
is generally $d=3$
in real materials although simulations are often made in $d=2$
since they are easier. In two-dimensional DLA~\cite{meakin}, $d_f$ is  
close to 1.7. The radius $R$ can be defined at any time as the distance
from the center such
that a diffusing atom has a probability lower than a certain value
(e.g. 0.1) to reach it.

One can wonder whether continuous theories based on   (\ref{dif}) 
can also produce a fractal.
To our knowledge, there is no clear theoretical answer, but 
Arneodo et al.~\cite{couder} performed experiments which can reasonably
be described  by continuous models and were able
to produce fractal objects quite analogous to DLA
clusters~\cite{couder}.

The various patterns which arise from diffusive instabilities result from
differences in the details. For instance, in directional solidification
there is a temperature gradient which has an important stabilizing
effect and 
has no equivalent in the experiments of Arneodo et al.~\cite{couder}
which, as reported above, yield fractal
objects. However, the onset of the instability results from equations 
which,  for a given initial condition (e.g. a plane front $z=vt$) 
are very similar in all the cases. 
At the beginning of the instability, the deviation $\delta z(x,y,t)$
from the planar shape   is  indeed very small and the equations can
be linearized with respect to  $\delta z(x,y,t)$. 
A Fourier transformation in the ($x,y$) space yields 
$d z_k(t)/dt=\omega_k z_k$, where the coefficients $\omega_k$ 
depend on the model, and there is an instability if at least 
one of them is positive. This method is called  a
{\it `linear stability analysis'}.
It is sufficient to prove that a plane front is unstable, but does
not predict the shape of the resulting dendrites which can
only be deduced from a nonlinear treatment. 

Before leaving diffusion instabilities, we come back to snow.
The formation of snow crystals in clouds proceeds in three steps \cite{hobbs}:
i) Heterogeneous nucleation of an ice particle by deposition from
the vapour phase on an impurity called `ice nucleus' or by nucleation
in a supercooled water droplet by an ice nucleus.
ii) Growth of the ice particle by deposition of water molecules 
evaporated from liquid droplets and diffusing 
through the vapour phase, until the diameter reaches
about 0.2 mm. iii) Growth by sticking of 
supercooled liquid water droplets
on the ice particle or by aggregation  of ice particles.
Examples of ice nuclei are small crystals of AgI. The typical diameter 
of a water droplet in a cloud is between 10 and 100 $\mu$m. 
Although diffusion plays a part in these processes, the simple 
scheme of a diffusion instability is not applicable. 

In view of the considerable amount of work devoted to diffusion 
instabilities, 
nothing more will be said about this topic in the  present review.
Instead, we shall describe instabilities which arise when the 
atoms do not diffuse toward the crystal, but
travel `ballistically', i.e. without collisions, from a source, and then
form epitaxial layers. 

We have nevertheless devoted this short chapter
to diffusion instabilities for
several reasons.  First, for the completeness of this review on
growth instabilities. Second,  instabilities
which arise from diffusion on the moving front
will be  encountered in chapter \ref{PP2}. Third, the methods
which are useful to investigate diffusion instabilities 
can often be used also for other instabilities. For instance,
linear stability analysis will be used in parts \ref{PP1}, \ref{PP2} and
\ref{tsf2}.


\section{Growth from diluted vapour and MBE}
\label{GG1}

\subsection{MBE, what is that?}

The remainder of this review is devoted to  
techniques where the deposition is ballistic rather
than diffusive, so that no instability can arise from
diffusion  in the fluid phase. Apart from the short chapter 
\ref{shadow}, the attention will be concentrated on Molecular Beam
Epitaxy (MBE) and similar techniques.

Among all the advanced semiconductor growth techniques, 
Molecular Beam Epitaxy provides the  greatest ease
 of growing complex semiconductor multilayer structures 
with a precise control of the thickness (up to the monolayer scale
accuracy), the composition and  
the doping of the involved layers~\cite{hand}. 
In this technique, growth proceeds 
under ultrahigh vacuum conditions (pressures
$\leq~10^{-8}$ Pascals)
by the condensation of thermal energy molecular beams on an underlying
single crystal substrate. 
The substrate has a strong influence on the 
growth.  Generally, the deposited film 
adopts as far as possible
the orientation and crystallographic
characteristics of
the substrate. As a matter of fact, when thermal atoms or molecules arrive
 on a substrate,
 they diffuse until they reach an adequate crystallographic site where
they can be incorporated (see Fig.~\ref{fig1gg}). 
One can distinguish two main cases. 
i)
On singular surfaces, atoms or molecules can
 meet each other and aggregate to form two-dimensional
 nuclei. Then, these two-dimensional 
 nuclei grow by incorporating further atoms or molecules and finally
coalesce to form new epilayers. This growth mode 
called `two-dimensional nucleation growth mode'
 can be monitored in-situ on an atomic scale by Reflection High
 Energy Electron Diffraction (RHEED) intensity oscillations. A RHEED
intensity maximum
 corresponds to a single atomic or molecular layer 
completion because it is the surface roughness minimum. 
ii)
On vicinal surfaces (see definition in chapter \ref{PP2}), this growth
mode can  only exist 
 if the terraces are large enough 
with respect 
 to the diffusion length $\ell_D$ (see App.~\ref{ell_D}).
Otherwise, atoms and molecules are incorporated at existing steps,
whose forward motion results in the growth of the crystal. This
second growth mode is called `step flow'. In both cases, most
 of the  incorporation sites are supplied by steps
which result either from two-dimensional  nuclei edges or from vicinal
surface steps and their number and distribution  
are  important features of the growth front.

MBE growth can be influenced  by many parameters, e.g. 
the incident beam rates which determine the relative
concentration of adatoms 
at the growth front, the adatom re-evaporation rates,
 the substrate temperature which 
 determines the adatom migration velocity
 but also the surface chemical and/or structural properties 
such as roughness, composition or surface reconstruction. 


\subsection{MBE of metals}

The case of metals is conceptually the simplest. The naive representation
of atoms as hard spheres is not too bad, and the surface is in
many cases not reconstructed. 

Thin metallic layers and multilayers  of pure elements or alloys are
interesting in several respects: magnetic properties,
mechanical and elastic anomalous 
behaviour~\cite{Jank93}, enhanced
catalytic properties~\cite{Herm96}. A great deal of attention has been
devoted in recent years to magnetic properties of metallic thin layers
and multilayers. Among them, Giant Magneto-Resistance 
and
Perpendicular Magnetic Anisotropy 
may lead to technological
improvement of the magnetic recording techniques. 
The possibility of a Perpendicular Magnetic Anisotropy 
as a consequence of the reduced symmetry, 
was first pointed out by N\'eel.
Strong experimental evidence for this anisotropy was found by
Gradmann's group~\cite{Grad86}. In multilayers, composed of magnetic
and non-magnetic metals, interface anisotropy is expected, 
too~\cite{Broe91}. The magneto-elastic effect, which involves the same
fundamental energy, i.e. the spin-orbit coupling, may be the leading
factor~\cite{Hug96}.  Strong 
magneto-crystalline anisotropy has also been observed in
 metallic alloys layers grown by sputtering or Molecular Beam Epitaxy
 when an anisotropic structure is obtained with the easy magnetic axis
 perpendicular to the layer (hexagonal, tetragonal, etc.)~\cite{Ceb94,Geha97}.
It has also been shown that antiferromagnetic coupling
may occur between magnetic layers through non-magnetic layers, and the
magnitude of this effect oscillates with the non-magnetic layer
thickness~\cite{Park90}. This magnetic coupling gives rise to 
Giant Magneto-Resistance~\cite{Baib88}. 
The occurrence and magnitude of these properties depend
dramatically on the quality (roughness, etc.) of the layer.
The instability of the growing front may alter this quality.


\subsection{III-V compounds}

In the remainder of this chapter,  attention will
be focussed on III-V semiconductors which have been most extensively
studied because of their technological interest.
In this  case, the substrate temperature 
and the V-III ratio are
chosen in such a way that 
the growth rate is controlled by the group-III element 
incorporation rate (close to one), the group-V element being
incorporated
 only if chemically reacting with group III-elements. Thus, increasing
the group-III element 
deposition rate increases the growth 
velocity.
On the other hand
an increase of the  V-element  beam rate decreases the surface 
diffusion constant.
In this case,  the choice  of a particular V/III ratio 
associated with a particular
 substrate temperature corresponds to the choice of
particular growth kinetics. Since
the group V element evaporates when it is not at once incorporated,
the MBE growth of III-V compounds can be described
 with a reasonable 
approximation by  models in which a atoms of a single 
species are considered~\cite{hand,ES-GaAs,ME_Z,nucl_Vv}. This species is
the 
group III element, whose deposition and diffusion on the surface
determines
the growth rate and the type of growth. Such models will be used in the
following 
chapters and are applicable to GaAs, for instance, if the word `atom' is
everywhere replaced by `Ga atom'.  However, one has to be  careful 
when using this approximation. For instance, as noted by Tersoff
et al.~\cite{teretal} the temperature dependence of the
Ga adatom density $\rho_0$
at equilibrium is not given, as it would be in a one-component system, 
by 

\begin{equation}
 \rho_0= a^{-2}e^{-\beta W_0}
\label{rho0}
\end{equation}
where $a^2$ is the area per atom and $W_0$ a constant. The reason is the
following: terraces contain an equal number of Ga and As atoms,
while Ga adatoms move presumably alone, without the company of an As
atom. Therefore, if a Ga
adatom is released from a terrace, a consequence
is (apart from the energy cost) the release of an As atom from the same 
terrace into the vapour. Therefore, there is an entropy gain which
modifies 
$W_0$ by a temperature-dependent amount. 
Treating the As vapour as an ideal gas, 
Tersoff et al.~\cite{teretal} obtain
\begin{equation}
 \rho_0= a^{-2}(P/P_0)^{-1/m}e^{-\beta W'_0}
\label{rho0'}
\end{equation}
where $W'_0$ is a constant, $P$ is the As pressure, 
$m=2$ for the diatomic vapour As$_2$, and~\cite{JVAP}
\begin{equation}
P_0(T) = \sqrt{\frac{M^3(k_BT)^5}{2\pi^3\hbar^6}}
\label{P_0}
\end{equation}
where $M$ is the mass of the molecules of the vapour (150 times the
proton mass 
in the case of As$_2$). 
Thus, if (\ref{rho0'}) is replaced by (\ref{rho0}) in a single
component
model, $W_0$ depends on temperature. For the values used by Tersoff et
al. 
($T=850$ K and $P=10^{-4}$ Pascals) formula
(\ref{P_0}) yields $(P/P_0)^{-1/m} \approx 10^{9}$, a remarkably large 
value. If, instead of being at equilibrium, the system is subject to MBE 
growth, an effective pressure can be defined~\cite{teretal} as
``that pressure which would give the same As desorption rate in
equilibrium
as actually occurs''. Experimentally, Tersoff et al. find a factor 
$ 10^{15}$ unfortunately far from the above mentioned factor $10^{9}$.


\subsection{MBE, why?}

The merit of MBE is the possibility to prepare
a wide range of III-V  semiconductor materials\footnote{II-VI
semiconductors can also be prepared and are currently studied in 
laboratories, but their industrial use is not so wide as that of III-V
materials.},
with tailor-made lattice constants or electronic  properties. 
For instance, in a ternary compound A$_{x}$B$_{1-x}$C, the choice of $x$
allows for tuning the lattice constant so as to make possible epitaxy
on a given substrate. Instead of the lattice constant, one can tune the 
electronic band-gap in order to obtain the desired optic or electronic 
properties. Examples are provided by the ternary compounds with A=As,
B=P and C=Ga.
Pure  GaAs ($x=1$) 
has a direct band gap in the infrared, while GaP  an indirect band gap
in the green, and their mixture  GaAs$_{0.6}$P$_{0.4}$ has a direct
 band gap emitting in the red.
 This film growth technique provides thus a flexible and
 powerful tool to select a particular wavelength by choosing the alloy
 composition (see Fig.~\ref{fig2gg}) and/or to engineer the band
structure 
by growing layers of predetermined and well controlled alloy
composition.

These materials (binary compounds or alloys) can 
be combined  in single or multilayer `heterostructures'
made of different semi-conductors which
have different electronic properties (e.g. band gap energy) 
but closely matching lattice parameters.

The simplest heterostructures are quantum wells, 
which are used in optoelectronics 
and constitute the active zone of optoelectronical devices (lasers or
optical amplifiers). In a quantum well,
the charge carriers (electrons and holes)
are confined between two planes  (see Fig.~\ref{fig5gg}a). 
Examples of materials appropriate to make a quantum well are 
Ga$_{1-x}$In$_x$As and Al$_{1-x}$In$_x$As. These two
 alloys are lattice-matched on 
an InP substrate for $x=0.53$ and $x=0.52$,
 respectively, but their band gap energy varies from 0.76eV 
(1.65$\mu$m) for GaInAs to 1.46eV (0.85$\mu$m) for AlInAs. 

These quantum wells themselves can be organized in complex
superlattices. 
It should be noted that band discontinuity resulting from the alignment 
of different band structures can induce another kind of interesting
carrier  confinement:  the junction often includes a step and a notch  
in the valence or the conduction bands. A typical example is the
heretojunction between highly n-type AlGaAs and lightly doped GaAs
(Fig.~\ref{fig5gg}b). For most 
technological applications, the important point is to make perfect
interfaces between materials 
free of structural or chemical defects. However,
 it is actually  a difficult task because of the surface 
segregation~\cite{MGHBV} of the binary
 with the lower surface energy which acts against the chemical
abruptness of the interface.
 
\subsection{Instabilities}

MBE allows to make chemically unstable alloys.
This challenge to chemistry is not free from any danger.
The most usual danger is a `chemical instability', i.e.
composition modulations (alternating AC rich and BC rich domains) 
appear during the growth, either
parallel or perpendicular to the growth plane. As an example,  
Fig.~\ref{fig3gg} shows the surface of unstable alloys of
composition close to Al$_{0.5}$In$_{0.5}$As, 
which are unstable and tend to decompose
into two phases of composition close to AlAs and InAs.

 From a technological point of view,
 a chemical instability leads to  a 
non-uniformity of the band-gap and thus induces an undesirable
broadening of
 the photoluminescence line width and a lowering of the electron mobility.
 But, on the other hand, the  chemical instability
of the alloy can be used to obtain 
spontaneous nanoscale strained structures (wires or dots) if correctly
controlled. The deviation from randomness can also take the form of an 
ordered phase. Both  composition modulation (Fig.~\ref{fig3gg}) 
and long range ordering (Fig.~\ref{fig4gg}), 
have been observed
and sometimes they coexist in the same sample.

Once it has been grown, the bulk material can survive during centuries
at room temperature, because bulk diffusion is very slow. 
In contrast, surface diffusion is pretty fast at the growth temperature,
and must be so to ensure a good crystal quality.

Even in the absence of any chemical instability, a multilayer or a double
layer
can be unstable if the lattice constants of the pure materials are too 
different. 
When a semi-conductor is grown lattice-mismatched with the substrate, a
thin
 pseudomorphic layer is expected to grow first uniformly and
commensurably with 
the strain energy increasing linearly with thickness up to a critical
threshold.
 Below this critical threshold which drastically depends on the growth
conditions, a compressed or stretched  pseudomorphic layer is obtained, 
with a modified band structure 
which can be  useful for a wide range of practical 
applications. Beyond the threshold, dislocations or other types of 
instabilities usually appear.

While chemical instabilities will not be addressed in the
remainder of this review, elastic instabilities will be studied in 
detail in chapters \ref{Marty1} and following.


\section{The Ehrlich-Schwoebel instability on a high symmetry surface}
\label{PP1}
 
\subsection{Introduction}
\label{ES_int}

In the present chapter and in the next one we will give a description 
of kinetic instabilities
occurring in homoepitaxial growth, due to the so-called
Ehrlich-Schwoebel~(ES)
effect which makes the sticking process of an adatom to a step
asymmetric.
This effect was discovered thirty years ago by Ehrlich and
Hudda~\cite{EH},
who observed $-$through a field ion microscopy technique$-$ that
tungsten atoms 
diffusing on a terrace were repelled by a descending step.
It was shown by Schwoebel and Shipsey~\cite{schwoebel} that this
effect is stabilizing on a vicinal surface, while its destabilizing 
character for a high symmetry surface was pointed out by Villain only 
a few years ago~\cite{VilJdP}.

For a quantitative analysis of this mechanism, microscopic details
are extremely important: lattice structure, surface orientation, step
orientation, sticking process. For example, Stoltze~\cite{Stolze}
has performed a detailed study of the surface energetics of several fcc
metals, based on the effective medium theory. 
According to this work,
an ES barrier exists for any surface and step orientation. The extra
energy barrier is of the order of 100/300~meV (a fraction of the
diffusion
barrier on a flat surface) and it depends on the sticking process,
which may take place through a hopping over the edge or through an
exchange mechanism with an edge atom. The former is almost always
favorable
for the (100) and (111) surface orientations, while the latter prevails
in several cases for the (110) orientation.

{\it Ab initio} calculations show that the `real' picture may be even
more
complicated. For Al(111) $-$which has an extremely low diffusion
barrier: 40~meV$-$ Stumpf and Scheffler~\cite{stumpf} have shown that
the
exchange process is favorable, but the extra energy barrier is very
small,
and perhaps smaller than the `long-range' attraction energy between
step and adatom. For Au(111)~\cite{Ferrando} the static barriers
(at $T=0$K) are fairly large and almost equal for the two mechanisms,
but
simulations at $T>450$K show that exchange does not require an
extra energy.

The experimental picture  is not universal. Field ion microscopy
studies by Ehrlich and collaborators (see the paper~\cite{KE} by Kyuno
and Ehrlich and references therein) show that an
adatom approaching a descending step may feel a step-edge trap rather
than
a step-edge barrier, which may reduce the diffusion bias 
(the up-hill current induced by the ES effect: see below). 

Several authors~\cite{seb} have proposed and/or performed an 
indirect evaluation of the additional step-edge barrier through the 
study of the nucleation rate on top of islands. 
The method is applied to metal epitaxy (Fe, Ag, Pt) and the results 
generally attest the existence of such a barrier.
For instance, Morgenstern et al.~\cite{morgen} find 
a Schwoebel barrier equal to 0.13 $\pm$ 0.04 eV for Ag(111).

In contrast,  the existence and the importance of step-edge barriers in
semiconductor materials is a much more debated question.
These barriers have been invoked to explain the roughening of Si(100)~%
\cite{EagGil} before the crystal-amorphous transition;
on the other side a recent and detailed study~\cite{ES-Si} of the growth
of Si(111) shows a morphology which does not agree with the unstable
evolution due to ES barriers.
Nevertheless, some calculations~\cite{dassarma} based on empirical
potentials~\cite{desjon} do give a finite
barrier even for the (111) orientation.

A surface evolution in agreement with step-edge barriers  is found for 
Ge(001)~\cite{Nostrand,ES-Ge}, even if
the barriers are evaluated to be fairly weak.
Also for GaAs there is some evidence~\cite{ES-GaAs,Orme,Joh} supporting
their existence.

The efficiency of the ES barrier clearly depends on the temperature:
the ES instability $-$which hinders the attainment of the equilibrium$-$
manifests itself at low $T$.

The name Ehrlich-Schwoebel instability is generally used in relation
to growth of a high symmetry surface: in this case it gives rise to
the formation of three-dimensional structures rather than to a 
layer-by-layer growth. 
This instability is studied in the present chapter.

On a vicinal surface, where
growth proceeds via step flow, the same microscopic mechanism
(the ES effect) determines meandering of the steps, usually
referred to as Bales-Zangwill~(BZ) instability~\cite{BZ}, 
which resembles the
fingering instabilities discussed in chapter~\ref{snow}. 
It will be studied in the next chapter.

\subsection{Continuum description of the surface}
\label{c_d}

The evolution of the surface shape may be attributed to two types
of mechanisms. i) Deterministic ones, which tend for instance to
reduce the free energy and to restore thermal equilibrium.
ii) Stochastic processes related to the randomness of the deposition, of
atom
diffusion, of nucleation\,\dots~. The stochastic part is usually called
`noise',
in analogy with noise in electric circuits for instance, which is caused
by a stochastic motion of electrons.
Thus, if $z(\vec x,t)$ denotes the local height of the surface profile
in
$\vec x=(x,y)$,
its temporal variation is expected to satisfy an equation of the form:
\be
\partial_t z(\vec x,t)= {\cal A}(\{z\}) + \hbox{noise}
\label{eqA}
\ee
where ${\cal A}(\{z\})$ is a deterministic functional of $z(\vec x,t)$
and $\partial_t\equiv\partial/\partial t$.

If volume diffusion is neglected, the dynamics of the surface is
governed by
three microscopic processes: deposition, surface diffusion and
desorption. By definition, surface diffusion conserves matter on the
surface;
deposition from the vapour phase is the source of new matter, but
generally
it is a process which does not depend on the surface profile: 
it contributes to the local velocity $\partial_t z$ through
a term $F(\vec x,t)=F_0 +\delta F(\vec x,t)$. We can get rid of the
average value $F_0$ by simply redefining the local height:
$z(\vec x,t)\to (z(\vec x,t)-F_0t)$, while $\delta F(\vec x,t)$
contributes to the noise. 
In some cases the deposition rate does depend on the surface profile,
for example when the average flux impinges sideways upon the surface. 
This may give rise to shadowing instabilities, which will be discussed
in
Chap.~\ref{shadow}. Another possibility is a rate of attachment of the
impinging atom which depends on the local environment.

Finally, desorption can be important, especially in
semiconductors, whose typical growth temperatures are higher than for
metals. The description of a high symmetry surface will be worked out 
mainly under assumption that evaporation is negligible.

Once that matter is conserved on the surface, the next step is to
assume that overhangs, voids and interstitials are not relevant,
so that volume as well is conserved.
If it is so, the functional ${\cal A}(\{z\})$ must be the divergence of
the surface current and therefore: 
\be
\partial_t z(\vec x,t)=-\nabla\cdot\vec j + \hbox{noise}
\label{Lan_eq}
\ee
where the current $\vec j$ is a functional of the surface profile 
$z(\vec x,t)$, similarly to ${\cal A}(\{z\})$ in Eq.~(\ref{eqA}). 
An important question is whether $\vec j$ is a local or nonlocal
functional.
In other words, is it sufficient to know the surface profile in $\vec x$
($z(\vec x,t),\nabla z,\nabla^2 z,\dots$) to determine the growth
velocity 
$\partial_t z$, or is it determined by the surface profile in
a finite space around $\vec x$~?
In principle, epitaxial growth is not a local process, mainly because
diffusion is not. In this sense, Eq.~(\ref{Lan_eq}) must be understood
as an
equation which is valid on large space and time scales: space scales
larger than the typical distance between steps and time scales larger
than the filling time of one monolayer ($=1/F_0$).
Nevertheless, even so a nonlocal description may be necessary: we will
come across one case later on.

Because of the translational invariance in the $z$-direction, the
current
$\vec j$ as well as the quantity ${\cal A}$ in the general case of non
conserved volume, does not depend explicitly on $z(\vec x,t)$, but only
on its derivatives. So, it may be useful to think of $\vec j$ (or ${\cal
A}$)
as a functional of the local slope $\vec m=\nabla z$.
By taking the gradient of (\ref{Lan_eq}), we obtain:
\bea
\partial_t \vec m &=& -\nabla (\nabla\cdot\vec j)
+\nabla(\hbox{noise})\\
&=& -\nabla^2\vec j - \nabla\wedge(\nabla\wedge\vec j)
+\nabla(\hbox{noise})
\label{eq_m_2}
\eea

In 1+1 dimensions the current and the slopes are scalar quantities and
therefore the curl term in the right-hand-side disappears
$(\partial_x\equiv\partial/\partial x)$:
\be
\partial_t m(x,t) = - \partial_x^2 j + \partial_x(\hbox{noise})
\;~~~(1+1~\hbox{dim})
\label{eq_m_1}
\ee

If the current is derivable from some effective free energy ${\cal F}$
\be
j(x,t) = - {\delta {\cal F}\over\delta m}
\ee
the evolution equation for $m$ may take or not the form of the 
Cahn-Hilliard~\cite{CH} equation, depending on ${\cal F}$.

In 2+1 dimensions the curl term vanishes if the current is
nonrotational: $\nabla\wedge\vec j=0$. 
Since $\vec j$ is a functional of $\vec m$ and $\nabla\wedge\vec m=
\nabla\wedge (\nabla z) \equiv 0$,
the current is expected to be nonrotational only if it is linear in
the slope, but this is generally true only in the earliest stages of
growth, when $|\vec m|\ll 1$. Therefore $\nabla\wedge\vec j\ne 0$
as a rule.
All the same, the evolution of the surface may have many similarities
with  a phase separation process (see chapter~\ref{2dm}).

\subsection{Different sources of noise}
\label{sss1}

Till now the possible sources of noise have not been indicated
explicitly.
In Molecular Beam Epitaxy there are three: i)~Fluctuations in the
incoming flux (the so-called shot noise); ii)~Fluctuations in the
diffusion
current (conserved noise); 
iii)~Fluctuations in the nucleation process of new islands.

The first two have been well studied in the context of several different
models~\cite{JVAP,book_BS}:
generally speaking, their effect is to roughen a surface which would be
flat in the absence of noise. Here the typical situation we want to
study is
different: the surface is deterministically unstable and therefore the
`deterministic roughness' may overcome the stochastic one, induced by
noise.
Furthermore, if the dynamics resembles a phase separation process,
noise is known to be an irrelevant parameter, at least in
2+1 dimensions.
A second remark on shot noise is in order: roughness induced by shot
noise is
in part healed by surface diffusion. It is therefore important to 
understand on which time and length scales roughness is relevant.
Here we report a qualitative argument due to Villain~\cite{VilJdP},
which
will also be used in the context of unstable growth of a high symmetry
surface (see App.~\ref{app-coarsening}).

The average number of atoms falling on a region of size $\ell$ during
time $t$
is: $\overline{N}=F_0 t\ell^d$ 
and its fluctuation due to a Gaussian shot noise is:
$\Delta N\approx \sqrt{\overline{N}}=\sqrt{F_0 t\ell^d}$. 
Such `volume' fluctuations determine  
a `height' fluctuation $\Delta h=\Delta N/\ell^d\approx
\sqrt{\bar h/\ell^d}$, where $\bar h=F_0t$ is the average number of
deposited layers. The central question is: What is $\ell$~?
In the context of a surface which is deterministically stable, $\ell$ is
the diffusion length $\ell_D$ (see App.~\ref{ell_D}): 
diffusion is supposed to heal shot noise
just on this scale. If $\ell_D\approx 10^2\div 10^3$, in two dimensions
we will have $\Delta h\approx 1$ after the deposition of 
$10^4\div 10^6$ layers, which is quite a large value.

The contributions of shot noise and diffusion noise to
Eq.~(\ref{Lan_eq})
are generally written as $\delta F(\vec x,t)$ and
$\eta_D(\vec x,t)$ respectively, with:

\bea
\langle \delta F(\vec x,t)\rangle =0~~&&~~
\langle \delta F(\vec x,t)\delta F(\vec x',t')\rangle =
2F_0\delta(\vec x-\vec x')\delta(t-t')
\label{bruit}\\
\langle \eta_D(\vec x,t)\rangle =0~~&&~~
\langle \eta_D(\vec x,t)\eta_D(\vec x',t')\rangle =
-2D\nabla^2\delta(\vec x-\vec x')\delta(t-t')
\eea
where $D$ is the surface diffusion constant.

Concerning nucleation, it is due to an encounter of two (or more) 
adatoms during the
diffusion process. So, we might ask why nucleation noise is something
different from diffusion noise. It is indeed possible to separate the
two contributions in the following way: on a region of high slope,
nucleation is almost negligible (see App.~\ref{ell_D}), 
but diffusion noise still exists; 
conversely, in the continuum limit $a\to 0$ we will see that diffusion
noise is negligible while nucleation noise survives.

A further feature which characterizes nucleation noise with respect to
shot and diffusion noise is that for the latter ones it is clear the
meaning of `absence of noise'. In other words, it is obvious which is
the corresponding deterministic model: 
$\delta F(\vec x,t)\equiv 0$ and $\eta_D(\vec x,t)\equiv 0$.
But which is
the deterministic model of nucleation? We will show that even if a model
can be proposed, it may give rise to a nonanalytic profile in the
nucleation sites.

An important question is whether nucleation noise is enough to roughen 
the surface by itself. The answer, according to the works by Elkinani
and  Villain~\cite{EV} and Wolf et al.~\cite{noise_Wolf}, is `no'. 
Even more important, there is numerical evidence~\cite{PV} that 
nucleation noise is able to heal surface defects at small length scales.
However, a deeper analytical comprehension of nucleation noise
would be necessary. Preliminary studies will be found  in  
Refs.~\cite{noise_Wolf,Kallabis}.

\subsection{The surface current $\vec j$}
\label{the_current}

This chapter addresses a problem which is common to several branches of
physics:
How to derive a continuum equation, valid at large length scales,
starting
from a microscopic and discrete point of view? Here the situation is 
particularly difficult because the surface is made up of two
different microscopic objects $-$steps and adatoms$-$ whose
characteristic
time scales are fairly different: if $\tau_{ad}$ is the typical
diffusion
time to a neighbouring site and $\tau_{step}$ is the time 
necessary to advance the
step of one lattice constant, it is easily found that $\tau_{ad}\approx
a^2/D$
and $\tau_{step}\approx 1/(F_0\ell a^{d-1})$, 
where $\ell$ is the typical distance
between steps. The condition $\tau_{step}\gg\tau_{ad}$ reads (we put
$a=1$)
$\ell\ll(D/F_0)=\ell_D^{1/\delta}$, with $1/\delta > 2$
(see App.~\ref{ell_D}).  Since $\ell$ must be smaller 
than $\ell_D$ and $\ell_D\gg 1$, the above condition is certainly
fulfilled.
Therefore adatoms see stationary steps during their motion and steps
feel adatoms only as an `average' current: 
the Burton, Cabrera and Frank theory~\cite{BCF} of crystal growth is
indeed
based on this quasi-stationary hypothesis.

A possible approach to the problem of determining a continuum
Langevin equation is to start from the microscopic rates for
the single processes (deposition, diffusion, evaporation~...) in a 
solid-on-solid model and write down a master equation for the discrete
variables $h_i$ (the height of the $i$-th column in the SOS model).
This procedure generally requires the truncation of an infinite
hierarchy of
equations to arrive to a set of discrete Langevin equations, one for
each $h_i$; afterwards, it is necessary to find a continuum limit in the
space
variables and this requires a regularization scheme for step functions.
This method has been used by several authors: Plischke et
al.~\cite{ME_P}
for Metropolis dynamics, Zangwill et al.~\cite{ME_Z}, Vvedensky et
al.~\cite{ME_V} and more recently by
P\u redota and Kotrla~\cite{ME_K} for Arrhenius dynamics.

As pointed out by the same authors this approach suffers from
important problems, the major one regarding the regularization scheme
for the step function: no rigorous procedure exists, different schemes
give different results and in the final results some coefficients
entering
in the regularization procedure which should depend on the flux $F_0$
remain undetermined. A critical discussion of the master equation
approach
in the case of equilibrium relaxational dynamics can be found in
Ref.~\cite{KDM}.

A completely different approach
uses the method of Burton, Cabrera and Frank~\cite{BCF}: 
one  writes 
the  diffusion equation for
adatoms on a given terrace and solves it with suitable boundary
conditions at steps: this 
gives rise to a current of adatoms, from which
step velocities can be evaluated. 
Once they are known, one can study numerically
the evolution of the surface or one can try to obtain a continuum
evolution equation through some `coarse graining' procedure. This method
is
generally restricted to sufficiently large slopes: 
in fact it is not easy to introduce nucleation in this kind of approach.

Instead of giving further details on the different methods which can be
employed to obtain some expression for $\vec j$, we prefer to introduce
the different terms phenomenologically and justify them.

\subsubsection{The Ehrlich-Schwoebel current}
\label{ch4D1}

The first place clearly deserves the term which is responsible for
the instability treated in the present chapter: the current induced by
the
Ehrlich-Schwoebel effect. 
Let us imagine a surface which has a pyramidal structure and therefore
top,
bottom and vicinal terraces (see Fig.~\ref{tbv}).

Such a `three levels' structure exists even in the {\it absence} of any
step-edge barrier (see, for example, the simulations reported in Fig.~7a
of Ref.~\cite{EV}). If an ES barrier exists, it has a double effect:
i)~on a top terrace, since atoms are hindered from descending, the
probability of nucleation is increased and this makes pyramids higher;
ii)~on a vicinal terrace, adatoms prefer sticking to the ascending step
and therefore existing holes are not filled.  Both effects destabilize 
the flat surface, giving rise to  three-dimensional growth. From a 
continuum point of view, 
the above mechanisms give rise to a nonequilibrium current
$\vec j_{ES}$ which depends on the slope of the surface. This is of
extreme
importance, because all the other terms in the current will be seen to
vanish on a surface of constant slope.

Let us start by considering a 1+1 dimensional surface, which is supposed
to have a positive slope. 
On a terrace of size $\ell$ we solve the diffusion equation: 
$\partial_t\rho=F_0 +D\partial_x^2\rho$
in the quasi-static approximation ($\partial_t\rho=0$),
which is justified by the relation $\tau_{step}\gg\tau_{ad}$. This gives
$\rho(x)=A+Bx-(F_0/2D)x^2$, where constants $A$ and $B$ are determined
through
boundary conditions at steps in $x=\pm\ell/2$. Here, the sticking
coefficients
$k_+$ and $k_-$ (respectively from the upper and lower terrace) are
introduced: an infinite coefficient means that the adatom is
automatically
incorporated as it reaches the step, while a vanishing one means the
adatom is unable to stick. The existence of a current $\vec j_{ES}$ is
due to
an asymmetry between $k_+$ and $k_-$: in fact, such an 
asymmetry makes asymmetric
the density of adatoms with respect to the center $x=0$ of the terrace,
i.e. $B\ne 0$. Since the microscopic current on the terrace is:
$j_{mic}=-D\partial_x\rho=-D[B-(F_0/D)x]$, its average value is simply:
$j_{ES}=\langle j_{mic}\rangle_{ter}=-DB$. 

If the equilibrium density is neglected with respect to the density due
to the external flux, boundary conditions read:
$D\rho'(\pm\ell/2)=\pm k_\mp\rho(\pm\ell/2)$.
In the simplifying hypothesis that $k_-=\infty$ we can introduce a
single
parameter, the so-called ES length: $\ell_{ES}=D/k_+$, and the boundary
conditions read $\rho(\ell/2)=0$ for the ascending step and
$\rho'(-\ell/2)=-\rho(-\ell/2)/\ell_{ES}$ for the descending one.
Under these conditions it is found that 
$B=-F_0\ell_{ES}\ell/[2D(\ell_{ES}+\ell)]$ and therefore\,%
\footnote{In this chapter 
and in the following, $z$ is  the height divided by the monolayer
thickness
$c$, so that the slope $m$ has the dimension of the reciprocal of a
length.}
\be
j_{ES}={F_0\ell_{ES}m\over 2(1+|m|\es)|m|}
\label{j_es_2}
\ee

The previous relation can be easily read in the two limiting cases of
a strong ES effect ($\es\gg\ell$) and a weak one ($\es\ll\ell$).
In the former case all the atoms falling on the terrace contribute to
the
current and therefore $j_{ES}$ is proportional to $\ell$ and independent
of
$\es$; in the latter one, only the atoms falling within 
a distance $\es$ from the
ascending step contribute to the current and $j_{ES}$ is now
proportional to
$\es$.

Eq.~(\ref{j_es_2}) is clearly not applicable in the limit $m\to 0$,
because the 
relation $|m|=1/\ell$ is valid as far as no nucleation takes place on
the
terrace itself. The simplest way to obtain the correct expression for
any
value of $m$ is to observe that the current must vanish on the high
symmetry orientation $m=0$; so, it is expected to be linear at small
slopes:
$j_{ES}=\alpha m$ and the constant $\alpha$ is determined by matching
the
two expressions for $|m|\simeq 1/\ell_D$. This gives 
$\alpha =F_0\es\ell_D/[2(1+\es/\ell_D)]$. 
The simplest function interpolating between
the two limiting expressions is:
\be
j_{ES} = {F_0\es\ell_D m\over
2(1+\es/\ell_D +|m|\es)(1+|m|\ell_D)}
\label{j_es}
\ee
which reads, for weak and strong step-edge barriers:
\bea
j_{ES} &=& {F_0\es\ell_D m\over
2(1+|m|\es)(1+|m|\ell_D)}~~~~\es\ll\ell_D\\
j_{ES} &=& {F_0 \ell_D^2 m\over
2(1+|m|\ell_D)^2}~~~~~~~~~~~~\es\gg\ell_D
\eea

A more rigorous derivation of $j_{ES}$ at small slopes is given in
Ref.~\cite{PV}: for $|m|<1/\ell_D$ the surface is made up of a
`sequence' of top, bottom and vicinal terraces of typical width $\ell_D$,
whose average slope is just $m$: since vicinal terraces are the only
ones to contribute both to the (average) slope and 
to the ES current, and both
quantities are linear in the number of vicinal terraces in the
`sequence' under study, we  conclude that the ES current and
the slope must be proportional. 
This method  gives the same expression as before for
the constant $\alpha$.

Before going on, let us comment on different methods for determining
$j_{ES}$. In the spirit of the procedure just explained, we could
modify the starting differential equation for the density $\rho$ in
order
to take into account nucleation at {\it that} level. This procedure has
been
used by Myers-Beaghton and Vvedensky~\cite{nucl_Vv}, who have introduced
a term proportional to $\rho^2$ to simulate the formation of a dimer. In
this
case the diffusion equation reads: $F_0 + D\rho''(x)
-\varsigma\rho^2(x)=0$.
This equation is not solvable in closed form, but in the limiting cases
$\ell\to 0$ and $\ell\to\infty$ it displays~\cite{rev_Krug} the same
behaviour as before. In the limit $\ell\to 0$ this is obvious, since in
such
a limit the quadratic term is negligible. 

We want to emphasize that two
different lengths are relevant for the problem: the ES length
$\ell_{ES}$
and the diffusion length $\ell_D$. 
The latter one is generally defined as the typical distance between
nucleation centers on a high symmetry surface (see App.~\ref{ell_D}), 
and the value of $\es$
does not enter in $\ell_D$. Nevertheless, when nucleation can take place
on vicinal and/or top terraces, the ES barrier affects the adatom
density and therefore the typical size of a vicinal or top terrace.
It is possible~\cite{EV,Psolo} to introduce a `nucleation' length
$\ell_n$
for a vicinal terrace ($\ell_n^V$) and for a top one ($\ell_n^T$)
(for a bottom one we simply have $\ell_n^B=\ell_D$): a new terrace
nucleates on an old one when $\ell$ becomes of the order of $\ell_n$.
When $\es=0$, $\ell_n^V=\ell_n^T=\ell_D$ and for a finite ES effect the
nucleation lengths decrease (because the adatom density increases):
$\ell_n^V$ has a weak dependence on $\es$ and $\ell_n^V(\es=\infty)=
\ell_D/\sqrt{2}$~\cite{Psolo}, while $\ell_n^T$ goes to zero when
$\es\to\infty$ (see Fig.~\ref{ln-fig}). 
Strictly speaking, the quantity $\ell_D$ in Eq.~(\ref{j_es})
should be replaced by $\ell_n^V$, but we have just seen that
$\ell_n^V\simeq\ell_D$ for any value of $\es$. Because of this,
the nucleation term (proportional to $\rho^2(x)$) in the diffusion
equation
can be neglected if $\ell\ll\ell_D$ whatever is the value of $\es$:
in this limit, Eq.~(\ref{j_es_2}) should be recovered.

Finally, we will mention the previously discussed `microscopic' approach
which makes use of the master equation: in the presence of step-edge
barriers
it gives~\cite{ME_V} an ES current proportional to the slope, but no
expression beyond the linear approximation is known so far. Nevertheless
the 
expression for the proportionality coefficient $\alpha$ found
by Vvedensky et al.~\cite{ME_V} has 
the merit to clearly show that the ES current vanishes 
when detailed balance is satisfied. Our expression~(\ref{j_es})
trivially
satisfies the detailed balance: since thermal detachment from step has
been
neglected there, it is sufficient that $j_{ES}$ vanishes for $F_0=0$.

The expression we have found for $j_{ES}$ has two main limits:
it is valid in 1+1 dimensions and it does not take into account possible
discrete effects in the $z$ direction. For example, if we are still in
one
dimension and $m=0$ means the high symmetry (10) orientation of a square
lattice, then $m=1/a$ necessarily means the (11) orientation, which is
also
of high symmetry. Does the ES current vanish as well on this
orientation?
If so, Eq.~(\ref{j_es}) is wrong because it  only vanishes for 
$m=0,\pm\infty$. The fact that $j_{ES}(\pm 1/a)$ should vanish is 
not obvious since flux could break the symmetry.%
\,\footnote{Siegert and Plischke (see note [11] of
Ref.~\protect\cite{SP})
have pointed out that it should not be relevant.
Indeed, the problem is equivalent to have an incoming
flux forming an angle of $45^\circ$ with the surface (10): the ES
current
vanishes anyhow, since shadowing effects are completely negligible.}

Concerning  real, 2+1 dimensional surfaces, we can observe
that here as well as for other terms in the current we must accept some
reasonable generalizations of the expression found in 1+1 dimensions
and nothing more. Microscopic approaches can not help here, because
they give expressions at the lowest order (for example, the linear one
for
the ES current), in which case the generalization is generally trivial.

Nevertheless, since the evolution of the surface depends only on some
very
general features of the current, we should not be worried too much about
this.
We will see in the following that the only relevant features of 
$\vec j_{ES}$ are: i)~A linear behaviour at small slopes:
$\vec j_{ES}=\alpha\vec m$, with a positive $\alpha$, 
so that the surface will 
be linearly unstable. ii)~The existence or the absence of finite zeros
in the
current. iii)~Possibly, the behaviour of the current at large slopes, if
such zeros do not exist.
iv)~The in-plane symmetry of the current.

The simplest expression for the ES current in the absence of the above
mentioned
zeros is the one proposed by Johnson et al.~\cite{Joh}:
\be
\vec j_{ES} = {\alpha\vec m\over 1+m^2\ell_D^2}
\label{j_Joh}
\ee
We observe that this expression may be valid only for $\es\gg\ell_D$.
To take into account discrete effects in the $z$-direction and therefore
possible zeros in the current, Siegert and Plischke~\cite{SP} have
replaced $\vec m$ with $\vec m/(1-m^2a^2)$: this corresponds to the
substitution of $\tan\theta\equiv ma$ by $(1/2)\tan 2\theta=\tan\theta/
(1-\tan^2\theta)$,

i.e. a function of period $\pi$ by a function of
period $\pi/2$. The resulting current is~\cite{SP}:

\be
\vec j_{ES} = {\alpha\vec m(1-a^2m^2)\over
(1-a^2m^2)^2 + m^2\ell_D^2 }\equiv\vec m f(m^2)
\label{j_s_p}
\ee

The previous expression vanishes on the `circle' $|m|=1/a$: there is a
continuous in-plane symmetry. Discrete effects {\it in} the plane should
break this symmetry. For example, for a cubic lattice we could 
write~\cite{SP,S97}:
\be
(j_{ES})_{x,y} = m_{x,y} f(m^2_{x,y})
\label{j_a_p}
\ee
Now, the number of zeros is finite and equal to four:
$\vec m= (\pm 1,\pm 1)/a$.
On the basis of the `relevant features' of $\vec j_{ES}$ discussed
above,
Eqs.~(\ref{j_s_p}) and (\ref{j_a_p}) can be replaced by the following
simpler expressions:
\bea
\vec j_{ES} &=& \alpha\vec m(1-a^2m^2) ~~~~\hbox{Same qualitative
properties as
(\protect\ref{j_s_p})}\label{j_s_p_2}\\
(j_{ES})_{x,y} &=& \alpha m_{x,y} (1-a^2m^2_{x,y}) ~~~~\hbox{Same 
qualitative properties as (\protect\ref{j_a_p})}\label{j_a_p_2}
\eea
A model with $\vec j_{ES}$ given by Eq.~(\ref{j_s_p_2}) has been studied
by
Stroscio et al.~\cite{Stro}.

To conclude this part, we can 
wonder whether some other mechanisms can contribute to
a slope dependent current. It is indeed so: 
generally these mechanisms are not thermally activated and are due
$-$on one side$-$ to the release of the condensation heat when the
adatom lands on the surface, and $-$on the other hand$-$ to the fact
that
the adatom does not necessarily land on a hollow site.
These processes are called (see Ref.~\cite{Evans}) downward funneling, 
knock-out, transient mobility, and they are illustrated in 
Fig.~\ref{non_the_pro}.

Their effect is a down-hill current which depends on the density of
steps
and therefore on the slope of the surface. In the simplest picture they 
give rise to a current $j_{NT}=-\tilde\alpha m$ which must be added to
$j_{ES}$. Now, two possibilities exist: if $\tilde\alpha>j'_{ES}(0)$
the total current $(j_{NT}+j_{ES})$ is downhill and no instability
takes place; conversely, if $\tilde\alpha<j'_{ES}(0)$ the total current
is
uphill at small slopes but downhill at large slopes: therefore it
vanishes
at a finite slope $m_0$ even if $j_{ES}$ does not! For example, if
$j_{ES}$ is given by the one dimensional version of Eq.~(\ref{j_Joh})
then $m_0=\sqrt{(\alpha/\tilde\alpha)-1}/\ell_D$.

A final remark is in order: The current $j_{NT}$ and therefore
$\tilde\alpha$
depend on the crystal structure, but there is no reason to think that
the
slope $m_0$ at which it counterbalances $j_{ES}$ has some special value.

\subsubsection{Mullins-type current}
\label{Mul_cur}

Let us consider a grooved surface in the absence of any
external flux, so that it is close to equilibrium. Furthermore, we
suppose 
that the temperature is sufficiently low
so that evaporation is negligible and
the surface relaxes to equilibrium through surface diffusion.
This problem is a `classical' one (see the recent review~\cite{groove})
and for a non singular surface it has been solved forty years ago by
Mullins~\cite{Mullins}: atoms thermally detach from steps and feel a
surface chemical potential which is larger in the region of negative
curvature. This determines a current from the top to the bottom of
grooves,
which relaxes the surface. More precisely, we have a current:
$\vec j_M=-\Upsilon \nabla\mu$ where $\Upsilon$ is a kinetic
coefficient.
The expression given by Mullins is $\Upsilon=\beta\rho_0 D$ where
$\rho_0$ is the equilibrium adatom density 
and $\beta=1/(k_BT)$, where $k_B$ is the Boltzmann constant,
set equal to one in this chapter and in the following one.
The chemical potential $\mu$ is the variational derivative 
of the surface free energy\,\footnote{The correct 
definition~\protect\cite{JVAP}
is that $\mu$ is the Gibbs free energy per particle.
The pressure will be assumed to be low enough to allow the replacement
of the Gibbs free energy by the Helmholtz free energy. 
In the present chapter, the solid  is assumed incompressible
and the free energy reduces to
the surface free energy ${\cal E}$, apart from  a constant part,
so that $\mu=\delta {\cal E}/\delta z$. 
In other chapters the elastic bulk energy
will be taken into account. }:
$\mu=\delta {\cal E}/\delta z$, 
where ${\cal E}$ has the general expression:
\be
{\cal E}=   \int dxdy \varphi(z_x,z_y)~~~\hbox{with}~~~
\varphi=\sigma(\nabla z) \sqrt{1+c^2(z_x^2+ z_y^2)}
\label{muuu}
\ee
where $z_{x,y}\equiv\partial_{x,y} z$, $c$ is the out-of-plane lattice
constant and $\sigma(\nabla z)$ is the (slope-dependent) surface
tension.

In the limit of small slopes ($|\nabla z|\ll 1/a$) and for in-plane
isotropy,
$\varphi=(c^2\tilde\sigma/2)(z_x^2+ z_y^2)$ and $\mu$ is simply
proportional
to the curvature of the surface: $\mu=-a^2c^2\tilde\sigma\nabla^2 z$, 
$\tilde\sigma=(\sigma(0)+c^{-2}\sigma''(0))$ 
being the surface stiffness.  Finally, the current reads
\be
\vec j_M = K\nabla(\nabla^2 z) = K\nabla^2\vec m~~~~~~~~K\equiv
\Upsilon a^2c^2\tilde\sigma=\beta D(\rho_0 a^2)(\tilde\sigma c^2)
\label{j_k}
\ee
where we have used the fact that $\nabla\wedge\vec m=0$.

If we use $\vec j_M$ as the surface current inside 
Eq.~(\ref{Lan_eq}), we easily find that a sinusoidal profile of
wavelength $\lambda$ relaxes to the flat surface in a time of the order
of $\lambda^4$. This current, as previously mentioned, 
has a {\it stabilizing} character. 

After this digression on the current for a surface which is 
i)~non singular and ii)~close to equilibrium, we can ask
what all this has got to do with a growing surface, which also may be
a high symmetry one. A first answer is that flux is just what makes
Mullins' theory applicable even to a singular surface: in fact, the
surface
is roughened by the flux and therefore behaves as a non singular one,
even if $T$ is lower than the roughening transition temperature.
Nonetheless, the basic mechanism of Mullins' current is thermal
detachment from
steps: a process which may be hindered by the flux, if the typical time
for detachment is longer than the typical time $(1/F_0a^2)$ to fill
in one layer.
For example, Stroscio and Pierce~\cite{Str_Pie} assert that thermal
detachment is negligible at room temperature during the
homoepitaxial growth of Fe. Because of this, in Ref.~\cite{Stro} they
replace $\vec j_M$ with a current of the form
$\vec j=K\nabla^2\nabla^2\vec m$.
As we shall see, in the presence of a flux the thermal detachment
mechanism 
can be dominated by other ones.

In Ref.~\cite{Rodi} two of us have tried to study directly the growing,
far from equilibrium surface.
We started  from the assumption that the relation $\vec j_{ne}=
-\Upsilon_{ne}\nabla\mu_{ne}$ is still valid, where the subscript ${ne}$
stands for `non equilibrium'. $\mu_{ne}$ must be read as a fictitious
`step chemical potential': steps at distance $\ell$ exchange atoms
because 
of the existing gradient of $\mu$: $\delta\mu_{ne}/\ell$. 
The result is $\Upsilon_{ne}=p_0\beta a^{2-d}$, 
where $p_0$ is the emission rate for step site. The pseudo
surface free energy ${\cal E}$ giving rise to $\mu_{ne}$ is due both to
step
energy and to entropy. The result is that Eq.~(\ref{j_k}) is still 
valid at small slopes, with $K=K_{ther}$ given by~\cite{Rodi}:
\bea
K_{ther}&=&p_0a^2\ell_D~~~~~\hbox{1+1 dim}\label{K_ther}\\
K_{ther}&=&p_0\beta a^2\gamma\ell_D~~~~~\hbox{2+1 dim}\nonumber
\eea
where $\gamma$ is the step stiffness (equal to the step free energy per
unit length in the simplest case, when it is isotropic).

At larger slopes the picture is more complicated because ${\cal E}$
is no more quadratic in $\nabla z$. For example, in 1+1 dimensions
$\mu_{ne}$ is not proportional to the curvature $z''$, but rather
to $(z''/|z'|)$: the same result has been obtained by Krug et 
al.~\cite{KDM} for a solid on solid model, close to equilibrium.

The main result of this part is that a current of the form
(\ref{j_k}) is expected even if the surface is strongly out of
equilibrium,
provided that thermal detachment is allowed (i.e. $p_0\ne 0$). We want
now to 
suggest that a Mullins-like current is expected even if thermal
detachment is forbidden: in this case, $\vec j_M$ is due to nucleation
and 
diffusion noise; Let us start with the latter one. 

Randomness in
diffusion means  that the number of adatoms reaching the step of a
terrace
of size $\ell$ per unit time fluctuates around the average value (equal
to
$F_0\ell/2$ in the absence of step-edge barriers).
Nonetheless adatoms feel the curvature only close to a step and once
a freshly landed adatom has approached a step it can no more be
distinguished from a freshly detached atom from the step:
therefore the contribution of diffusion noise to the kinetic coefficient
$K$ is entirely similar to $K_{ther}$ provided that $p_0$ is replaced
by ($F_0\ell/2$):
\bea
K_{dif} &=& F_0 a^2\ell_D^2/2~~~~~~~\hbox{1+1 dim}\label{K_dif}\\
K_{dif} &=& F_0 a^3\ell_D^2\beta\gamma~~~~~~~\hbox{2+1 dim}\nonumber
\eea

The discussion of nucleation noise starts from the analysis of
Eq.~(\ref{Lan_eq}) in the linear approximation:
\be
\partial_t z = -\nabla\cdot(\alpha\vec m +K\nabla^2\vec m)
\ee
where the first term in brackets is due to the ES effect. The solution
is $z(\vec x,t)=z_0\exp(-i\vec q\cdot\vec x +\omega t)$, with
$\omega=\omega(\vec q)=\alpha q^2-Kq^4$.
The flat surface is therefore unstable ($\omega>0$) if
$q<q^*=\sqrt{\alpha/K}$
and the most unstable mode (which maximizes $\omega(q)$) corresponds to
$q_u=\sqrt{\alpha/2K}$. In other words, the Mullins-like current
stabilizes the
surface on scales smaller than $\lambda^*=2\pi/q^*=2\pi\sqrt{K/\alpha}$
and the instability develops after a finite time of the order of 
$t_u=1/\omega(q_u)\simeq K/\alpha^2$ and with a typical wavelength
$\lambda_u=2\pi/q_u=\sqrt{2}\lambda^*$.
This stabilization is evident in the homoepitaxial growth of
GaAs(001) as reported by Orme et~al.~\cite{Orme}: the substrate is
rough at small length scales ($<0.1\mu$m) and the preexisting defects
are healed before the ES instability develops on larger scales. In the
case of Ge(001) (Ref.~\cite{Nostrand}) another feature appears: at
higher
temperatures, larger thicknesses must be attained in order to observe
the
instability. This is easily explained by the expression of $t_u$: at
higher
temperatures the ES barriers are less effective and therefore $\alpha$
decreases, whilst $K$ increases (see Eqs.~(\ref{K_ther},\ref{K_dif})
and Eq.~(\ref{K_nuc}) here below).

In Ref.~\cite{PV} we have studied a one dimensional model of growing
surface where the only source of noise is nucleation of new islands and
thermal detachment from step is forbidden. In such a model, according
to the above discussion the $K$-term should be absent; nevertheless
numerical simulations clearly show that a finite $\lambda_u$ does
exist; even more importantly, $\lambda_u$ does not exist if we adopt a
(somewhat artificial) deterministic model for nucleation.\footnote{In
this 
model a new terrace is nucleated on an old one as soon as its length
equals the nucleation length $\ell_n$ (see App.~\protect\ref{ell_D}).
The artificial character of this model 
determines a nonanalytical form of the profile in the nucleation sites.}
These results prove that
nucleation noise is also a source of a Mullins-like current: Which is
the relevant value of $K=K_{nuc}$ in this case?
If $\es=0$, the only parameters entering our model are the
intensity of the flux $F_0$ and the diffusion length $\ell_D$.
Since $[K]=$length$^{4-d}$/time, dimensional analysis suggests that
\be
K_{nuc}\approx F_0\ell_D^4
\label{K_nuc}
\ee
both in 1+1 and in 2+1 dimensions.

In conclusion, we have found that in a far from equilibrium surface,
thermal detachment, diffusion and nucleation noise all contribute to
a Mullins-like current:
\be
K = K_{ther} + K_{dif} + K_{nuc}
\ee
In 2+1 dimensions, since $p_0=(D/a^2)\exp(-\beta W_a)$ where $W_a$
is the energy to transform a kink atom into an adatom, we will have
that:
\bea
K_{nuc}\gg K_{ther}~~~~~~&\hbox{if}&~~~~~~\ell_D^3(\beta\gamma a)\ll
a^3\exp(\beta W_a)\\
K_{nuc}\gg K_{dif}~~~~~~&\hbox{if}&~~~~~~\ell_D^2\gg (\beta\gamma a)a^2
\label{Knd}
\eea

Since $\beta\gamma a\approx 1$, expression (\ref{Knd}) confirms that
in the continuum limit $\ell_D\gg a$, the 
`nucleation noise' contribution to $K$ prevails over the
`diffusion noise' contribution.
Conversely, the contribution due to thermal detachment will be
negligible
at sufficiently low temperatures: $\exp(\beta W_a)\gg(\ell_D/a)^3$.

We conclude by pointing out the main limit of this part: ES barriers
have been neglected in the evaluation of the various contributions to
$K$.
The reason is that $K_{ther},K_{dif}$ and $K_{nuc}$ are present even in
the limit $\es=0$ and therefore for a weak ES effect we can neglect it.
Conversely it is not certainly guaranteed that the given expressions
for $K$ are still valid for a strong ES effect. 
For instance $K_{ther}$ must vanish in such a limit, because in general
adatoms
will stick only to the same step they have detached from and therefore
there is no exchange of atoms between steps: $K_{ther}(\es=\infty)=0$.
For the same reason we also expect that $K_{dif}(\es =\infty)=0$.

\subsubsection{Symmetry breaking current}
\label{sbcu}

According to the terms introduced up to now for the surface current
$\vec j$,
Eq.~(\ref{Lan_eq}) has two symmetry properties: in-plane reflection
symmetry
($\vec x\to -\vec x$) and up-down symmetry ($z\to -z$). 
The general validity of the former depends on the in-plane symmetry
of the surface. In the case of a three-fold symmetry, which is
pertinent for epitaxial growth of fcc(111) metals, such symmetry is
absent and the relevant current~\cite{S97} will be given in
Eq.~(\ref{3fold}).
Conversely, irrespective of crystal structure and surface
orientation, there is no basic reason to expect that the up-down
symmetry 
survives when the surface is subject to an external flux.

The simplest expression for a current which changes sign with $\vec x$
but
does not change sign with $z$ is: 
$\vec j_{SB} =\nabla G(m^2)$, where $G$ is any even
function of the slope $\vec m$ and the subscript {\footnotesize SB}
stands for
symmetry breaking. Before proceeding we can make some general 
considerations on $\vec j_{SB}$. If we consider a sinusoidal profile:
$z(\vec x) = z_0\sin(\vec q\cdot\vec x)$ it is straightforward to
write down:
\bea
\vec j_{M}  &=& -Kq^2\vec m\\
\vec j_{SB} &=& 2G'(m^2)[\partial_{xx}^2 z +\partial_{yy}^2 z]\vec m
\eea

The ES current (see Eqs.~(\ref{j_Joh}-\ref{j_a_p_2}))
has the same sign of the slope and for this reason
it has a `destabilizing' character. In contrast, the
Mullins-like current $\vec j_{M}$ is stabilizing; finally,
the symmetry breaking current has a different behaviour
according to the
sign of the curvature: $[\partial_x^2 z +\partial_y^2 z]$.
We will see that the physically relevant case corresponds to
$G'(m^2)>0$: this means that $\vec j_{SB}$ is destabilizing (resp.
stabilizing) in a region of positive (resp. negative) curvature and
therefore the bottoms of a grooved profile are narrower
than the tops and in some models they even look like crevasses.

The first analytical derivation of $\vec j_{SB}$ has been done by
Politi and Villain~\cite{PV} in the case of a 1+1 dimensional surface.
Afterwards, this result has been confirmed by Krug with a different 
method~\cite{Korea}.
The existence of a curvature-dependent term in the current which has not 
the form of $j_{M}$ is easily deduced by the following consideration:
a surface profile whose slope is not constant is surely
time-dependent, even if $\es=0$ and all the sources of the Mullins-like
term (thermal detachment and noise) are switched off:
in this case the velocity of each step $-$in the BCF theory$-$ is 
proportional to half of the sum of the lengths of the upper and 
lower terraces. This means that $\partial_t z\ne 0$ even if
$\vec j_{ES}=\vec j_{M}=0$.

A very simple, but not rigorous argument to justify $\vec j_{SB}$ is now
given: the `microscopic' current on a terrace is $\vec j_{mic}=
-D\nabla\rho$. In the paragraph on the ES current we have evaluated
$j_{ES}$ as an average of $j_{mic}$ on a terrace of size $\ell$:
step-edge barriers make $\rho$  asymmetric with respect to the center of
the
terrace and therefore $j_{ES}\ne 0$. Nonetheless, even if $\es=0$
the density $\rho$ varies from one terrace to another, if there is a
curvature in the surface profile: in fact, the solution of the
diffusion equation $D\nabla^2\rho +F_0=0$ on a
terrace of size $\ell$ with $\rho$ vanishing at steps gives an
average value $\bar\rho\approx (F_0\ell^2/D)$. The terrace width $\ell$
is of
of the order of $\ell_D$ at small slopes and of the order of
$1/|m|$ at high slopes.
The simplest function interpolating between these two limiting
behaviours~\cite{PV,Korea} is: $\ell^2\approx \ell_D^2/(1+m^2\ell_D^2)$.
Therefore we obtain:
\be
\vec j_{SB} \approx -D\nabla\left({F_0\over D}{\ell_D^2\over
1+m^2\ell_D^2}\right)= -\nabla\left({F_0\ell_D^2\over
1+m^2\ell_D^2}\right)
\label{j_sb}
\ee
So, the function $G(m^2)$ appears to be:
\be
G(m^2)\approx -{F_0\ell_D^2\over 1+m^2\ell_D^2}
\ee
whose derivative $G'(m^2)$ is always positive. At large slopes:
$\vec j_{SB}\approx -F_0\nabla(1/m^2)$ which agrees $-$in its 1+1
dimensional version$-$ with the results of Refs.~\cite{PV,Korea}. 
The numerical prefactor is (1/8) for Politi and Villain and (1/12) for
Krug. 
It is important that both methods give a constant prefactor, not
depending
on the ES length. If the rough argument given above to justify
$\vec j_{SB}$ is slightly modified to take into account finite
step-edge barriers, the function $G$ appears to be $\es$-dependent,
but the dependence is extremely weak: the prefactor simply varies
between 
(1/12) and (1/3) for $\es$ varying from zero to infinity.

Anyway, we should realize that such argument is far from being
rigorous, because the diffusion between terraces  must be understood
as an effective one, since steps capture atoms.
In this respect, it is important to point out that Eq.~(\ref{Lan_eq})
$-$and so the current $\vec j-$ is the result of a
spatiotemporal average and therefore $\vec j_{SB}$ may appear even if no
interlayer diffusion is allowed by the microscopic dynamics.
Indeed, more microscopic approaches based on the Master
equation~\cite{ME_V}
confirm that such term is present even if thermal detachment is
forbidden.

At small slopes we have 
\be
\vec j_{SB}\approx F_0\ell_D^4\nabla(m^2)\equiv \lambda\nabla(m^2)
\label{j_sb_1}
\ee

It is noteworthy that $\lambda\approx K_{nuc}$: this is not 
surprising because $[\lambda]$=$[K]$ and since $F_0$ and $\ell_D$
are the only parameters entering the model, there is no other way
to obtain a quantity which is dimensionally correct.
The current (\ref{j_sb_1}) appears in the study of the conserved
Kardar-Parisi-Zhang equation~\cite{SGG}; in the more specific context of
MBE, it was introduced by Villain~\cite{VilJdP}.

We conclude by observing that a rigorous derivation of the surface
current should give all the terms we have introduced:
the ES current, which depends on the slope $\vec m$; the
symmetry-breaking
term, which depends also on the curvature $\nabla\cdot\vec m$;
the Mullins-like current, which depends on higher order derivatives
$(\nabla^2\vec m)$. Nonetheless, at the moment
a systematic derivation is lacking.

\subsection{Experimental results}
\label{PPexpp}

In this paragraph we want to present and discuss some experimental
results 
concerning unstable MBE growth on a high symmetry surface, which are
thought to be due to step-edge barriers. Let us start by describing the
main features of the ES instability on a singular surface.

After deposition of a certain thickness of material, which depends
on the material itself and on the growth conditions, a three
dimensional mound structure sets in. These mounds can be
characterized by a typical size $L$ and a typical slope $m_0$:
it is often the case that $L$ increases with time via a
coarsening process, where bigger mounds eat neighbouring smaller ones.
Experimental data for $L(t)$ are plotted in a log-log scale in order
to obtain the so-called coarsening exponent $n$: $L(t)\sim t^n$.

In Tab.~\ref{exp_table} we summarize some experimental data,
concerning metals (Fe, Cu and Rh) and semiconductors (GaAs and Ge).
For any experimental system we give: 
the growth temperature $T$ in Kelvin; 
the coarsening exponent $n$ with its error (when indicated);
the range of sizes $L_{min}\div L_{max}$ of the mounds over which $n$ is 
calculated or the typical values of $L$ which are displayed; 
the slope of the mounds;
the range of thicknesses $N$ corresponding to the sizes given in the
$L$ column 
or the maximal thickness reached in the articles referenced
here; the intensity of the flux; the experimental techniques.

 From the third column we see that $n$ varies between $0.16\pm 0.04$ and
0.56 and that three values out of seven are in the range $0.23/0.26$.
It is not always possible to know on which range of values of $L$ the
exponent refers to; if it is possible, such range $(L_{min}\div
L_{max})$
is indicated in column ``$L$". It is therefore seen that the ratio
$L_{max}/L_{min}$ varies of a factor
from two~\cite{Zuo} up to ten~\cite{Tsui}. 

It is important to point out that for all metals a mound structure
already appears after the deposition of a very few layers, or even in
the
submonolayer regime. The case of semiconductors is different: for
example
let us consider the epitaxial growth of germanium~\cite{Nostrand}.
Here $-$as well as for silicon$-$ the main problem is that
the system undergoes a `crystalline/amorphous growth' phase
transition at a critical thickness $h_{cr}$ which strongly depends
on temperature: this implies for example that it is difficult to
track the ES instability at different thicknesses and it may be
necessary to increase $T$ in order to obtain mounds of larger size.
In the case of Ge, the size of the first mounds 
increases with $T$. Their
typical size, much larger than in most of metals, indicates a weaker 
ES barrier.

A second point we want to stress regards the slope of mounds:
Is it constant or does it increase with time?
Experimentally, the question puts additional problems for two reasons:
slope
may increase because the `constant slope regime' has not yet been
attained, or slope may look to be constant in the range
$L_{min}\div L_{max}$  because the increase rate
is too slow. These possibilities should be taken into account for
Refs.~\cite{Zuo} and \cite{Stro}, because in both cases larger and
constant slopes are found in the same~\cite{Ernst,Jorri} or similar~%
\cite{thurmer} systems, respectively.\footnote{Clearly we must take care
in comparing experimental results obtained at different temperatures
(Ref.~\protect\cite{Stro} at room temperature and
Ref.~\protect\cite{thurmer} 
at $T\sim 150^\circ$C).}

We finally observe that large slopes, corresponding to well
defined orientations, are found only for metals, while
unstable growth in semiconductors seems to be characterized by fairly
small slopes $(<3^\circ)$.

\subsection{Continuum theory of the instability}
\label{ch4F}

In chapter~\ref{the_current} we have discussed the surface current $\vec
j$
which should enter in the evolution equation for the surface profile.
Here we will concentrate on the few models for which some well
established results exist and we will discuss how `robust' such results
are
with respect to modifications in the model. 

\subsubsection{One dimensional models}
\label{1-d}

We have said that $j$ is generally made up of three terms:
\be
j = j_{ES} + j_M + j_{SB}
\ee
Concerning the ES current there are two main classes, according to the
existence or not of finite zeros in $j_{ES}$. Anyhow $j_{ES}(m)$ can
always be derived from an effective potential energy $U(m)$:
\be
j_{ES}(m) = -{\delta\over\delta m}\int dx U(m(x))~~~~\hbox{with}~~
U'(m)=-j_{ES}(m)
\ee
The Mullins-like current has the expression $j_M=Km''(x)$ at
sufficiently
small slopes, which can also be written:
\be 
j_M = -{\delta\over\delta m}\int dx {K\over 2} (m')^2
\ee
Nevertheless there is no basic reason to use the small-$m$ expansion
for $j_M$ if the surface develops an instability with large slope
regions. Starting from Eq.~(\ref{muuu}) it is easily found that a more
general
expression is: $j_M = \Upsilon\partial_x [\tilde\sigma(m)m']$, 
where $\tilde\sigma (m) =[\sigma(m)/\sqrt{1+c^2m^2} + \sigma'(m)
\sqrt{1+c^2m^2}/c^2m]$ is
an even function of $m$. In this case $j_M$ is no more
derivable from a free energy: some considerations will be done
at this regard in the following. 

Finally, the term $j_{SB}=\partial_x G(m^2)$ can not be derived from 
a free energy either. This is not true for any symmetry breaking terms,
but it is necessary to pass to a higher order term~\cite{SS}.
For example:
\be
\tilde j_{SB} =\tilde\lambda\partial_x[(m')^2] =
-{\delta\over\delta m}\int dx {\cal F}_{SB}
\ee
with ${\cal F}_{SB}=(\tilde\lambda/3)(m')^3$.

If we neglect for the moment all the nonpotential
terms, we are left with the expression:
\be
j = j_{ES}(m) + Km''(x)
\label{j_s_1d}
\ee
where $j_{ES}$ may have (model I) or not (model II) zeros at finite $m$.

Before starting to study dynamics, it is important to consider 
stationary configurations of Eq.~(\ref{Lan_eq}), corresponding to a
vanishing current: $j\equiv 0$. This equation is immediately
recognized as the equation of motion of a particle of mass $K$ moving
in the potential $V(m)=-U(m)$. To be definite let us consider the
following expressions for the ES current:
\bea
j_{ES} &=& \alpha m[1-m^2/m_0^2]~~~~~\hbox{model I}\label{j_I}\\
j_{ES} &=& {\alpha m\over [1+m^2\ell_D^2]^\gamma}~~~~~\hbox{model II}
\label{j_II}
\eea

Model II reduces for $\gamma=1$ to the current introduced by Johnson
et al.~\cite{Joh} in 2+1 dimensions (Eq.~(\ref{j_Joh})) and by
Hunt et al.~\cite{Sander} in 1+1 dimensions. 
All microscopic models yield $\gamma= 1$.
Nonetheless it is useful to
study the generalization to $\gamma > 1$~\cite{PT}.
The corresponding expressions for $V(m)$ are:
\bea
V(m) = {\alpha\over 2} m^2 - {\alpha\over 4m_0^2} m^4
\;~~~~&&\hbox{model I}\label{I}\\
V(m) = \left({\alpha\over 2\ell_D^2}\right)\ln(1+m^2\ell_D^2)
\;~~~~&&\hbox{model II,}~\gamma=1\label{II1}\\
V(m) = -\left[{\alpha(\gamma -1)\over 2\ell_D^2}\right]
{1\over(1+m^2\ell_D^2)^{\gamma -1}}
\;~~~~&&\hbox{model II,}~\gamma>1\label{II2}
\eea

In Fig.~\ref{V(m)} we plot these functions.

In our mechanical analogy $m$ plays the role of space, $x$ of time
and stationary configurations correspond to oscillatory motions
around the equilibrium position $m=0$. 
Such motions have different behaviours in the three cases: the important
feature is how the period of the oscillations (i.e. the wavelength
$\lambda$
of the stationary configuration) depends on the amplitude
(i.e. the maximal value $m^*$ of the slope). In the harmonic regime
$[V(m)=\alpha m^2/2]$ the period does not depend on the amplitude: this
corresponds to the linear regime of Eq.~(\ref{Lan_eq}), with a well
defined wavelength and an `undefined' amplitude (depending on the real
time, 
in the
dynamical problem). When the amplitude is large enough we enter in the
nonlinear regime and the question is: Does $\lambda$ increase or
decrease with $m^*$? 
For both models I and II, as well as for the current given in
(\ref{j_es}), $\lambda$ increases with $m^*$ because the `force'
$j_{ES}$
decreases with respect to the harmonic regime, i.e. $j_{ES}/m$
decreases with $m$. This condition can be found by writing:

\be
|j_K|=|j_{ES}|~~~\Longrightarrow~~~K{m^*\over\lambda^2}\approx j_{ES}(m^*)
\; ~~~\Longrightarrow~~~\lambda\approx\sqrt{Km^*\over j_{ES}(m^*)}
\label{L_M}
\ee

This means that stationary solutions of arbitrarily large 
wavelength exist and they correspond to increasing values of $m^*$:
this common feature is one of the necessary ingredients to have a
coarsening phenomenon. However, important differences exist between
the models: in model~I, $\lambda\to\infty$ when
$m^*\to m_0^-$, while in models~II, $\lambda\to\infty$ when $m^*$ also
diverges. In model~I it is possible to define a stationary solution
which corresponds to the boundary conditions: $m(x)\to\pm m_0$
when $x\to\pm\infty$. This solution is called `domain wall' or `kink'
because $m(x)$ is always constant (and equal to $\pm m_0$) except in a
narrow region around $x=0$. The width of this region is
\be
L_{dw}={2m_0\over|m'(0)|}=\sqrt{8K\over\alpha}
\ee
which (up to a numerical factor of order unity)
is nothing but the lower critical wavelength $\lambda^*$:
The surface is dynamically stable at smaller scales.

In the limit $m^*\to m_0^-$ stationary configurations are made up of
domains of increasing size $L$ where the slope is alternately equal
to $\pm m_0$, separated by domain walls. This picture breaks down for
model~II because now $m^*\to\infty$ and it is no more possible to
define domains of constant slope. Nonetheless a major difference
exists between the cases $\gamma=1$ and $\gamma>1$:
for $\gamma=1$ there is no stationary solution corresponding to the
boundary conditions $m(x)\to\pm\infty$ when $x\to\pm\infty$, while this 
solution does exist if $\gamma>1$. This also implies that for $\gamma=1$
(the physically relevant case) the maximal velocity in $m=0$ diverges
when $m^*\to\infty$: in other words stationary solutions are
characterized by a curvature which diverges in the minima and maxima of
the
profile (i.e. where $m(x)\equiv dz/dx=0$).

Stationary solutions are important for the dynamics for the following
reasons: coarsening means that these stationary configurations are
stable 
with respect to amplitude fluctuations, but unstable with respect to
wavelength fluctuations. The evolution of the system is such that the
surface profile passes through stationary configurations of increasing
wavelength. 

Let us now come back to the expression~(\ref{j_s_1d}).
Since our one dimensional current $j$ is derivable from the effective
free energy:
\be
{\cal F} = \int dx\left[ {K\over 2}(m')^2 + U(m)\right]
\ee
the evolution equation for the order parameter $m(x)$ is given by
Eq.~(\ref{eq_m_1}):
\be
\partial_t m(x,t) = \partial_x^2\left({\delta F\over\delta m}\right)
+\partial_x(\hbox{noise})
\ee

For model~I we exactly obtain the so-called Cahn-Hilliard
equation~\cite{CH}, which has been extensively studied.
First we give the results and afterwards some qualitative explanations:
\bea
&&L(t)\sim \ln t~~~~\hbox{noise}\equiv 0~~~
\hbox{Ref.~\protect\cite{Langer}}\\
&&L(t)\sim t^{1/3}~~~~\hbox{shot noise}\ne 0~~~
\hbox{Ref.~\protect\cite{KO,KM}}
\eea

The best way to explain these results is to recall the kink picture.
A system with only one kink corresponds to a surface profile with only
one mound: $m(x)\equiv z'(x)\to\pm m_0$ when $x\to\mp\infty$. This
stationary profile is stable, but if we put more kinks the profile is
no more stable because kinks interact through the tails of their
profiles.
The shape of the tail can be easily determined linearizing the
equation $j=0$ around $m_0$: $m(x)=m_0-\epsilon(x)$. We obtain:
\be
j_{ES}(m) + Km''(x) = -j'_{ES}(m_0)\epsilon(x) -K\epsilon''(x) =0
\ee

The function $\epsilon(x)$ has therefore an exponential behaviour:
$\epsilon(x)=\epsilon_0\exp(-\kappa x)$ with 
$\kappa=\sqrt{|j'_{ES}(m_0)|/K} = \sqrt{2\alpha/K}$. The inverse of
$\kappa$ can be recognized (up to a constant) as the width of the domain
wall:
$L_{dw}=1/\kappa$.

It is therefore reasonable to expect that kink interaction and
kink velocities (in the absence of noise) decrease exponentially with
the interkink distance $L(t)$. Coarsening means that kinks feel an
effective
{\it attractive} interaction:
if a kink has a velocity $v\sim\exp(-\kappa L)$ where $L$ is the
distance with its nearest kink, then the time $t(L)$ necessary for the
annihilation of a pair of neighbouring kinks is $t(L)\sim L\exp(\kappa
L)$,
which gives: $L(t)\sim L_{dw}\ln t$.

The effect of noise is  more difficult to take into account, 
because $m(x,t)$ is a conserved order parameter:
$\partial_t \int dx m(x,t)=0$. On one side noise would make kinks move
independently through a sort of random walk; on the other side the
constraint of conservation couples the motion of the different kinks.
In the absence of conservation we would simply have $L(t)\sim t^{1/2}$.
In fact, in a random walk the time necessary to travel a distance $L$
is $t\sim L^2$. If $m(x)$ is conserved the constraint slows down the
kink motion and we pass from $t^{1/2}$ to $t^{1/3}$.

\vskip 0.5cm
Now let us comment on the case $\tilde\sigma(m)\ne$~constant.
If $\tilde\sigma$ depends on the slope $m$ the current reads:
\be
j = \Upsilon\partial_x(\tilde\sigma(m)m') + j_{ES}(m)
\ee

What about stationary configurations? We can define the
function $M(m)=\int_0^m du \tilde\sigma(u)$, so that:
\be
j = \Upsilon M''(x) + j_{ES}(m(M))
\ee
where $M(m)$ is an odd function which always increases with $m$,
since $M'(m)=\tilde\sigma(m)>0$.  We have seen that 
\be
{j_{ES}(m)\over M} = {j_{ES}(m)\over m}\cdot{m\over M}
\label{sigma_coar}
\ee
must be a decreasing function of $m$, in order to have coarsening.
The first ratio on the right hand side is a decreasing function of $m$,
because we suppose that coarsening does exist if $\tilde\sigma(m)$
is constant (i.e. if $M=m$). On the other side,
if $\tilde\sigma(m)$ decreases with $m$, the ratio $(m/M)$ increases
because $M/m$ is the average value of $\tilde\sigma(u)$ in the interval
$(0,m)$. According to this, the right hand side of
Eq.~(\ref{sigma_coar})
is the product of a decreasing function and of an increasing function of
$m$ and therefore no general conclusion can be drawn from it, 
if a specific form for $\sigma(m)$ [and therefore for $\tilde\sigma(m)$] 
is not assumed.

\subsubsection{Two-dimensional models}
\label{2dm}

In chapter~\ref{c_d} we already remarked that a continuum
description of crystal growth in 2+1 dimensions differs from a standard
phase
separation process even if the current can be derivable from an
effective free
energy. Let us discuss this point.

The generalization of Eq.~(\ref{j_s_1d}) is:
\be
\vec j = \vec j_{ES}(\vec m) + K\nabla^2\vec m
\ee
This equation is almost always derivable from a free energy (it is
indeed so
for Eqs.~(\ref{j_Joh}-\ref{j_a_p_2})) 
but not for any form of
$\vec j_{ES}$, in opposition to the one dimensional case.
The reason is that
\be
\vec j_M = K\nabla^2\vec m = -{\delta\over\delta\vec m}
\int d\vec x {K\over 2}\left[(\nabla m_x)^2 + (\nabla m_y)^2\right]
\ee
while $\vec j_{ES}$ can be written as
\be
\vec j_{ES}(\vec m) = -{\delta\over\delta\vec m} \int d\vec x U(\vec m)
\ee
if and only if:
\be
{\partial^2 U\over\partial m_x\partial m_y} =
{\partial^2 U\over\partial m_y\partial m_x}~~~~~~~\hbox{i.e.}~~~~~~~
{\partial\over\partial m_y} [(j_{ES})_x] =
{\partial\over\partial m_x} [(j_{ES})_y]
\label{j_der}
\ee

Eqs.~(\ref{j_Joh}-\ref{j_a_p_2}) and similar ones have the general form:
\be
\vec j_{ES} = \vec m g(m^2)~~~~~~\hbox{or}~~~~~~
(j_{ES})_{x,y} = m_{x,y} g(m_{x,y}^2)
\ee
where the first one is valid for in-plane symmetry and the second for a 
square symmetry. In both cases the relation (\ref{j_der}) is fulfilled,
as well as for their generalization (\ref{jS}), see below. 
For a three-fold symmetry, Siegert~\cite{S97} 
has suggested the following modification of Eq.~(\ref{j_a_p_2}):
\bea
&&(j_{ES})_x = \alpha (1+2m_x)[m_x(1-m_x) -m_y^2]\label{3fold}\\
&&(j_{ES})_y = \alpha m_y (1-2m_x -2m^2)\nonumber
\eea
Again, this equation fulfills the condition (\ref{j_der}).
The conclusion of this part is that $-$in practice$-$ the passage
from 1+1 to 2+1 dimensions does not put additional problems for the
derivation of $\vec j$ from a free energy.

We now state the two major features which are introduced in 2+1
dimensions: i)~The constraint $\nabla\wedge\vec m=0$, and ii)~The
impossibility for the time evolution of $\vec m$ to be governed by the
Cahn-Hilliard equation.
It is important to point out their common origin: the fact that the
order parameter $\vec m$ derives from the local height $z(\vec x,t)$
[$\vec m=\nabla z \Rightarrow \nabla\wedge\vec m =0$] and therefore the
fact that its evolution equation derives from Eq.~(\ref{Lan_eq}).
So, we have
\be
\partial_t \vec m = -\nabla (\nabla\cdot\vec j)
\ee
and not
\be
\partial_t \vec m = -\nabla^2 \vec j
\label{58}
\ee
Conversely, the phase separation process for a conserved order
parameter is governed by Eq.~(\ref{58}), in order to implement
the conservation law
\be
\partial_t \int d\vec x m_i = -\int d\vec x \nabla^2 j_i
\ee
which vanishes for the divergence theorem.

Siegert and Plischke~\cite{SP,S97,SP96}
have solved numerically Eq.~(\ref{Lan_eq}) with
$\vec j = \vec j_{ES} + K\nabla^2\vec m$, where $\vec j_{ES}$ is
given by Eq.~(\ref{j_s_p}) [continuous symmetry in the plane] or
Eq.~(\ref{j_a_p}) [four fold symmetry in the plane], 
and in the presence of noise.

For a `true' phase separation process, the two currents give different
results for the coarsening law $L(t)\sim t^n$ (see the
review~\cite{bray}):
$n=1/4$ if the potential $U(\vec m)$ has a continuous infinity of
minima [in $|m|=1/a$, for Eq.~(\ref{j_s_p})] and $n=1/3$ if
$U(\vec m)$ has a finite number of minima [in $\vec m=(\pm 1,\pm 1)/a$,
for Eq.~(\ref{j_a_p})]. So, the result by Siegert and Plischke 
for the surface growth
models (Eqs.~(\ref{j_s_p}-\ref{j_a_p})) is surprising: $n\simeq 1/4$ in
both cases~\cite{SP}.

A deeper comprehension of the difference between the surface growth
evolution and a `real' phase separation process can be gained by
comparing
the different pictures of Fig.~\ref{2d_growth}.

They refer, respectively, to the following models:
\bea
&&\partial_t \vec m = \nabla^2\left({\delta {\cal F}_{ani}\over
\delta\vec m}\right) + \nabla\eta \;~\left(
\begin{array}{c}
\hbox{\small Four-state}\\ \hbox{\small clock model}
\end{array}
\right)
\hbox{~~Fig.~\protect\ref{2d_growth}a
}\label{MS_clock}\\
&&\partial_t \vec m = \nabla\left[\nabla\cdot
\left({\delta {\cal F}_{ani}\over \delta\vec m}\right)\right] +
\nabla\eta ~\left(
\begin{array}{c}
\hbox{\small Surface growth:}\\ \hbox{\small four-fold in-plane
symmetry}
\end{array}
\right)
\hbox{~~Fig.~\protect\ref{2d_growth}b
}\label{MS_growth}\\
&&\partial_t \vec m = \nabla\left[\nabla\cdot
\left({\delta {\cal F}_{iso}\over \delta\vec m}\right)\right] 
\;~~~~\left(
\begin{array}{c}
\hbox{\small Surface growth:}\\ \hbox{\small in-plane invariance}
\end{array}
\right)
\hbox{~~Fig.~\protect\ref{2d_growth}c
}\label{MR_growth}
\eea
where:
\bea
&&{\cal F}_{ani} = \int d\vec x\left\{{K\over 2}\left[ (\nabla m_x)^2 +
(\nabla m_y)^2\right] +\alpha\left[ -{m^2\over 2} + {m_x^4 + m_y^4\over
4}
\right]\right\}\\
&&{\cal F}_{iso} = \int d\vec x\left\{{K\over 2}\left[ (\nabla m_x)^2 +
(\nabla m_y)^2\right] +\alpha\left[ -{m^2\over 2} + {m^4\over 4}
\right]\right\}
\eea

We stress that (b) and (c) are both growth models, but with different 
symmetries of the order parameter while (a) and (b) have the same
symmetry 
for $\vec m$, but the operators `gradient' and `divergence' are
interchanged.

A point of basic importance is that in surface growth a domain wall must
be 
straight, because it corresponds to the intersection of two planes.

In this respect Siegert, Plischke and Zia~\cite{SPZ} have also
proven  that domain wall relaxation is much faster than coarsening, so
that
domain walls look straight on the time scale of coarsening.

More precisely, if
$\vec m_1$ and $\vec m_2$ are any two minima of ${\cal F}_{ani}$ or
${\cal F}_{iso}$: $\vec m_{1,2}=|\vec
m|(\cos\theta_{1,2},\sin\theta_{1,2})$,
then their intersection takes place along a line forming an angle
$\theta=(\theta_1+\theta_2)/2$ with the horizontal axis.

This means that in principle model (b) allows for only four distinct
kinds of domain wall, forming respectively angles
$\theta=0,\pm\pi/4,\pi/2$
with the $x$-axis, but domain walls parallel to the $x,y$ axes
$(\theta=0,\pi/2)$ cost less energy (because one of the two components
of 
the order parameter keeps constant) and therefore dominate the others.
On the other hand, while the domain walls parallel to 
the Cartesian axes can
intersect, giving rise to a pyramid, the other two
walls cannot intersect. The constraint $\nabla\wedge\vec m=0$ requires
that each `diagonal' domain wall $(\theta=\pm\pi/4)$ $-$if present$-$
terminates at the
same ending point as a horizontal {\it and} a vertical domain wall.
In Fig.~\ref{2d_growth}b all the possible intersections are visible.

In the isotropic growth model (Fig.~\ref{2d_growth}c) the energy
cost of a domain wall does not depend on its orientation, but the
constraint
deriving from $\vec m=\nabla z$ implies a relation between the values of
$\vec m$ in 
neighbouring domains and the direction of the separating
domain wall. The consequence is  a
periodic surface profile which is made of a network of square-based or
hexagonal-based pyramids. Only in the former case the up-down
symmetry
is fulfilled. The statistics of the orientations of the different
facets/domains
indeed show~\cite{MR_private} that they are not randomly distributed,
but four orientations $\vec m_i=(\cos\theta_i,\sin\theta_i)$, with
$\theta_{i+1}=\theta_i+\pi/2$ are more likely to take place.
We speculate that this phenomenon would be even more evident 
in the presence of noise 
(as in Fig.~\ref{2d_growth}b).

In a very recent paper~\cite{S98}, Martin Siegert has proposed a more
complex scenario for coarsening in 2+1 dimensions.
He stresses the importance of having two classes of domain walls
according to the values of $\theta$ (see above): parallel to the axes 
($\theta=0,\pi/2$) and diagonal ones ($\theta=\pm\pi/4$).
This fact should prevent from the applicability of any theoretical
approach based on the existence of a single length scale, because the
densities of the two different classes of kinks 
($\rho_0=\rho(\theta=0,\pi/2)$
and $\rho_1=\rho(\theta=\pm\pi/4$)) have different time dependences.
Even if diagonal domain walls are more energy-expensive, they play a
basic role
in the dynamics because the surface would not coarsen in absence of 
them~\cite{SPZ,S98}.

The two relevant length scales are $R=1/\rho_0$ and $D=1/\rho_1$:
The latter one has a fast increase: $D(t)\sim t^{1/3}$, while the former 
one increases more slowly, with an exponent close to $1/4$.

The previous picture can be completed
with an expression for the
ES current which is more general than
Eqs.~(\ref{j_s_p_2},\ref{j_a_p_2}):
\be
(j_{ES})_x = m_x (1-m_x^2 -bm_y^2)~~~~~~ (j_{ES})_y = m_y (1-m_y^2
-bm_x^2)
\label{jS}
\ee

Eqs.~(\ref{j_s_p_2},\ref{j_a_p_2}) are recovered respectively for $b=1$ 
and $b=0$. The parameter $b$ (whose relevant range of variability for
a quadratic symmetry is $-1<b<1$) gives the dependence of each
component $(j_{ES})_{x,y}$ of the current on the relative perpendicular
component of the slope $m_{y,x}$. Such dependence is mainly due to
edge diffusion, whose importance has been experimentally
shown in metal epitaxy by Jorritsma et al.~\cite{Jorri}.
The picture given above is applicable for $b>-3/4$, when diagonal domain
walls
are suppressed by energetics. Conversely for $-1<b<-3/4$ they are
favored  over horizontal/vertical domain walls. 
Since coarsening requires the nucleation of roof tops, for $b=-3/4$ this 
process does not cost anything and a larger coarsening exponent 
$(n=1/3)$ is found~\cite{S98}.

The same result should be valid for 
a triangular symmetry, relevant for (111) surfaces of fcc crystals,
where only one type of domain walls is present. This prediction agrees
with
the experimental results by Tsui et al.~\cite{Tsui} on the growth of
Rh(111).
\vskip 0.5cm

We conclude this part with some considerations on different
continuous growth models.

\subsubsection{Symmetry-breaking terms} 
\label{sbt}

In 1+1 dimensions it has been recently shown~\cite{kinks} that
a current of the form $j_{SB}=\partial_x G(m^2)$ does not change the
coarsening laws, at least for the model~I.
Let us discuss briefly this point.

Kawasaki and Ohta~\cite{KO} studied the dynamical evolution of model~I
in
one dimension by reformulating the problem of phase separation for the
order parameter $m$ as a problem of nonlinear dynamics of kinks.
A kink represents a (zero dimensional) domain wall between regions
where $m$ takes different constant values, corresponding to the zeros
of the ES current ($=\pm m_0$ for the current (\ref{j_I})). 
The kink profiles $M_\pm(x)$ are the solutions of the equation $j=0$
with
$M_+(\pm\infty)=\pm m_0$ for the positive kink, and
$M_-(\pm\infty)=\mp m_0$ for the negative one. Because of the up-down
symmetry, $M_-(x)=-M_+(x)$ and it is easily found that
\be
M_\pm(x)=\pm m_0\tanh (\kappa_0 x/2)~~~\hbox{with}~~~
\kappa_0=\sqrt{2\alpha/K}
\ee

Because of kink interaction, kink dynamics has a deterministic component 
governed by the minimization of the pseudo surface free-energy,
and a random component due to shot-noise.

Since the kink picture does not depend on the derivability of the
surface
current from any free-energy, it has been applied~\cite{kinks}
to the case of symmetry
breaking induced by $j_{SB}=\lambda\partial_x(m^2)$. 
We can summarize the main results
as follows: 
(i)~The functional form of the kink profiles does {\it not} change.
(ii)~It is no more true that $M_-(x)=-M_+(x)$. In fact now:
\be
M_\pm(x)=\pm m_0\tanh (\kappa_\pm x/2)
\ee
with $\kappa_+>\kappa_-$ (positive kinks are narrower than negative
kinks)
and $\kappa_+\kappa_-=\kappa_0^2$ (the product of their widths is
constant).
(iii)~The total effective interaction between kinks is {\it not}
qualitatively modified.
(iv)~The coarsening laws are the same, but the crossover time between
the logarithmic one (at short times) and the power-low one (at longer
times)
may be significantly increased.
(v)~Positive kinks may be so narrow that their width is comparable to
the
lattice constant. In this case the positive kink should be replaced by
an
angular point, with a discontinuity of the surface slope.

Concerning the symmetry-breaking in 2+1 dimensions, the only study of a
surface growth model in the presence of such term is a numerical
analysis
due to Stroscio et al.~\cite{Stro}, who do not use the Mullins term: 
$\vec j_M = K\nabla^2\vec m$,
but a higher order one: $\vec j = K\nabla^2(\nabla^2\vec m)$. Their 
numerical integration gives a coarsening exponent $n=0.18\pm 0.02$ which 
is told to be insensitive to the presence or the absence of $\vec
j_{SB}$.

\subsubsection{Constantly increasing slope} 
\label{slop}

At the moment there is
no rigorous study $-$not even in 1+1 dimensions$-$ for a model where the
ES current has not zeros at finite slope. Hunt et al.~\cite{Sander}
have studied numerically model~II with $\gamma=1$ (see Eq.~(\ref{j_II}))
in 1+1 dimensions, and found $n\approx 0.22$ in the presence of a weak
shot 
noise: therefore, a result well different from the values 
obtained for model~I: $n=1/3$ in presence of noise and $n=0$
(logarithmic coarsening) in absence of noise.
The qualitative argument based on noise effects and given in 
App.~\ref{app-coarsening}, if applied 
to the case $\gamma=1$ gives~\cite{Tang,Privman} $n=1/5$ in 1+1
dimensions.

The kink picture is manifestly not applicable for model~II, since
the current has no zeros at finite $m$ and therefore regions of constant
slope cannot exist. Even more important, the equation $j=0$ has no
solution with $M(\pm\infty)=\pm\infty$ as boundary condition:
this means that periodic stationary solutions do not converge to a
self-similar limit.

Preliminary results~\cite{PT} on models~II (with $\gamma=1$ and
$\gamma>1$)
in 1+1 dimensions show that steepening makes coarsening faster in
absence of noise and slower in presence of it. The reason of the
deterministic ``acceleration" is the following: even if a kink 
picture is not suitable to describe the interface, at least for
$\gamma>1$ the kink can be replaced by the wave function of the
ground state of a potential well, whose profile depends on the exact
form of the unstable current. Since the kink profile has an exponential
tail while the wave function decays algebraically for models~II,
potential wells interact more strongly than kinks, and as a consequence
of
this coarsening is faster.
The reason of the noisy ``slowing down" has not a simple explanation,
but it is probably due to the competition between steepening and
coarsening (see also the next paragraph on Monte~Carlo simulations).

\subsubsection{The Zeno model and the need of a nonlocal evolution
equation}
\label{zenom}

The `Zeno~model'~\cite{PV} deserves some attention.
it is a
one-dimensional model for a high-symmetry surface which evolves through
instantaneous attachment of adatoms to preexisting steps and creation
of new steps via nucleation of new terraces (a process which is
intrinsically random). In this model only fluctuations in the nucleation
process contribute to the Mullins-like term $(K=K_{nuc})$; since
nucleation
is unlikely on small terraces, the $K$-term is expected to decrease when
the slope increases. This explains the strong up-down asymmetry
displayed by the model (see Fig.~\ref{Zeno_mound}): nucleation is not
effective to counterbalance the ES current in a region of high slope and
the SB term is destabilizing in a region of positive curvature.
In the Zeno model the asymmetry has a striking consequence:
the regions of positive curvature are replaced by angular points.
Because of the discontinuity of the slope $m$ in an angular point, the
SB term in the current $(j_{SB}=\partial_x G(m^2))$ gives a divergent
contribution to the velocity of the interface, if a local
evolution equation is assumed: $\partial_t z = -\partial_x j_{SB}$.%
\,\footnote{The Mullins term gives an even more divergent
contribution (because of the higher derivative), but nucleation (and
therefore
$K_{nuc}$) is completely negligible in a large slope region.}
The solution~\cite{PV} is to replace $\partial_t z(x,t) = -\partial_x j$
with
the following nonlocal evolution equation:
\be
\partial_t z(x,t) = -\partial_x \int dx' \chi(x-x') j(x',t)
\ee
where $j(x',t)$ is the local current evaluated in the
point $x'$ and the kernel $\chi(x-x')$ has the form:
$\chi(x)=\exp(-|x|/x_0)$.

The Zeno model has a second major feature: the existence of an upper
critical wavelength $\lambda_c^{sup}$ at which coarsening stops. 
The reason is that
stationary configurations $(\partial_t z\equiv 0)$ exist only for
$\lambda<\lambda_c^{sup}$. At larger scales mounds are unstable even for
height fluctuations, and therefore deeper and deeper crevasses form
in between.

\subsection{Monte Carlo simulations}
\label{mcar}

This final paragraph is devoted to kinetic Monte Carlo (MC) simulations
of a 
singular surface, in the presence of step-edge barriers. We will limit
ourselves to some representative works in 2+1 dimensions.

An important preliminary statement concerns the implementation of a
constant slope in a simulation where crystal structure effects are 
difficult to take into account. The adopted solutions may differ:
In a simple cubic SOS model, the simplest way is to allow the freshly
landed
adatom to hop to a neighbouring lower site (see Refs.~\cite{SP96,Smil}).
For more realistic cubic lattices (bcc/fcc) the typical solution is a
downward funneling mechanism~\cite{FA,thurmer_sim} according to which
a higher coordination site has a larger `capture area'.

\v Smilauer and Vvedensky~\cite{Smil} 
obtain different time dependences for the slope
by varying the intensity of the down-hill current. Two limiting cases
are noteworthy: constant slope and absence of the down-hill current. 
In the former case, the relevant model should give a coarsening exponent
$n\simeq 1/4$. Simulations give a value varying between 0.21 and 0.26.
The latter case (absence of a down-hill current)
is interesting because slope is certainly expected not
to saturate (slope\,$\sim t^{\vartheta'}$), and a na\"\i ve application
of 
the condition $|\vec j_{ES}|\sim|\vec j_M|$, with $|\vec j_{ES}|\sim
1/|m|$
would give $\vartheta'=n$. 
Simulations~\cite{Smil}, in one case out of two, give a result which
agrees: 
$n=0.19,\vartheta'=0.21$. 
An important feature which emerges from their simulations is that size
growth is slowed down by an existing slope increase, that is to say 
steepening and coarsening of mounds are two competitive processes.
This fact is supported by the qualitative analyses of 
Refs.~\cite{Tang,Privman}, discussed in App.~\ref{app-coarsening}: 
the exponent $n$ decreases, if the slope passes
from being constant to increase with the mound size.

Siegert and Plischke~\cite{SP96} consider,
in addition to  isotropic   ES barriers, anisotropic ones for which
$(j_{ES})_y\equiv 0$
and the resulting mound structure has a one dimensional profile
($m_y=0$).
Nonetheless the anisotropic 2+1 dimensional model differs from the same
model in one dimension (at least if noise is present) because the latter
does not order at finite temperature. So, while the isotropic case gives
$n=0.26$ (very close to $1/4$) the anisotropic case does not give
$n=1/3$
$-$as expected for the noisy one dimensional model$-$ but a much smaller
value: $n\approx 0.18$.

Finally, we want to comment on the works 
by Th\"urmer et al.~\cite{thurmer_sim}
and by Amar and Family~\cite{FA} on bcc/fcc(100) lattices.
The models seem to be fairly similar, but the results differ
substantially: 
Th\"urmer et al. find constant slopes $-$corresponding to
the facets (012)$-$ and a coarsening exponent $n\approx 0.24/0.26$.
On the other side Amar and Family find a much broad range of variability
for $n$ ($n\approx 0.16/0.25$ for weak barriers and $n\simeq 1/3$ for
strong ones) and the slope has various behaviors.
A possible explanation is that even minimal differences in the model
may influence step-edge diffusion, and --after Refs.~\cite{Jorri,S98}--
the coarsening behaviour.\footnote{%
We should also observe that in Ref.~\protect\cite{thurmer_sim} fits are 
generally carried out at larger thicknesses that in
Ref.~\protect\cite{FA}.}


\section{Kinetic instabilities of vicinal surfaces}
\label{PP2}

\subsection{Introduction}
\label{chh}

A vicinal surface is a surface which is cut along an orientation
close to a high symmetry one. It is made up of low index terraces
(e.g.~(001) or (111)) separated by atomic steps. 
Steps are unavoidably present in any real
surface: for example, because the surface is always slightly miscut, but
also
because steps may be thermally activated or because a screw
dislocation ends up in a step at the crystal surface; finally, steps may
be created during the growth through nucleation. The first factor is the
only one to produce steps all of the same sign (ascending or descending
ones),
and the second factor is completely negligible at small temperatures.

In the presence of an external flux,
a step acts as a sink by capturing adatoms diffusing in the vicinity
and therefore advancing. In average a straight step (as 
those which result from cutting the crystal) remains parallel to 
its original direction, while the steps
due to screw dislocations give rise to a spiral growth.
In this chapter, dislocations as well as thermally activated steps will
be ignored.

A growing  vicinal surface can only be stable if three conditions are
fulfilled.
i) Diffusing adatoms stick to  preexisting steps rather than
nucleate new islands on the terraces. ii) Terraces keep
the same width. iii) Steps remain straight. 
When one of these conditions is not fulfilled, it gives rise to a well
defined instability, called respectively i) three-dimensional growth,
ii) step-bunching and iii) step-meandering. 

Since the typical distance between nucleation centers on a high symmetry
surface is the so called diffusion length $\ell_D$
(see App.~\ref{ell_D}), we would be led to
conclude that the condition for avoiding three-dimensional growth is
simply $\ell<\ell_D$, where $\ell$ is the terrace width.
Nonetheless nucleation is a stochastic process and therefore nucleation
events take place, even if rarely, on a terrace smaller than $\ell_D$.
So, an important issue is the asymptotical stability of step-flow
growth against nucleation: this question will be discussed at the end of
the chapter, and till then nucleation will be ignored.

Let us start by showing what the ES effect has to do with step-bunching
and meandering instabilities.
To make a clear distinction between the two phenomena, 
we will consider the following limiting cases: 
i)~a one dimensional vicinal surface (in order to 
avoid the possibility of 
meandering) and ii)~a single step (in order to avoid step-bunching). 

\noindent i)~Let us suppose that adatoms prefer to stick to the 
ascending step, as they do in the presence of an ES effect: this means
that
a terrace which is larger than its neighbours will reduce its width
because the ascending step proceeds faster than the descending one.
The opposite is true for a smaller terrace. The conclusion is that
the ES effect stabilizes the width of the terraces {\it during
growth}.\,%
\footnote{It is important to recall that impurities can pin the 
steps and therefore play a major role in step-bunching 
during growth~\protect\cite{SBI}.}
Conversely, if the crystal surface evaporates, an ES effect means that
an adatom will detach preferentially on the lower terrace and therefore
a larger terrace will become even larger: a step-bunching instability
takes place {\it during evaporation}. This scenario was predicted
thirty years ago by Schwoebel and Shipsey~\cite{schwoebel}. 

\noindent ii)~Now let us consider a single step. 
This makes sense only if evaporation is present, otherwise 
an infinite current of adatoms
would stick to the step and it would move with infinite velocity.
In the presence of desorption an adatom has a lifetime $\tau$, which
means that
it travels through a linear distance $x_s=\sqrt{D\tau}$ before
evaporating: so, only adatoms landing within a distance $x_s$ from the
step will contribute to the step advancement. If a train of steps is
correctly considered (as in case~i), the limit of vanishing desorption is still
physical, because sinks at distance $\ell$ do exist.

The meandering instability is conceptually similar to the diffusion 
instabilities treated in chapter~\ref{snow}: 
the growth front develops some protrusions which capture more atoms
and therefore grow even more. The main difference is that also atoms
coming from the `inner side' of the front contribute to step motion,
and the meandering instability can arise only if the inner contribution
is too weak to compensate the outer contribution, in other words 
if an ES effect is present.
However, Saito and Uwaha~\cite{SaitoLB} have shown that a meandering
instability can take place even in the absence of an ES effect, if the
step
velocity exceeds a critical value. In fact, during the growth,
in the reference system of the
step the adatoms of the lower terrace have a higher component of the
velocity
in the direction perpendicular to the step and therefore the `flux' of
the
incoming adatoms from below is greater than from above.\footnote{In this
case
the instability may take place even during evaporation; however a
finite ES effect stabilizes~\protect\cite{SaitoLB} the profile of the
receding step.}
However, the required flux is so high that the diffusion
length $\ell_D$ is very small and nucleation can no more be neglected.
Finally, the drift of adatoms may be even increased and controlled {\it
via}
application of an external electric field, which produces the
electromigration phenomenon~\cite{EM}.

We want to stress that the conditions for the meandering instability
and those for step bunching are opposite: the former takes place if
the surface grows, while the latter if the surface sublimates.
Therefore, the consequence of points i) and ii) is that a two
dimensional 
train of steps is generally expected to be unstable.

The study of the dynamics of a vicinal surface will proceed as follows:
linear stability analysis of a single step~\cite{BZ} and of a train of
steps~\cite{PEKMV}; study of the weakly nonlinear regime, close to the
threshold of instability, for a single step~\cite{mis} and for a train
of steps~\cite{Karma,OPL}; Monte Carlo simulations~\cite{SaitoLB,RSK};
continuum description of
the vicinal surface~\cite{OPL,RSK,tesiOPL,OPLMB}. We will conclude with
some
remarks on three-dimensional growth and step-flow~\cite{RSK,KS}.

\subsection{Single step: linear stability analysis}
\label{ch5B}

Before starting we would like to point out the qualitatively different
approach
which is required by a vicinal surface, with respect to a high symmetry
one.
On a singular surface, steps are continuously created (through
nucleation)
and destroyed (through coalescence): at least in the case of stable
growth,
the lifetime of a step is given by the typical time necessary to fill
one layer: $t_{ML}=1/(F_0a^2)$. 
Since the continuum description is valid on time scales
larger than $t_{ML}$, steps do not enter `directly' in such description.
Conversely, steps have an infinite lifetime on a vicinal surface (if
nucleation is neglected) because they all have the same sign (descending
or
ascending one). This means that a step motion picture is better than a
`surface current' picture to describe the dynamics of the surface.

A second remark concerns the microscopic dynamics: 
the density $\rho$ of adatoms on a
terrace is governed by the diffusion equation in the quasi static
approximation:
\be
D\nabla^2 \rho - {\rho\over\tau} + F_0 = 0
\label{eq_det}
\ee
which is a {\it linear} equation. The nonlinear character of the step
evolution derives from boundary conditions at steps.

In the following, $x$ will be the average direction of the step and $y$
the in-plane perpendicular direction, so that the step may be described
by a single-valued function $\zeta(x,t)$ with the lower terrace
corresponding to $y>\zeta$. The normal component to
the step of the gradient of concentration $(\partial \rho/\partial n)$ is
proportional to the difference between the nonequilibrium $(\rho)$ and
equilibrium
$(\rho_{eq})$ concentrations at the step itself:
\bea
 D\left({\partial \rho\over\partial n}\right)_+ &=& k_+ (\rho-\rho_{eq})_+\\
-D\left({\partial \rho\over\partial n}\right)_- &=& k_- (\rho-\rho_{eq})_-
\eea
where the signs $+/-$ correspond to the lower/upper terrace 
(i.e. to $y=\zeta_\pm$)
and the equilibrium concentration is given by:
\be
\rho_{eq} = \rho_{eq}^0 (1+\tilde\Gamma\kappa)
\ee
The second term depends on the curvature\,\footnote{In the following the
curvature is understood to be positive if $\partial^2_x\zeta <0$.} 
$\kappa$ of the step: 
\be
\kappa = - {\partial^2_{xx}\zeta\over [1+(\partial_x\zeta)^2]^{3/2}}
\ee
and reflects the Gibbs-Thomson effect: close to a region of positive
curvature 
the concentration is higher because of the higher rate of thermal
detachment 
from the step.
The coefficient $\tilde\Gamma=\Omega\beta\gamma$ depends on the surface
area per atom $\Omega$ and the ratio 
between the step stiffness $\gamma$ and the temperature
$T=1/(k_B\beta)$.
 
The existence of an ES barrier means that $k_+>k_-$: if the barrier is
perfectly reflecting (one-sided model), then $k_-=0=(\partial_n \rho)_-$; 
on the other side, if the step is in equilibrium with the lower terrace,
then
$(\rho-\rho_{eq})_+=0$ which corresponds to $k_+=\infty$.
In the case of a single step, the boundary condition to infinity
$(y\to\pm\infty)$ is that
the concentration goes to a constant value, which corresponds to the 
balancing between deposition and evaporation:
\be
\rho(x,y,t) \to \rho_\infty = \tau F_0  \hbox{~~~~single step}
\label{666}
\ee

Once the concentration profile is known as a function of the step
profile,
the step velocity $v_n$ may be determined via the relation:
\be
(\Delta \rho_s) v_n = D [(\partial_n \rho)_- - (\partial_n \rho)_+]
\label{777}
\ee
where $\Delta \rho_s$ is the difference of surface concentration between
the 
solid phase
and the `adatom gas' phase. Since the latter is negligible
with respect to the former: $\Delta \rho_s = 1/\Omega$. 

Finally, $v_n$ and $\zeta(x,t)$ are linked by a kinematical relation:
\be
v_n = {v_0 + \partial_t\zeta\over\sqrt{1+(\partial_x\zeta)^2}}
\label{c0}
\ee
where $v_0$ is the `drift' velocity, pertaining to a straight step.
It is noteworthy that for the linear stability analysis, the 
formulas: $\kappa=-\partial^2_x\zeta$ and $v_n =v_0 + 
\partial_t\zeta$ can be used.

Let us now consider the simplest case: a single step in the one-sided
model (i.e. an infinite ES barrier).
We will follow the treatment given by Bena, Misbah and
Valance~\cite{mis}.

The concentration profile and the `drift' velocity $v_0$ are given by:
\be
\rho_0(y) =  \rho_\infty - (\rho_\infty - \rho_{eq}^0)\exp(-y/x_s)~~~~~~~
v_0 = \Omega x_s (F_0-F_{eq}) \equiv \Omega x_s \Delta F
\label{v0}
\ee
where $F_{eq}=\rho_{eq}^0/\tau$. Their interpretation is simple:
only adatoms arriving within a 
distance $x_s$ from the step are able to stick to
it before evaporating and therefore contribute to the step velocity.
If thermal detachment is forbidden, $v_0$ is simply given by the
number of atoms sticking to the step per unit time and unit length, i.e.
by the
term proportional to $F_0$ in (\ref{v0}). The term proportional to
$F_{eq}$
takes into account the rate of
detachment from the step. Finally, $\rho_0(y)$ is the 
one-dimensional solution of Eq.~(\ref{eq_det})
`interpolating' between $\rho_{eq}^0$ in $y=0$ and $\rho_\infty$ at infinity.

The linear stability analysis proceeds by considering a small deviation
from the stationary concentration:
\be
\rho(x,y,t) = \rho_0(y) + \rho_1(y)\exp(iqx +\omega t)
\ee
and from the straight step profile:
\be
\zeta(x,t) = \zeta_1\exp(iqx +\omega t)
\ee
and by solving the differential equation at the first order in
$\rho_1,\zeta_1$.
The existence of a nontrivial solution $(\rho_1\equiv\zeta_1\equiv 0)$
requires a dispersion relation between $\omega$ and $q$:
\be
{\omega(q)\over\Omega D} = -\tilde\Gamma \rho_{eq}^0 \Lambda_q\cdot q^2 
+ {\tau\Delta F\over x_s} \left(\Lambda_q - {1\over x_s}\right)
\label{omega}
\ee
with $\Lambda_q=\sqrt{q^2+1/x_s^2}$. It is indeed sufficient to expand
the
previous expression to order $q^4$:
\be
{\omega(q)\over\Omega D} = \left[ -{\tilde\Gamma \rho_{eq}^0\over x_s} +
{\tau\Delta F\over 2}\right] q^2 - \left[ {\tilde\Gamma \rho_{eq}^0
x_s\over 2}
+ {\tau\Delta F x_s^2\over 8}\right]q^4
\label{q4}
\ee

The straight step is dynamically unstable if $\omega(q)$ is positive in
some range of $q$-values. Since the $q^4$ term is always negative, there
will be an instability only if the coefficient of $q^2$ is positive, and
this
requires that $\Delta F$ is larger than a threshold value given by:
\be
(\Delta F)_{cr} = {2\tilde\Gamma \rho_{eq}^0\over x_s\tau}
\label{thre}
\ee
However, this threshold is physically unimportant, because it is much
smaller
than the value $F_{eq}$ the flux must overcome in order to allow the
crystal to grow instead to evaporate. In fact $(\Delta F)_{cr}/F_{eq}=
2(a/x_s)(a\beta\gamma)$, where $(a/x_s)$ is much smaller than one and
$(a\beta\gamma)$ is of order unity.

If $\Delta F > (\Delta F)_{cr}$ there is a band of unstable modes,
ranging from
$q=0$ to $q=q^*$. The general expression for $q^*$ is fairly
complicated,
because it is
derived from a cubic equation. Nonetheless, if we are very close to
the threshold instability the approximation (\ref{q4}) can be used and
we find:
\be
q^* = {1\over x_s}\sqrt{8\epsilon\over 3}~~~~\hbox{with}~~~~
\epsilon \equiv {1\over 2} - {\tilde\Gamma \rho_{eq}^0\over\tau x_s \Delta
F}
\equiv {1\over 2} -\xi
\label{epsilon}
\ee
The dimensionless parameter $\epsilon$ measures the distance from the
threshold. For $\epsilon\ll 1$ we are in the so called weakly nonlinear
regime, and its analysis will be done in the next paragraph.

Formula (\ref{omega}) clearly shows that stabilizing effects are entirely
due
to thermal detachment: In fact $\omega(q)\ge 0~\forall q$ if
$\rho_{eq}^0=0$. 
Such effects derive from the fact that thermal detachment is encouraged
in a  region of positive curvature and discouraged if the curvature is
negative: this process partly corresponds to an effective 
step-edge diffusion which 
resembles the Mullins-like term introduced in chapter~\ref{Mul_cur}.
As we will see in the next paragraph, it is possible to give a more
solid basis to this interpretation.

However, thermal detachment is also at the origin of the negative term
in the
coefficient of the $q^2$ term (see Eq.~(\ref{q4})): in the absence of
$\rho_{eq}^0$
we would simply have: $(\omega/\Omega D)\simeq (\tau F_0/2)q^2$.
This means that step detachment `renormalizes' and lowers the
coefficient $(\tau F_0/2)$, but with the basic contribution
of evaporation: 
in fact the threshold value vanishes in the limit $\tau\to\infty$.
In other words, step detachment is only effective  at short scales if
it must `use' surface diffusion, but becomes effective also at longer scales if
it can use evaporation.

Now let us discuss the more general case of finite ES barriers and
interacting steps.
In their pioneering  work, Bales and Zangwill~\cite{BZ} 
considered finite ES barriers for a train of in-phase steps;
afterwards, Pimpinelli et al.~\cite{PEKMV} extended their study to
arbitrary
phases, showing the importance of this
extension in the case of an evaporating surface.

The main result of the treatment of finite ES barriers is that the
qualitative
picture given in the previous chapter is confirmed: 
a nonvanishing $k_-$ implies a larger threshold value for the flux. 
The condition
$\Delta F>(\Delta F)_{cr}$ reads $\xi<\xi_c= 1/2$ for an infinite ES
barrier
(see Eq.~(\ref{epsilon})),
while in the opposite limit of very weak asymmetry 
$(k_+\simeq k_-)$ we have~\cite{PEKMV}: 
\be
\xi_c= {d_- - d_+\over 2(d_+/x_s +1)^2}~~~~~~~~\hbox{with}~~~~~~
d_\pm={D\over k_\pm}
\ee

Here and in the following we will suppose that $d_+=0$ (fast attachment 
from the lower terrace) and therefore $d_-$ is nothing but the ES length
$\es$.
In the case of a single step
it should be compared with $x_s$ ($\es\gg x_s$ meaning a strong ES
effect).
With these notations, the previous value of $\xi_c$ (valid for a weak ES
effect)
simply reads: $\xi_c=\es/2$.

\subsection{Linear analysis for a train of steps}
\label{chhk}

Now let us consider the stability of a train of steps. If
$\ell$ is the step-step distance, the case of a single step
is recovered in the limit $\ell\gg x_s$. In fact, in this case
there is a region between each pair of neighbouring steps 
where the concentration
$\rho$ of adatoms has the constant value $\rho_\infty$ and therefore 
the concentration  profile is the one given in Eq.~(\ref{v0}).
Since the case $\ell\gg x_s$ has been studied in the previous paragraph, 
we consider the opposite limit: $\ell\ll x_s$.

Using the results of Ref.~\cite{PEKMV}, the amplification rate can be written
as:
\be
\omega(q,\phi) = \Omega\Delta F [(\cos\phi -1) +\ell^2 q^2/2]
\left({\es\over \es+\ell}\right)^2
-\Omega D\tilde\Gamma \rho_{eq}^0 q^2
\left[{2(1-\cos\phi) +(q^2+x_s^{-2})\ell(\ell+\es)\over\ell+\es}\right]
\ee
where $\phi$, the phase shift between neighbouring steps, plays the role
of
the wavevector $q_y$ in the direction perpendicular to the steps.

Separating terms of different order in $q$, the previous formula reads:
\be
\omega(q,\phi)=\Omega\Delta F (\cos\phi -1)\left({\es\over
\es+\ell}\right)^2
+\left\{{\Omega\Delta F\over 2}\left({\ell \es\over \es+\ell}\right)^2
-\Omega D\tilde\Gamma \rho_{eq}^0\left[{2(1-\cos\phi)\over\ell+\es}+
{\ell\over x_s^2}\right]\right\}q^2 -\Omega D\tilde\Gamma \rho_{eq}^0\ell
\cdot q^4
\label{calcio}
\ee

The quantity $\omega(0,\phi)$ corresponds to a rigid translation of each
step
and therefore it allows to study the step-bunching instability:
\be
\omega(0,\phi) = \Omega\Delta F (\cos\phi -1)\left({\es\over
\es+\ell}\right)^2
\ee
This quantity is always negative (and therefore the train of rigid steps
is 
stable)
if $\Delta F > 0$, i.e. if the surface is growing. The opposite is true
(and therefore the train of rigid steps is unstable) during evaporation.
Furthermore $\omega(0,\phi)$ is maximized by $\phi=0$ (in-phase motion)
if
$\Delta F > 0$ and by $\phi=\pi$ if $\Delta F < 0$. This means that the
single step approximation is acceptable in the former case, 
but it is not in the latter one.

In the case of evaporation, the coefficient of the $q^2$-term 
in (\ref{calcio}) is always 
negative and therefore the $q^4$-term can be neglected:
\be
\omega(q,\pi) = 2\Omega|\Delta F|\left({\es\over \es+\ell}\right)^2
- {\Omega \rho_{eq}^0\over\tau}\left[ {1\over 2}
\left({\ell \es\over \es+\ell}\right)^2 + \tilde\Gamma
\left({4 x_s^2\over\ell +\es} +\ell\right)\right] q^2
\;~~\hbox{\small{evaporation}}
\ee
The sum in square brackets reduces to $[\ell^2/2 +\tilde\Gamma\ell]$ for
a 
strong ES effect $(\es\gg\ell)$ and to $[\es^2/2 +4\tilde\Gamma
x_s^2/\ell]$ 
for a weak one $(\es\ll\ell)$. 
It may also be interesting to evaluate the width of the
band of unstable modes, i.e. the value $q^*$ for which 
$\omega(q^*,\pi)=0$. It is easily found that $q^*=2/\sqrt{\ell^2 
+2\tilde\Gamma\ell}$ for $\es\gg\ell$ and $q^*=(\es/\ell x_s)
\sqrt{\ell/2\tilde\Gamma}$ for $\es\ll\ell$.

In the case of growth the most destabilizing mode corresponds to 
$\phi=0$ and the 
rate of increase
\be
\omega(q,0) = \left[{\Omega\Delta F\over 2}
\left({\ell \es\over \es+\ell}\right)^2 -\Omega\tilde\Gamma
{\rho_{eq}^0\over\tau}\ell\right] q^2 -\Omega\tilde\Gamma
{\rho_{eq}^0\over\tau}\ell x_s^2\cdot q^4
\;~~\hbox{\small{growth}}
\label{step_g}
\ee
is positive when the flux exceeds a threshold, just as in the
single-step case.
The value of the threshold is now 
\be
(\Delta F)_{cr} = {2\tilde\Gamma \rho_{eq}^0\ell\over\tau}
\left({\es+\ell\over\ell\es}\right)^2
\ee
which reads $(\Delta F)_{cr} = 2\tilde\Gamma \rho_{eq}^0/\tau\ell$ for a
strong ES effect, and $(\Delta F)_{cr} = 2\tilde\Gamma
\rho_{eq}^0\ell/\tau\es^2$
for a weak one.
The former expression is nothing but Eq.~(\ref{thre}), where $x_s$
has been replaced by $\ell$: this is reasonable, because the adatoms
contributing to the motion of a given steps are the ones landing within
a distance $x_s$ for a single step, and within a distance $\ell$ for a 
train of steps. A similar comparison is not possible for a weak ES
effect
because in this case there is a more subtle interplay between $\ell,\es$
and $x_s$.
Finally, in the
absence of evaporation $(\tau\to\infty)$ and for a strong ES effect
($\es\gg\ell$), Eq.~(\ref{step_g}) reads:
\be
\omega(q,0) = {\Omega F\ell^2\over 2} q^2 - \Omega D\tilde\Gamma
\rho_{eq}^0
\ell\cdot q^4
\ee
which is the same as Eq.~(4) of Ref.~\cite{OPL}.

\subsection{Single step: weakly nonlinear analysis}
\label{ssn}

The linear stability analysis is useful in two respects: because it
allows to
determine for which values of the different parameters step-flow growth
is
stable and because it allows to extract the time and space scales which
are relevant for the subsequent nonlinear evolution.
In this paragraph we will consider the dynamics of a single step, just
above the
threshold of the instability, i.e. for $\Delta F/(\Delta
F)_c=1+2\epsilon$,
with $\epsilon\ll 1$, where $\epsilon$ was defined in (\ref{epsilon}).
This analysis was carried out by Bena, Misbah and Valance in
Ref.~\cite{mis}.
The calculations are a bit lengthy but instructive and therefore they 
will be given in App.~\ref{app_mis}.
Here we will just explain the main lines of reasoning.

The first step corresponds to make $x,y,\zeta(x,t)$ and $t$
dimensionless,
by rescaling lengths with $x_s$ and time with $\tau$, and
to consider the concentration $u$ with respect to its constant value 
$\rho_\infty=F_0\tau$
at infinity, rather than the `absolute' concentration: $u=\rho-F_0\tau$.
This way, the diffusion equation simply reads: $\nabla^2 u -u=0$. In the
same units, close to the threshold the dispersion relation is
\be
\omega = \Omega\Gamma (2\epsilon q^2 - 3q^4/4)
\label{omega_adi}
\ee
where $\Gamma=\tilde\Gamma \rho_{eq}^0/x_s$. 

The previous relation clearly shows that the `active' modes 
(i.e.~the unstable ones) are of the order of
$q\simeq\sqrt{\epsilon}$ and 
$\omega$ is of the order of
$\epsilon^2$. In other words, the unstable structure has
an initial wavelength of the order of 
$\lambda^*=1/\sqrt{\epsilon}$ and it takes a time
$t^*=1/\epsilon^2$ to develop: so $\lambda^*$ and $t^*$ are the good
space and time units for the nonlinear analysis, and from $x,t$ we will
pass to $X=\sqrt{\epsilon}x,~T=\epsilon^2 t$. (Note that $y$ is not
rescaled,
because the step profile is only a function of $x$ and $t$.)

The next step is to expand both the concentration field $u(X,y,T)$ and
the
step profile $\zeta(X,T)$ in powers of $\epsilon$ with the idea to solve
the problem (differential equation + boundary conditions)
recursively. 
Since $\zeta$ must vanish in the limit $\epsilon=0$, the term
in $\epsilon^0$ in its expansion is not present and it is therefore
useful 
to put $\zeta(X,T) =\epsilon H(X,T)$. At this point, the functions
$u$ and $H$ can be written as: $u=\sum_{n=0}^\infty u_n\epsilon^n$ and
$H=\sum_{n=0}^\infty H_n\epsilon^n$. 
At each order of the integration of the
differential equation $\nabla^2 u -u=0$ a new `integration function'
$A_n(X,T)$ appears, which should be determined {\it via} boundary
conditions.
What is found~\cite{mis} is that $A_1$ and $A_2$ keep undetermined till
the third order. At this order,
the  boundary conditions imply that $H_0(X,T)$ 
fulfills the differential equation 

\be
(\Omega\Gamma)^{-1}\partial_T H_0 = -2\partial^2_X H_0 -{3\over
4}\partial_x^4
H_0 + (\partial_x H_0)^2
\label{KS}
\ee

This equation is called Kuramoto-Sivashinsky~\cite{eqKS} (KS) equation
and exhibits spatiotemporal chaos.
Saito and Uwaha~\cite{SaitoLB} have performed Monte Carlo simulations on
a simple cubic~(SC) lattice for different values of the parameters.
In particular they have pointed out the importance of the crystalline
anisotropy in the step stiffness $\gamma$: an effect which is not taken
into
account in the previous nonlinear analysis and whose study needs to get
to the
fifth order in the $\epsilon$-expansion~\cite{SaitoJ}.

If the step is oriented along a direction of maximal stiffness (for
example the [10] orientation for the SC lattice) the anisotropy does not 
play an important role and above the threshold $(\Delta F)_{cr}$ the
step
displays spatiotemporal chaos (Fig.~1c of Ref.~\cite{SaitoLB}):
grooves at average distance given by $\lambda_u$ are 
constantly created 
(when locally the distance is larger than $2\lambda_u$) and destroyed
(through collisions and annihilation).
Conversely, if the step is oriented along a direction of minimal
stiffness
(the [11] for a SC lattice) this `stabilizes' such orientation and a
more regular pattern appears (Fig.~4c of Ref.~\cite{SaitoLB}).
In the limit of strong anisotropy a periodic structure emerges.

The previous scenario has been explained by the same authors in a
subsequent paper~\cite{SaitoJ}. In the 
presence of an anisotropic step stiffness, $\gamma$ reads
\be
\gamma = \gamma_0 [1+\nu(1-\cos 4\theta)]\, ,
\ee
where $\theta$ is the angle between the step normal and the average
growth direction $y$. Accordingly, the coefficient of the 
lower-order linear term
$(\partial_X^2 H_0)$ in Eq.~(\ref{KS}) is modified by the factor
$[1-8\nu\epsilon^2 (\partial_X H_0)^2]$. This implies that the effect
of the anisotropy is vanishing at the threshold instability.
Secondly, a marked difference exists between $\nu <0$ (the stiffness
is maximal) and $\nu >0$ (the stiffness is minimal). In the former case 
the above factor is always positive and anisotropy effects are not
relevant. 
In contrast, in the latter case ($\nu>0$) the coefficient of
$(\partial^2_X H_0)$ may even change sign in regions of high slope
and therefore stabilize the minimal stiffness orientation.

Let us now come back to Eq.~(\ref{KS}). 
Its linear analysis gives the dispersion relation:
\be
\omega(q) = \Omega\Gamma (2q^2 - 3q^4/4)
\ee
After having reintroduced the old variables ($q\to q/\sqrt{\epsilon}$
and
$\omega\to\omega/\epsilon^2$) it takes exactly the form given in
Eq.~(\ref{omega_adi}). So, the linear terms of Eq.~(\ref{KS}) could be
`inferred' by the dispersion relation (\ref{omega_adi}). With regard to
the
nonlinear term in the KS equation, it is the simplest and lowest order
nonlinear
term which does not depend explicitly on $H_0$ (because of translational
invariance, the dynamics cannot depend on the absolute position of the
step).
The sign of the nonlinear term is irrelevant for the subsequent
analysis, 
because it may be changed passing from $H_0$ to $-H_0$.

It may be useful to display explicitly $\epsilon$ in the evolution
equation
for the interface. In order to avoid confusion let us come back to the
initial variables $(x,t,\zeta)$ and $-$in terms of these$-$
let us define: $x'=\sqrt{8/3}(x/x_s)$, $h=\zeta/x_s$ and
$t'=(16\Omega\Gamma/3)
t/\tau$. The evolution equation for $h(x',t')$ then reads:
\be
\partial_{t'} h = (1/2) (\partial_{x'} h)^2 -\epsilon\partial^2_{x'} h -
\partial^4_{x'} h
\label{KS2}
\ee
where $\epsilon$ (defined in Eq.~(\ref{epsilon})) has the form
$\epsilon={1\over 2}({\Delta F\over(\Delta F)_{cr}} -1)$.
When $\Delta F<(\Delta F)_{cr}$, 
$\epsilon$ is negative and the step is linearly 
stable. The quartic term is negligible and Eq.~(\ref{KS2}) takes the
form of
the deterministic Kardar-Parisi-Zhang equation: nonetheless, since
no deterministic instability is present it is necessary to add a noise
term $\theta(x',t')$ to its right hand side. Karma and
Misbah~\cite{Karma}
have therefore modified Eq.~(\ref{KS2}) to take into account stochastic
fluctuations; Furthermore, to make it valid beyond the small range
$\epsilon\ll 1$, the coefficients of the nonlinear $(\partial_{x'} h)^2$
and
quartic $(\partial^4_{x'} h)$ terms become dependent on the incoming
flux.
Finally, Eq.~(\ref{KS2}) is generalized in the following way:
\be
\partial_{t'} h = {\lambda\over 2} (\partial_{x'} h)^2
-\epsilon\partial^2_{x'} h -
\mu \partial^4_{x'} h + \theta(x',t')
\label{KS3}
\ee
where $\lambda=\Delta F/(\Delta F)_c$, $\mu=(2+\lambda)/3$ and:
\be
\langle\theta(x'_1,t'_1)\theta(x'_2,t'_2)\rangle = R\delta(x'_1-x'_2)
\delta(t'_1-t'_2)~~~\hbox{with}~~~R=\sqrt{3/8}(T/\gamma x_s)
\ee

This equation has been solved numerically by Karma and
Misbah~\cite{Karma}
and discussed in details in a recent paper by Pierre-Louis and Misbah~%
\cite{OPLMB}. Stochastic fluctuations are important below 
the instability threshold (when the straight step is linearly stable) 
while deterministic ones are important
above the threshold. In a small range around $\epsilon=0$ both are
relevant.
This may be seen~\cite{OPLMB} by rescaling $x'$ and $t'$ in order to
have
the coefficients
of the two linear terms of order unity. Afterwards, the rescaling of the
amplitude $h$ will fix the relative importance of the nonlinear term and
the noise. If we put:
\be
x'=\tilde x/\alpha_x~~~t'=\tilde t/\alpha_t~~~h(x,t)=\alpha_h\tilde h
(\tilde x,\tilde t)
\ee
we obtain:
\be
\partial_{\tilde t}\tilde h = -\alpha_x^2\alpha_t^{-1}\epsilon
\partial^2_{\tilde x}\tilde h -\alpha_x^4\alpha_t^{-1}\mu
\partial^4_{\tilde x}\tilde h +\alpha_h\alpha_x^2\alpha_t^{-1}
{\lambda\over 2}(\partial_{\tilde x}\tilde h)^2 + \Theta
\ee
where $\Theta =(\alpha_h\alpha_t)^{-1}\theta$ and therefore:
\be
\langle\Theta(\tilde x_1,\tilde t_1)\Theta(\tilde x_2,\tilde t_2)\rangle
=
R\alpha_h^{-2}\alpha_x\alpha_t^{-1}\delta(\tilde x_1-\tilde x_2)
\delta(\tilde t_1-\tilde t_2)
\ee

The condition on the linear terms requires that
$\alpha_x=\sqrt{|\epsilon|
/\mu}$ and $\alpha_t=\epsilon^2/\mu$. On the other hand $\alpha_h$ may
be
determined by fixing the coefficient of the nonlinear term
$((\alpha_h)_{NL}\alpha_x^2\alpha_t^{-1}\lambda=1)$ or the amplitude of
the noise $(R(\alpha_h)_{n}^{-2}\alpha_x\alpha_t^{-1}=1)$. The range of
values
of $\epsilon$ where both noise and instability are relevant is
determined by
the condition that $(\alpha_h)_{NL}\sim(\alpha_h)_{n}$, i.e.
$|\epsilon|\sim R^{2/7}$.

\subsection{Nonlinear analysis for a train of steps}
\label{chnl}

The first study on the dynamics of a train of steps in the nonlinear
regime is due to Pierre-Louis and Misbah~\cite{tesiOPL,OPLMB,OPLML}, 
who also take
into account the elastic repulsion between steps. They perform a
coarse-graining procedure to pass from the discrete evolution equations
for the steps (variables $\zeta_m(x,t)$) to its continuum version.
Here, we will limit ourselves to observe that the resulting equation has
the
form of an anisotropic Kuramoto-Sivashinsky equation~\cite{RK_KS} which
includes a propagative term. The relevant feature is that a
numerical study~\cite{OPLMB} shows a morphology resembling the late-time
evolution of a vicinal surface~\cite{RSK} in the absence of evaporation,
but in the presence of nucleation (see section~\ref{step-flow-nuc}).

In the treatment of the linear dynamics of a train of steps, evaporation
is not negligible in the sense that $x_s=\sqrt{D\tau}$ may be large
(larger than the step-step distance), but it is smaller than the
wavelength
of the emerging pattern. This implies that the limit $x_s\to\infty$
requires
a different study, which has been performed by O.~Pierre-Louis
et al.~\cite{OPL,tesiOPL}. The analysis of the most unstable mode (the
in-phase movement of the steps) close to the threshold of the
instability
(which now corresponds to a vanishing flux) gives the following
evolution
equation for the step $\zeta(x,t)$:
\be
\partial_t \zeta=-\partial_x
\left\{ {\tilde\alpha \; \partial_x\zeta  \over 1+ (\partial_x \zeta)^2}
+{\tilde\beta \over 1+ (\partial_x \zeta)^2}  \partial_x
\left[ {\partial_{xx}\zeta \over (1+ (\partial_x \zeta)^2)^{3/2}}
\right] \right\}
\label{eq_opl}
\ee
where $\tilde\alpha=\epsilon D/2$ and $\tilde\beta=D\Gamma\ell$. As
usual,
$\epsilon=\Omega F_0\ell^2/D$ measures the distance from the threshold:
since desorption is absent, it is equal to the P\'eclet
number~\cite{Tsao}
and represents the ratio of the
`step velocity' $(\simeq F_0\ell\Omega)$ to the diffusion velocity of
the
adatoms $(\simeq D/\ell)$.\,\footnote{It can also be read as the adatom 
density per lattice site, due to the incoming flux.}

Eq.~(\ref{eq_opl}) $-$which is valid for the one-sided model
(infinite ES barriers)$-$ has the form of a conservation law:
$\partial_t\zeta = -\partial_x J$.
In fact the evolution of each step is determined by the adatoms
falling and diffusing on the front terrace and by  rearrangement of the
step profile through detachment/reattachment of adatoms from the step itself.
The effect of these processes is that Eq.~(\ref{eq_opl}) looks like the
evolution of a one dimensional high-symmetry surface.
Nevertheless, major differences must be
stressed: the dynamics of a high-symmetry surface are only determined
(in the absence of desorption) by surface diffusion, which would
correspond
in the present context to step-edge diffusion, a process which is indeed
forbidden in the above model! 
Secondly, the form of Eq.~(\ref{eq_opl}) would be completely
different if only one step was considered (see chapter~\ref{ssn})
and the limit $x_s\to\infty$ would not be allowed.

It is now interesting to comment on the expression for the current $J$:
\be
J = {\tilde\alpha m\over 1 +m^2} + {\tilde\beta\over 1
+m^2}\partial_x\left[
{m'\over (1+m^2)^{3/2}}\right]
\label{curJ}
\ee
where we have introduced the local slope of the step $m=\partial_x\zeta$
in order to make the analogy with a one dimensional surface clearer.

The term proportional to $\tilde\alpha$ plays the role of the unstable
current
due to the ES effect, while the second term should represent a current
{\it \`a la} Mullins, but not in its `standard' linear form
$(\sim m'')$: the nonlinearity inside square brackets is nothing but
the nonlinear chemical potential, if the surface tension $\sigma$
does not depend on the orientation (see Eq.~(\ref{muuu})).
Conversely the prefactor $\tilde\beta/(1+m^2)$ represents a sort of
slope-dependent conductivity
which is due $-$in the present context$-$
to `kinetic' interaction between steps.\footnote{If $\ell$ is the
step-step distance in the direction perpendicular to straight steps,
the shortest path between steps with slope $m$ is $\ell/(1+m^2)$. In a
certain sense the adatom mobility must be reduced by the same factor in
order to keep the travel time between neighbouring steps constant.}

The latter nonlinearity plays a fundamental role, as easily seen by the
study of the stationary configurations $J\equiv 0$. In fact, the
reduction
of the mobility exactly cancels the same factor in the
$\tilde\alpha$-term
so that the unstable current never decreases. Introducing the variable
$M=m/\sqrt{1+m^2}$, the equation determining stationary configurations
takes the form $\tilde\beta M''(x) = 
F(M)\equiv -\tilde\alpha M/\sqrt{1-M^2}$.
Since $|F(M)|/M$ is an increasing function of the slope, the wavelength
$\lambda$ of the profile is expected to {\it decrease} with the maximal
slope (and therefore with the amplitude $A$).
This analysis is confirmed by the exact evaluation of $\lambda$ and
$A$~%
\cite{OPL}. As a consequence of this, the system does not 
coarsen and the step develops an unstable pattern whose wavelength
keeps constant and equal to the most unstable mode $\lambda_u=
2\pi\sqrt{2\tilde\beta/\tilde\alpha}$. The step width increases as
$\sqrt{t}$
without bound, as can be easily checked by replacing the ansatz
$\zeta(x,t)=A(t)g(x)$ in (\ref{eq_opl})~\cite{OPL,wed-cake}.
A numerical solution of Eq.~(\ref{eq_opl}) and full lattice gas
simulations
confirm the previous picture~\cite{OPL}.
Finally, an in-phase meandering of (1,1,17) Cu surfaces has been
recently 
observed by Schwenger et al.~\cite{SFE} at a sufficiently low
temperature.

Different and more phenomenological approaches~\cite{RSK} do not provide
the nonlinear prefactor $\tilde\beta/(1+m^2)$ in Eq.~(\ref{curJ})
and therefore coarsening along the step should be allowed.
However, the direct integration of the continuum equation does not seem
to
display coarsening, which is indeed found in Monte~Carlo 
simulations~\cite{RSK}.

\subsection{Continuum description and step flow}
\label{chsf}

In this chapter we address the following question: is it possible to
link the
step-motion picture for a vicinal surface, with a continuum description
based on the surface current?
In the step-motion picture we have a train of steps whose profiles are
given by $\zeta_n(x,t)$, where $x$ is the average direction of the
steps.
Let $y$ be the in-plane direction perpendicular to $x$ and $\ell$ the
average
distance between steps. If the steps are straight and equally spaced,
then
$\zeta_n\equiv 0$ and the height $z$ of the surface is simply given by:
$z(x,y,t) = -y/\ell$.

In the general case of a step profile depending both on $x$ and $n$ we
should
pass to a continuum function $\zeta(x,y,t)$. This is allowed
only if $\zeta_n(x,t)$ changes by a small amount from one step to the
next one:
this is clearly true for an in-phase step-motion, but for an
out-of-phase
step-flow $-$as the one which takes place during evaporation because of
step-bunching$-$ it is surely not true.

For in-phase motion, it is easily found that:
\be
z(x,y,t) = [\zeta(x,t) -y]/\ell
\ee
and
\be
\partial_t z =\partial_t\zeta/\ell~,~~~~~~\vec m=\nabla z=(1/\ell)
(\partial_x\zeta,-1)
\label{117}
\ee

Let us now suppose that evaporation is negligible so that the 
conservation equation $\partial_t z =-\nabla\cdot\vec j$ holds. Since
$\vec j$ depends on $\vec m$ and higher order derivatives,
$\nabla\cdot\vec j=\partial_x j_x$, and according to Eq.~(\ref{117}) 
we obtain:
\be
\partial_t\zeta(x,t) =\ell\partial_t z =-\partial_x(\ell j_x)\equiv
-\partial_x J
\ee

Therefore, the in-phase train of steps moves according to
one-dimensional
conserved dynamics via the current $J=\ell j_x$. To make this relation
more explicit we can write:
\be
J(\partial_x\zeta) = \ell j_x(m_x,m_y) = \ell j_x(\partial_x\zeta/\ell,
-1/\ell)
\ee

As an example, we will consider the current:
\be
\vec j = K\nabla(\nabla^2 z) + f(m)\vec m
\ee
which gives rise to the current $J$:
\be
J = \ell[K\partial_x(\nabla^2 z) +m_x f(m)] = K\partial_x^3\zeta(x,t) +
f(m_0[1+(\partial_x\zeta)^2]^{1/2})\partial_x\zeta(x,t)
\ee
where $m_0=1/\ell$.

Since $f(m_0)>0$ the solution $\zeta=0$ is unstable and the critical
wavelength is given by $\lambda_c=2\pi\sqrt{K/f(m_0)}$.
It may be interesting to compare the expression of $J$ with what is
found
from a step-motion approach in the linear limit $\zeta\to 0$. It is
clear
that the dispersion relation relevant for the current $J$ is
$\omega(q)=\alpha^* q^2 -\beta^* q^4$, with $\alpha^*=f(m_0)$ and
$\beta^*=K$.
The function $f$ can be derived from the BCF theory
in the limit $m_0\ell_D\gg 1$ (corresponding to the absence of
nucleation 
on the terraces, see Eq.~(\ref{j_es_2})). We find
$\alpha^*_1=F_0\es\ell^2/[2(\ell+\es)]$.
Conversely, the step-motion approach (Eq.~(\ref{step_g})), in the
absence of
desorption gives $\alpha^*_2=F_0\es^2\ell^2/[2(\ell+\es)^2]$ and
$\beta^*=\tilde\Gamma \rho_{eq}^0\ell D$, which
--for an infinite ES effect-- reduce to the quantities 
$\tilde\alpha,\tilde\beta$ introduced in Eq.~(\ref{eq_opl}).
Only in the same limit $\alpha^*_1$ and $\alpha^*_2$ are equal, while
they
differ for a weak ES effect.
We have not been able to clarify the reason of this disagreement.

\subsection{Step flow and nucleation}
\label{step-flow-nuc}

In this chapter we have seen that the steps of 
a growing vicinal surface 
are subject to a meandering instability, but do not undergo a
step-bunching
instability: the terrace width keeps constant and steps move in phase.
In the introduction to the part on vicinal surfaces we raised the
question
of step-flow stability against `three-dimensional growth'.
The transition may be induced by nucleation: between two steps a new
terrace comes up and $-$because of the ES effect$-$ a mound structure
tends to
develop; on the other side, the upper advancing step tends to capture
and
embed the new island, therefore restoring step-flow. In a certain sense,
the
faster of the two processes will establish if step-flow is stable or
not.
Competition between step-flow and two-dimensional nucleation
has been seen $-$for example$-$ by Tung and Schrey~\cite{Tung}
during Silicon epitaxy.

The destabilizing phenomenon is not the nucleation event by itself, but
the mound which can form on the new island.  For a weak ES effect
on a one dimensional surface,
the probability to nucleate a second island on it before the incoming
step
captures the first one is of the order of $(\ell/\ell_D)^4$, 
a vanishing probability
for $\ell_D\to\infty$. In the opposite limit of an infinite ES effect,
this
probability is of order unity and therefore the instability eventually
develops.  This is clearly shown in simulations by Krug and
Schimschak~\cite{KS}, who also evaluate the time $t_{ins}$
necessary for the instability to take place. They calculate $t_{ins}$
from the condition that a mound is formed whenever an atom is deposited
on
top of a mobile adatom and find~\cite{KS} $t_{ins}\approx
t_{ML}\ell^{-2}
(D/F_0)^{3/4}$.
We propose a different criterion: a mound is formed whenever a
nucleation
event takes place and therefore an island is formed.
Since the nucleation probability per unit time on a terrace of size
$\ell$
(see App.~\ref{ell_D})
is~\cite{EV,Psolo} $P(\ell)\approx (1/t_{ML})(\ell/\ell_D)^4$, $t_{ins}$
is
nothing but $1/P(\ell)$: $t_{ins}\approx t_{ML}(\ell_D/\ell)^4
\approx t_{ML}\ell^{-4}(D/F_0)$.

The previous estimates are not applicable to a two-dimensional vicinal
surface because they neglect step meandering. Indeed this phenomenon
has been shown~\cite{RSK} to be relevant for the late time behaviour of
the stepped surface in the absence of desorption, whose morphology looks
very similar to that obtained on singular surfaces.

This is clearly seen in Fig.~\ref{Fig-Smil} which shows the evolution of
a vicinal surface: points of the same step touch and join, and defects
are therefore formed. These ones give rise --because of the ES effect--
to the formation of three-dimensional structures and eventually to the
destabilization of the surface.


\section{Shadowing instabilities}
\label{shadow}

When a crystal is grown from a beam, the beam is generally not normal to
the surface, but has an oblique incidence. 
Then, if a bump is formed, it may have a `shadow' which influences the
growth and contributes to  making
the growing surface unstable (Fig. \ref{eheh}d).
The resulting profile is obviously asymmetric. In certain cases
of technological interest, the deposit 
takes a {\it columnar} structure (Fig. \ref{colon}).
The initial phase of column  formation
may be viewed as an instability, and has been treated by the same
analytical methods as other instabilities~\cite{bruin',bruin}, 
namely non-linear  equations of the KPZ type. 

Columnar growth is usually observed when an intense 
atomic beam is deposited on a fairly cold surface moving at a high
velocity.
This process is used in technological devices~\cite{leamy,hodg,gau}, 
for instance for the manufacture of magnetic
tapes. This technology  is 
quite 
different from MBE, 
since neither the substrate nor the deposit  
are single crystals, and columns are separated by voids.
However, in both cases   ballistic deposition takes place
from a  beam of incoming particles.
In MBE, one tries to avoid bump formation, so that shadowing effects are
negligible.
This result is obtained by a slow deposition rate at high enough
temperature
on a rotating sample.
If these conditions are not realized, bumps or crevaces can appear as
described in 
chapter \ref{PP1}, and then oblique incidence produces shadowing.

The columns have a
well defined direction which makes an angle $\beta$ with the normal to
the surface.
This angle is an increasing function of the incidence angle $\alpha$ 
(Fig. \ref{colon}).  Experimentally, 
a
simple relation 

\begin{equation}
\tan \beta = \frac{1}{2} \tan \alpha
\label{tantan}
\end{equation}
is sometimes observed~\cite{nieuw}. Its validity is limited to low 
temperatures~\cite{licht}.
Theoretical attempts to relate $\beta$ to $\alpha$ have been made, 
using either the continuous (linearized) equations valid in the 
incipient stage~\cite{licht}, or geometric considerations applicable to 
fully developed columnar growth~\cite{tubul3,tubul1}. The theoretical
results are more
complicated than formula (\ref{tantan}) and suggest that $\beta$ does
not
depend only on $\alpha$, but also on the ratio of the beam intensity to 
the diffusion constant~\cite{licht} or on the density of the deposited 
film~\cite{tubul1}. It is noteworthy that the theories do not take
the crystal structure into account, although the formation of columnar 
structures which can arise at normal incidence
as a consequence of the Ehrlich-Schwoebel effect~\cite{EV,PV} is
related to the crystal structure. 

As mentioned above, the deposited film is porous and polycrystalline.
However, 
the direction of the column is a symmetry axis of  the
crystals~\cite{hodg,tubul1}.
This property, which is of importance for technological use, does not
seem
to have received a theoretical explanation.


\section{Wetting and non-wetting}
\label{wet}

In previous chapters it was assumed that the growing crystal 
is semi-infinite in the growth direction. However, for many
applications,
the growing crystal (`adsorbate') is deposited on a substrate which is 
a chemically different material. One generally wishes that the adsorbate 
forms a homogeneous layer with a planar surface. If it is so, 
 it is said that the
adsorbate {\it wets} the substrate, and layer by layer growth is
possible.
It is not always possible. The conditions of stability of
a homogeneous  adsorbate  with a planar surface will be investigated in
this chapter and in the following ones. 

In contrast with the previous chapters, where the instability
was of kinetic nature, we shall now mainly discuss thermodynamics.

A planar surface is stable at equilibrium if it minimizes the 
free energy. Alternatively, one can minimize the difference between 
the free energies of 
a state with a deformed surface (with bumps and valleys) and of the 
state with a planar surface, and the same mass of adsorbate and
substrate. This difference will be called `free energy increment'
and must be positive for
all deformations if the plane surface is to be stable.
The forthcoming search for instabilities will be a search for
negative free energy increments.

A part of the free energy  density excess  is localized 
at the interfaces and surfaces. An obvious reason
is the fact that chemical bonds should be broken to make a surface. 
This localized part is called {\it capillary} free energy, because it is
responsible for the familiar phenomena which arise in thin (`capillary')
pipes. It is present in liquids as well as solids.

Another part  of the free energy 
density excess which results from the creation of interfaces and 
surfaces is due to elasticity. It is delocalized, in the sense that it
decays but slowly 
when the distance to the interface increases. It may be interpreted as
resulting from a distortion of the chemical bonds in the whole system. 
It exists only in solids, and is especially important when the 
adsorbate takes the lattice constant imposed by the substrate. 
In this case, called
{\it coherent } epitaxy, the topology of the crystal is 
preserved.\footnote{This statement applies to most of the cases 
considered in this review, but is an oversimplified generalization. For
instance, coherent epitaxy of bcc Fe on fcc Ag is possible although 
the topology of the two lattices is different.
Another example is MgO on Fe. The extension to these cases of the concept 
of coherence is straightforward.} Since the
misfit $\delta a/a$ defined in paragraph \ref{1.1} is never completely equal
to 0, the adsorbate is strained with respect to its natural size,
and this generates elastic effects which will be studied in the next chapters.
In the present chapter, this elastic contribution  will be neglected,
as it is correct in the case of liquids.
Elasticity is presumably not very important in the growth of a solid,
even on another solid, if this growth is incoherent (i.e. the crystal
topology is broken).\footnote{Elasticity can
be important for the mechanics of two solids glued together,
since temperature variations can produce bending or crack formation,
independently of epitaxy and coherence.}

If elasticity is ignored, the stability condition of a plane interface, 
established by Young at 
the beginning of the nineteenth century, is  that the free energy 
per unit area $\tilde\sigma_{sg}$ of the substrate-gas interface is larger
than the sum  of  the free energy 
per unit area $\tilde\sigma_{sa}$ of the substrate-adsorbate interface, plus
 the free energy 
per unit area $\tilde\sigma_{ag}$ of the adsorbate-gas interface, namely
\begin{equation}
\tilde\sigma_{sg} > \tilde\sigma_{sa} + \tilde\sigma_{ag}
\label{young}
\end{equation}

If (\ref{young}) is not satisfied, droplets form during growth. Their
contact angle 
$\theta$ is given by Young's formula
\begin{equation}
\cos \theta=(\tilde\sigma_{sg}-\tilde\sigma_{sa})/ \tilde\sigma_{ag}
\label{young'}
\end{equation}

In certain cases (Volmer-Weber growth)
droplets form directly on the substrate. This occurs for Pb on
graphite(0001).
In other cases (Stranski-Krastanov 
growth) a few complete layers are deposited before droplets form
(Fig. \ref{drop}). This occurs for Pb on Ge(111)~\cite{JVAP,chambro}.

The droplet size is determined by kinetic
mechanisms~\cite{chakra,avignon1,avignon2}. 
In MBE growth, it is mainly
limited by diffusion of the atoms deposited on the surface, which try
to go to the nearest forming droplet. Since diffusion is slow,
Volmer-Weber 
and Stranski-Krastanov droplets are small, much smaller than water
droplets which are familiar in our everyday life. The 
size of  liquid droplets is close to the equilibrium size,
determined by gravity. They  form through collective, hydrodynamic
motions, not 
through atom diffusion. In the small solid droplets which arise from
 Volmer-Weber and Stranski-Krastanov growth, 
gravity is negligible and the droplet shape is fully determined by
surface tension. In the case of a liquid, droplets would  
actually be parts of spheres, 
whose angle with the plane is determined by (\ref{young'}). 
In the case of a crystal,
the shape can be determined from the Wulff 
construction~\cite{Nozieres,JVAP}.
It is often rather spherical at high enough 
temperature, but it is faceted at low temperature, as mentioned in 
paragraph \ref{1.2}.

Droplet formation caused by the interface energy can be regarded as
a growth instability, but there is not much to say about it. If one
wishes to grow a smooth adsorbate, non-wetting materials are just 
excluded.
On the other hand, there are, to our knowledge, no technological 
applications of {\it incoherent} Volmer-Weber and Stranski-Krastanov
droplets.
However similar droplets can arise from a cause which is not the
surface tension, as will be seen in the next chapters. From 
now on, (\ref{young}) will be assumed to be satisfied,
but elastic effects will be taken into account.

%



\section{Coherent and incoherent epitaxy:  
misfit dislocations and critical thickness}
\label{Marty1}
\subsection{Misfit dislocations}
\label{Marty1a'}

  From now on, the attention will be focussed on the epitaxial deposit
of an adsorbate on a  
substrate which has a fairly small misfit, say $\delta a/a<0.1$.
As seen in the previous chapter, the adsorbate can be coherent or incoherent. 
For a sufficiently small misfit (to be precised below)
the ground state of an epitaxial adsorbate of given thickness $h$
is coherent if $h$ is small enough. In this state,
the crystal topology is that of a 
perfect crystal, in particular each atom has the same number of nearest and 
next-nearest neighbours which form the same geometrical figure with only 
slightly modified distances.

The coherent state can become unstable if $h$ is increased or for another 
reason. In the new, incoherent ground state, the crystal topology is perturbed 
at the interface. Far from the interface, a perturbation of the
crystal topology  involves a large energy (at least if $h$ is large).
Therefore, the crystal topology far from the interface 
will first be assumed to be the same as in  the 
coherent state, although this assumption will be seen not to be always 
correct.\footnote{In other words, the attention will first be focussed on 
perfect dislocations, while partial dislocations, to be defined below,
are often created when $h$ is not yet very large.} 

At the interface, it can be shown~\cite{Fran49} that, if $|\delta a/a| \ll 1$,
the crystal topology 
is but weakly perturbed in large domains separated by lines. These
lines, along which the perturbation is large and linear elasticity is not
applicable, are called `misfit dislocations'. 

These line defects are indeed dislocations of the coherent state.
This  can be understood 
if one measures the atomic displacements $\vec{u}$ 
and the strain $\{\epsilon_{\alpha\gamma}\}$ with respect to 
this coherent state. 
Let us consider the variation $\delta \vec{u}$ of $\vec{u}$
along a path 
joining the centers  $A$ and $B$ of two neighbouring domains on the  interface.
If the path is through the substrate, $\delta \vec{u}=0$.
However, along a path through the adsorbate, 
the variation of this displacement is the integral 
of the strain  from the coherent state, and the components
$\epsilon_{xx}$ and $\epsilon_{yy}$ of this strain
parallel to the surface 
have a well-defined  sign, that  of $\delta a/a$. The integral can therefore 
not vanish. Along a  closed circuit going from $A$ to $B$ through the adsorbate
and then from $B$ to $A$ through the substrate, the variation  
$\vec{b}=\delta \vec{u}$
does not vanish. This is the typical property of a dislocation and
$\vec{b}$ is the so-called `Burgers vector' of this dislocation.

An example of  a misfit dislocation imaged by high resolution
electron microscopy is shown in  Fig.~\ref{fig:HREMdisloc}. 

If the crystal topology is perturbed only near the dislocation line
(in the so-called dislocation `core') the Burgers vector has to be a 
lattice vector and the dislocation is called a perfect
dislocation. It turns out that, if $h$ is not very large, 
the state of lowest energy often involves `partial dislocations', 
characterized by a Burgers vector which is a rational fraction of a 
lattice vector.  Partial dislocations are
the edges of two-dimensional defects, e.g. stacking faults. 
For many crystal structures,
e.g. the face-centered cubic lattice,
stacking faults have a rather low energy, while the energy of a 
dislocation of large Burgers vector is high. Thus, it can be preferable 
to create partial dislocations. The low energy of stacking faults often results
from the fact that the environment of each atom is weakly perturbed  up to 
nearest neighbours, and the crystal topology is only modified if 
next-nearest neighbours are taken into account. 

An example is provided by intrinsic stacking faults 
generated by Shockley partial dislocations with $\vec{b}$=
(1/6)[112] in FCC lattice (Fig.~\ref{fig:disloc_partial}).

In an ordered alloy, the two-dimensional defect related to a partial 
dislocation can be an antiphase boundary
rather than a stacking fault.

Misfit dislocations have the topological properties of ordinary 
dislocations. An essential difference is that they can be present in 
the ground state because their energy can become negative. Actually, 
the free energy associated with a dislocation can be written as 
a sum of 4 terms,

\begin{equation}
{\cal F}_{tot}={\cal F}_{disloc}+{\cal F}_{step}+{\cal F}_{fault}
+{\cal F}_{disloc/film}
\label{mernoire}
\end{equation}

The first 3 terms are also present in a pure system and are usually positive.
The first term ${\cal F}_{disloc}$ is the volume integral of the elastic energy
associated, for $\delta a/a=0$, 
to the strain produced by the dislocation. The second term 
${\cal F}_{step}$ is the free energy of the step which is necessarily 
created (or, exceptionally, destroyed) if the Burgers vector
has a vertical component  (i.e. normal to the interface). 
The third term ${\cal F}_{fault}$,
only present in the case of a partial dislocation, is the 
free energy of the stacking fault or antiphase boundary created by the
dislocation.

The last term ${\cal F}_{disloc/film}$ is negative.
It is the elastic energy gain resulting from
the strain relaxation  in the adsorbate when the dislocation is created. 
 It is 
responsible for the presence of misfit dislocation in the ground state.
The free energy ${\cal F}_{tot}$ depends on the thickness $h$ of the adsorbate
and becomes negative   when $h$ is greater than a threshold
$h_c$ called `critical thickness',  calculated in the next paragraph.

\subsection{Critical thickness}
\label{Marty1a}

\vskip3mm
{\it Nucleated dislocations and threading dislocations}
\vskip3mm

Misfit dislocations can originate from two different mechanisms, i)
nucleation of dislocations on defects, surface
etc.  and ii) propagation of preexisting dislocations of the substrate
(threading dislocations). The density of threading dislocations may be
    very low 
if the substrate is of good crystalline and surface quality,
and the nucleation of new dislocations is not easy since it may involve
a  high energy barrier. Thus, the thickness of the coherent
adsorbate may be larger than the critical thickness calculated
from thermodynamics. Moreover, the
dislocations which are observed are not necessarily those which have
the lowest energy, but those which correspond to the lowest activation
energy, as will be seen later. 
 Nucleation of dislocations at the surface is of particular 
importance in this review.

\vskip3mm
{\it Climb and glide of dislocations}
\vskip3mm

In the case of   dislocations  which are
 nucleated at the surface, they have to propagate down to the 
substrate-adsorbate interface, and they have {\it a priori}
two possible ways to do that, which are called
`climb' and `glide'. Climb implies
mass transport by diffusion in the bulk and
is generally possible only at high temperature. 
In Fig.~\ref{fig:disloc_climb}, an edge
dislocation with a Burgers vector parallel to the interface (the
best way to relax the elastic energy) has been introduced by climb
from the free surface.

In usual growth conditions,
only glide of dislocations is possible. 
On the other hand, in the case of layer by layer growth,
the epilayer is continuous and the
glide plane cannot be parallel to the interface. 
A consequence, which might be
seen  from geometry,
and also results from the formulae given below, is that 
$\vec{b}$ is in the glide plane.  
On the other hand, in the case of layer by layer growth,
the epilayer is continuous and the
glide plane cannot be parallel to the interface. This implies that the
Burgers vector has a vertical component (Figs.~\ref{fig:disloc_parf} and
\ref{fig:disloc_partial}). 
As will be seen, this component  does not contribute to
the relaxation, but has a cost in energy.

\vskip3mm
{\it Critical thickness for the equilibrium state.}
\vskip3mm

One can imagine
two extreme definitions of the critical thickness. i)
In a strictly thermodynamic sense, one might state that    the critical
thickness
is reached as soon as one can find a dislocation of any kind  which has a 
negative energy per unit length. This approach would lead to 	a 
strongly underestimated value of $h_c$.
ii) A complete study would evaluate all energy barriers which correspond
to all possible dislocations. This would be rather tedious.

Our approach in this paragraph is intermediate. The critical thickness
$h_c$
will be defined as the thickness beyond which  the energy
per unit length  $U_{tot}(h)$ of a dislocation lying 
at the interface is negative, {\it provided this
dislocation can be created by glide}. This definition discards
dislocations
with a Burgers vector parallel to the surface plane, which,
as will be seen,  have
a lower energy but can only be created by climb, an extremely slow
process. As will be seen in chapter \ref{dots}, it may
happen 
that, on a planar, high symmetry surface,  dislocations do not appear,
presumably because their nucleation energy is too high. Instead, the
adsorbate forms bumps or clusters, and only then, dislocations appear,
probably because surface irregularities allow the formation of
dislocations
of lower energy.

The critical thickness $h_c$ is given by 

\begin{equation}
U_{tot}(h_c) = U_{disloc}(h_c)+ s\gamma +U_{fault}
+U_{disloc/film}(h_c)=0
  \label{eq:eqhc}
\end{equation}
where the  4 terms correspond to the 4 terms of (\ref{mernoire}), $\gamma$
is 
the step energy per unit length, which has to be respectively added 
($s$=1) or subtracted ($s$=-1) if creating a dislocation emerging 
at the surface creates or destroys an atomic step. 
 The case $s$=-1, where the dislocation has 
nucleated on a preexisting step, lowers the critical thickness. 
However this process seems unlikely because the step would have to be 
a straight line whose direction is imposed by the intersection of the
glide plane 
and the surface. A schematic representation of a dislocation propagating
in 
the glide plane is shown on Fig.~\ref{fig:disloc3d}. The glide 
system is defined by the angle 
$\beta$   between the Burgers vector and the dislocation line lying 
on the interface between the substrate and the epilayer, and the angle 
$\phi$ between the glide plane and the surface of the epilayer. The procedure
is i) to evaluate the dislocation energy and the nucleation barrier for
all possible\footnote{Here, `possible' means that these parameters 
correspond to an actual glide system.} values of $\beta$ and 
$\phi$, and ii)  to look for the values of these parameters which yield the
lowest energy for a reasonable barrier  height. The step (ii) will not be 
carried out in detail here.

In this paper, we will present only the case where 
the substrate and the epilayer 
are isotropic elastic media with identical
Poisson ratio $\nu$ and Young modulus $E$.  The 
self-energy per unit length of a dislocation lying at the 
interface may be written~\cite{Houg90} as

\begin{equation}
  U_{disloc}(h)=E 
  b^2\frac{1-\nu\cos^2\beta}{8\pi(1-\nu^2)}\ln(\frac{\alpha h}{b})
\label{eq:udislocc}
\end{equation}	
where $\alpha$ depends on $\beta$ and $\phi$, and
on the core of the dislocation, a region 
where  linear, continuous elasticity does not
hold, since it would predict infinite strains and stresses. The value 
of $\alpha$ is controversial. The estimates for  $\beta= 60^{\circ}$ and 
$\vec{b}=(1/2)<110>$ lie between 0.6~\cite{Per92}  
and 2.0~\cite{Hull89}. In works related 
to Si-Ge strained layer structures, $\alpha$ is often
taken~\cite{Houg90} 
equal to 4. The expression of the dislocation energy in Matthews' 
paper~\cite{Matt75} corresponds to $\ln \alpha=1$. A precise expression
of the energy of a 
dislocation near a free surface has been obtained by
Freund~\cite{Freu90}. It relates $U_{disloc}(h)$ to $\phi$, 
the core radius $r_c$ and the core  energy $U_{core}$
which can be evaluated  in semiconductors, using 
atomistic simulations based on
empirical interatomic potentials~\cite{Bour83,Nand90}.

The third term of (\ref{eq:eqhc}), present in the case of a
partial dislocation only  (Fig.~\ref{fig:disloc_partial}) is
\begin{equation}
U_{fault}= \tilde\sigma_{fault} \frac{h}{\sin\phi}
\end{equation}	
where $\tilde\sigma_{fault}$ is the energy per unit area of the stacking fault 
or antiphase boundary associated with the dislocation.

The fourth term of (\ref{eq:eqhc}) is negative and proportional to the misfit,
\begin{equation}
  U_{disloc/film}(h)=\frac{-E}{1-\nu} \left|\frac{\delta
a}{a}\right| hb \sin\beta\cos\phi
\label{eq:udisloc}
\end{equation}	

The  factor $b \sin\beta\cos\phi$  can be justified as follows.
Let the $y$ axis be chosen parallel
to the dislocation, $z$ normal to the interface, and  $x$  perpendicular 
to $z$ and $y$.  With respect to the free
material, the coherent adsorbate is
compressed (if $\delta a/a>0$) or stretched (if $\delta a/a>0$) 
in the directions parallel to the plane. The effect of a misfit
dislocation is to relax a part of this constraint. The amplitude of the 
relaxation is  measured by $b_x=b \sin\beta\cos\phi$. The components
$b_z$ and $b_y$ measure respectively  a strain perpendicular 
to the interface, and a shear, which both do not contribute to the 
relaxation.\footnote{A
detailed calculation would make use of formula (\protect\ref{2.a})
of chapter \ref{tersoff}.}

Using formulae (\ref{eq:udislocc}) to (\ref{eq:udisloc}), 
the equation for the critical thickness (\ref{eq:eqhc}) can be
rewritten  as\footnote{If the denominator in (\ref{mutin}) is negative, the 
critical thickness for the particular type of defect of interest
is infinite, and that type of defect is discarded because the energy cost of 
the fault is larger than the gain due to relaxation.} 
\begin{equation}
   h_c= \frac
     {\displaystyle b(1-\nu\cos^2\beta)\ln(\frac{\alpha    h_c}{b})+
     \frac{8\pi(1-\nu^2)s \gamma}{b E }}
     {\displaystyle 8\pi(1+\nu)\sin\beta \cos\phi 
     \left(\left|\frac{\delta
a}{a}\right|-2\frac{1-\nu}{E}\frac{\tilde\sigma_{fault}}
     { b\cos(2\phi)\sin\beta}\right)}
\label{mutin}
\end{equation}

In Fig.~\ref{fig:hcrit}, the dependence of the critical thickness on
the misfit is 
displayed in the case of a perfect dislocation (1/2)[101] with the glide
plane (111) in a
FCC lattice (case of Fig.~\ref{fig:disloc_parf}). For this glide
system, 
$\beta=60^{\circ}$ and $\phi=54.7^{\circ}$. The step energy has been
neglected. 
The  two values of $\alpha$ used in Fig.~\ref{fig:hcrit} 
correspond to two typical values of the core energy, which depends on
the 
material. The 
critical thickness is seen to decrease rapidly with 
increasing misfit. For a misfit of
about 4\% the 
critical thickness  reaches the lattice parameter and
continuous  elasticity theory  must be replaced by 
calculations taking into account the discrete nature of atomic layers.

In metals the predicted and experimental values of $h_c$ agree fairly
well and are not very large in most of the cases which have been studied. 
However  the critical thickness may be much larger than the 
prediction in some materials. 
For example, under certain growth conditions, 
Ge/GaAs remains coherent up to 2 $\mu$m, while the 
theory gives 300 nm~\cite{Herm96'}. This discrepancy is presumably 
related to the large value of $h_c$.


\subsection{Nucleation of a dislocation at the growing surface}
\label{Marty1b}

If the density of preexisting dislocations in the substrate is not
sufficient 
to relax the stresses during growth by the extension of 
threading dislocations, dislocations must be  nucleated, 
and this involves an activation barrier.
Possible sources for dislocations are:

\noindent - Homogeneous nucleation of half-loops 
(whole or partial) at the free 
    surface of the epilayer. 

\noindent - Homogeneous nucleation of half-loops 
at the substrate/epilayer 
    interface. 

\noindent- Heterogeneous nucleation of complete 
loops at a growth defect 
    in the bulk of the epilayer.

\noindent- Nucleation of half-loops at the 
edges of islands in 3d growth or at 
the free surface defects. 

\noindent- Multiplication of dislocations, for
instance by cross-slip.

The homogeneous nucleation of a dislocation half-loop at the free
surface 
(Fig.~\ref{fig:loop}) is energetically more favourable than the
homogeneous 
nucleation of a complete loop at the substrate/epilayer
interface~\cite{Hirt82}. 
The formation energy ${\cal F}_{tot}$ 
of a half-loop  of radius $r$  on a plane inclined 
at an angle $\phi$ with 
respect to the surface has the form (\ref{mernoire}), where~\cite{Mare87} 

\begin{itemize}

\item 
The self-energy is approximately equal\footnote{There is a close
parallelism between the formulae of this paragraph and those of
paragraph \ref{Marty1a}. For instance, formula (\ref{eq:udislocccc}) can
be obtained from (\ref{eq:udislocc}) by multiplying by the half-loop length 
$\pi r$ and replacing $cos^2\beta$ by 1/2 and $h$ by $r$.}
to
\begin{equation}
{\cal F}_{disloc} = \frac{E \,\left( 1 - \nu/2  \right) \,r\,b^2}{8\,
\left( 1 - \nu^2  \right) }\,
   \ln (\frac{\alpha \,r}{b})
\label{eq:udislocccc}
\end{equation}

\item
The step energy is
\begin{equation}
{\cal F}_{step}= 2\,r\,s\gamma 
\end{equation}
where $s=1$ if a step is created, $s=-1$ if a step is  annihilated.

\item
The stacking fault free energy (in the case of a partial dislocation) is

\begin{equation}
{\cal F}_{fault}=  \frac{\pi \,{r^2}}{2}\,\tilde\sigma_{fault}
\end{equation}

\item
The elastic energy relieved by the loop is the work done by the 
biaxial stress $\{\sigma_{\alpha \gamma} \}$  
    applied to the area 
of the loop for a 
displacement equal to the Burgers
vector $\vec{b}$ : 

\begin{equation}
{\cal F}_{disloc/film}
 = -\hbox{Area} \times \sum_{\alpha \gamma}
 n_{\alpha, \gamma}\sigma_{\alpha \gamma}b_{\gamma} = 
   -{\frac{\pi \,{r^2}}{2}}\,b\,\sigma \,\sin \beta\,\sin \phi\,\cos
\phi
\label{eq:f_el_loop}
\end{equation}
where $\vec{n}$ is the unit vector 
perpendicular to the area of the half loop. 
Near the surface, the non-vanishing  stress components are 
only $\sigma_{xx}=\sigma_{yy}=\sigma$, proportional to
$\delta a/a$ as seen below from formula (\ref{minerve}) . 

\end{itemize}

In addition to the factor $\sin \beta\,\cos \phi$, already present
in (\ref{eq:udisloc}), formula (\ref{eq:f_el_loop}) 
contains the factor    $\sin \phi$. Thus, only inclined glide planes
($\phi \neq 0$ or $\pi/2$) can  be involved in the nucleation process. 

As $r$ increases from 0,  the total 
energy (\ref{mernoire}) of the half-loop increases,
 reaches a maximum for $r=r_c$ and then 
decreases. The critical radius $r_c$ is then given by 
$d{\cal F}_{tot}/dr=0$ or

\begin{equation}
r_c = {\frac{E \,\left( 1 - \nu/2  \right) \,{b^2}\,
       \left[ 1 + \ln (\alpha r_c/b) \right]  + 
       16\,\left( 1 - \nu^2  \right) \,{s\gamma}}{8\,\pi \,
       \left( 1 - \nu^2 \right) \,
      \left( \sigma \,b\,\sin \beta\,\sin \phi\,\cos \phi - 
      \tilde\sigma_{fault}  \right) }}
\label{dynna}
\end{equation}

This gives the energy barrier for dislocation nucleation,

\begin{equation}
{\cal F}_c = r_c\,{s\gamma} + \frac{E \,r_c\,b^2(1 - \nu/2 )\,
       \left[  \ln (\alpha r_c/b) -1 \right] }
       {16\,\left( 1 - \nu^2  \right) }
\label{dynna'}
\end{equation}

At the beginning of the relaxation, the biaxial stress for an
isotropic material is :

\begin{equation}
\sigma  = {\frac{-E }{1 - \nu }}\,
\frac{\delta a }{a}
\label{minerve}
\end{equation}

If $\gamma$ and $\tilde\sigma_{fault}$ are neglected
as well as logarithmic factors, insertion of 
(\ref{dynna}) into (\ref{dynna'}) shows that
${\cal F}_c$ varies as $b^3$. 
This favours nucleation of {\it partial} dislocations, which  have 
smaller Burgers vectors. For 
instance, Dynna et al.~\cite{DYNN96} calculated the loop energy as a
function of 
the loop radius for the epitaxial system Au$_{80}$Ni$_{20}$/Au(001) 
for Shockley
partial 90$^{\circ}$ dislocations (1/6)[112] and perfect 
60$^{\circ}$ dislocation (1/2)[101] 
    (see Fig.~\ref{fig:loop_energy}). 
At this composition, the misfit is 2.3\%. The energy barrier for the
nucleation of 
partial dislocations is clearly lower than that of perfect ones. This
result may 
explain why stacking faults and twins have been observed 
in this system and similar ones.

\subsection{Relaxation by an array of dislocations}
\label{Marty1c}

Above the critical thickness, it is energetically favorable to introduce 
dislocations. The number of dislocations  is limited by their interaction and
depends on the substrate thickness. 
The epitaxial strains are progressively relaxed by
 the formation of a lattice of dislocations located at or near the
interface between 
the epilayer and the substrate. The equilibrium density of dislocations
is reached 
when the total elastic energy ${\cal F}_{tot}$ is minimum. 
If, for the sake of simplicity, the step and fault energies are neglected,
${\cal F}_{tot}$  can be written as the sum of
the elastic energy ${\cal F}_{film}$
of the homogeneously strained film, the self-energy of 
the dislocation lattice ${\cal F}_{disloc}$, the interaction ${\cal F}_{disloc/film}$
between the dislocations and
the film, 
and the interaction ${\cal F}_{disloc/disloc}$ between  dislocations. Thus
\begin{equation}
  {\cal F}_{tot}={\cal F}_{film}+{\cal F}_{disloc}+{\cal F}_{disloc/film}+
{\cal F}_{disloc/disloc}
  \label{eq:eqrelax}
\end{equation}

In homogeneous and isotropic materials, the first three terms are rather
simple, and 
can be written readily. Indeed, ${\cal F}_{film}$ 
is the elastic energy 
of a film strained biaxially by the substrate : 

\begin{equation}
{\cal F}_{film} = \frac{E }{1 - \nu }\,
   \left(\frac{\delta \,a}{a} \right)^2\,h {\cal A}
\label{eq:efilm}
\end{equation}

The self-energy and the dislocation/film interaction energy 
are proportional to the quantities $U_{disloc}$ and $ U_{disloc/film}$
given by 
(\ref{eq:udislocc}) and (\ref{eq:udisloc}).

In  the case of an epilayer relaxed by two 
orthogonal periodic arrays of dislocations, one has
\begin{equation}
{\cal F}_{disloc}+{\cal F}_{disloc/film}={\frac{2{\cal A}}{\bar{d}}}
\,\left( U_{disloc} + U_{disloc/film} \right)
\label{eq:utoe}
\end{equation}
where $\bar{d}$ is the average distance between dislocations
and the factor 2 appears because of the 2 dislocation arrays. 
The most complicated term in (\ref{eq:eqrelax})
is the interaction between dislocations, ${\cal F}_{disloc/disloc}$. 
It depends on the distribution of the dislocations at the interface,
which can be periodic or not. 
Furthermore, due to oblique glide plane, the Burgers vector has a 
component perpendicular  to the surface. This component
does not contribute to the relaxation. Two possible arrays of 
misfit dislocation may be considered:

\begin{enumerate}
\item
An array of
dislocations characterized by the two angles $\beta$ and $\phi$ 
defined in paragraph \ref{Marty1a}, all dislocations having the same 
orientation.
\item
An array of 
dislocations characterized by $\beta$ and $\phi$ in which those 
components of the Burgers vector not relieving misfit strain alternate
in sign. This case is represented in Fig.~\ref{fig:dislocarray}.
\end{enumerate}

Exact expressions have been derived for the energy of an infinite array
of identical 
periodic dislocations taking into account the interactions between 
them~\cite{Will90}. Using this theory, Jain et al.~\cite{Jain92}
calculated the 
strain relaxation. More recently, Dynna et al.~\cite{Dynn94} gave a
closed form 
of the total energy for an alternating dislocation array. Their 
evaluation was 
based on  Head's  analytical calculation~\cite{Head53}  of the stress 
field for a dislocation 
in a semi-infinite, isotropic body. Their results will be presented
below, but let 
us begin with the earlier theories~\cite{Herm96',Ball83,Kasp86} 
which do not take into account the inhomogeneous part of the stress and
strain fields 
associated with dislocations interactions. The total energy is written
as the sum of 
the elastic energy of the epilayer homogeneously strained to
$\epsilon_{res}$ (residual 
elastic strain) and the energy of the set of
dislocations.

\begin{equation}
{{\cal F}_{tot}} = {\cal F}_{homo}(\epsilon_{res}) + 
\rho_{disloc}\,U_{disloc}{\cal A}
\label{eq:etprev}
\end{equation}
where ${\cal F}_{homo}$ is the energy of the homogeneously strained
epilayer 
and
\begin{equation}
{\rho }_{disloc} = 
  \frac{2\,|\,{\frac{\delta \,a}{a}} + \epsilon_{res} \,|}
    {b\,\sin \beta\,\cos \phi} 
\label{eq:rhodisloc}
\end{equation}
is the dislocation density. 
 ${\cal F}_{homo}$ is similar to ${\cal F}_{film}$ in 
(\ref{eq:efilm}),
but the  misfit $\delta a/a$  must be
replaced by the residual strain $\epsilon_{res}$. 
The difference between ${\cal F}_{homo}$ and ${\cal F}_{film}$ takes
into account both 
interaction energies ${\cal F}_{disloc/film}$ and 
${\cal F}_{disloc/disloc}$ 
in a mean field way i.e. only the homogeneous part of the strain 
and stress  due the dislocation
array is considered. 
Then using 
    (\ref{eq:etprev}, \ref{eq:rhodisloc}) and (\ref{eq:udisloc}) 
the total energy 
can be written
as :
\begin{equation}
{\cal F}_{tot} = \frac{E }{1 - \nu }\,h\,
 {\cal A}\,   \epsilon_{res}^2 + 
\left| \,\frac{\delta a}{a} +  \epsilon_{res}\, \right| \,{\cal A}
\,\frac{E \,b\,\left( 1 - \nu \,\cos ^2\beta  \right) }
      {4\,\pi \,\left( 1 - \nu^2  \right) \,\sin \beta \,\cos \phi }\,
    \log \frac{\alpha \,R_{cut}}{b}
\label{eq:etotball}
\end{equation}
where the cut-off radius, $R_{cut}$ is the minimum value of 
    the thickness $h$ 
and the interdislocation average distance $\bar{d}$.
Fig.~\ref{fig:energy_strain} displays the energy versus the residual strain
(lower axis) and dislocation density (upper axis) for several
adsorbate thicknesses. 
Below the critical thickness, the energy 
    exhibits a singular 
minimum for a residual strain equal to the misfit and zero dislocation
density. Above the critical thickness, the equilibrium occurs at
$\epsilon_{eq}$ for 
    a finite 
dislocation density. The critical thickness corresponds to the curve 
with a horizontal tangent when the residual strain is equal to the
misfit.

   When the inhomogeneous part of the stress and strain fields generated
by the 
dislocation array is taken into account to calculate the interaction
energy, the expression of the total energy is much more complicated. We
present here an 
example
of
 such an expression for the case of an epilayer relaxed by two 
orthogonal periodic arrays of alternating
dislocations~\cite{DYNN96,Vass96}
(Fig. \ref{fig:dislocarray}). 
The interaction energy is :

\begin{equation}
   {\cal F}_{disloc/disloc}=\frac{2{\cal A} }{d}
 (U_{disloc,array}^{para}+U_{disloc,array}^{perp})          
   \label{eq:Edislperiod}
\end{equation}
where the factor 2 arises from the two sets of orthogonal dislocations, 
$U_{disloc,array}^{para}$ is the interaction per unit length of one
dislocation 
with all other parallel dislocations and $U_{disloc,array}^{perp}$  is
the 
interaction of one dislocation with all perpendicular dislocations.
Their expressions are

\begin{eqnarray}
\lefteqn{U_{disloc,array}^{para}}\hbox{\hskip 2cm}
=  \nonumber\\
&\quad &
  \  {\frac{E \,{b^2}\,{{\sin }^2}\beta\,{{\cos }^2}\phi}{8\,\pi
\,\left( 1 - \nu^2  \right) }}\, \times
\nonumber\\&\quad&\quad
    \left[ \ln ({\frac{d}{2\,\pi \,h}}\,\sinh ({\frac{2\,\pi \,h}{d}}))
+ 
      {\frac{2\,\pi \,h}{d}}\,\coth ({\frac{2\,\pi \,h}{d}}) - 
      {\frac{2\,{{\pi
}^2}\,{h^2}}{{d^2}}}\,{\mbox{csch}^2}({\frac{2\,\pi \,h}{d}}) -
{\frac{1}{2}} \right] 
   \nonumber\\
&\quad &
   + \frac{E \,b^2\,\sin ^2\beta\,\sin^2\phi}
      {8\,\pi \,\left( 1 - \nu^2  \right) }\, \times
\nonumber\\&\quad&\quad
\left[ \ln ({\frac{d}{\pi \,h}}\,\tanh ({\frac{\pi \,h}{d}})) + 
      {\frac{\pi \,h}{d}}\,\left( \tanh ({\frac{\pi \,h}{d}}) - \coth
({\frac{\pi \,h}{d}}) \right)  
\right.\nonumber\\&\quad&\quad\quad\quad\quad\quad\quad\left.
       - \quad
      {\frac{{{\pi }^2}\,{h^2}}{2\,{d^2}}}\,\left(
{\mbox{sech}^2}({\frac{\pi \,h}{d}}) + 
         {\mbox{csch}^2}({\frac{\pi \,h}{d}}) \right)  + {\frac{3}{2}}
\right]
   \nonumber\\ 
&\quad &
     +  \frac{E \,b^2\,\cos^2\beta}{8\,\pi \,
     \left( 1 - \nu^2 \right) }\,
    \left[\ln ({\frac{d}{\pi \,h}}\,\tanh ({\frac{\pi \,h}{d}}))\right]
   \label{eq:Udislarraypara}
\end{eqnarray}
and

\begin{equation}
  U_{disloc,array}^{perp} = 
  \frac{E \,\nu \,h\,b^2\,\sin^2\beta\,\cos ^2\phi}
  {2\left(1 - \nu^2  \right) \,d}          
   \label{eq:Udislarrayperp}
\end{equation}

Such an expression has been used to calculate the relaxation of MgO
grown on  the (001) surface of bcc iron at room temperature
\cite{Vass96}. 
In this system the misfit is about 3.8$\%$. The results of the
measurements and the calculations are presented on
Fig.~\ref{fig:relaxmgo}. The open circles correspond to the experimental
values of the relaxation as measured by electron diffraction (RHEED)
during the growth. The glide system consists of perfect edge
dislocations of Burgers vector $(1/2)<011>$ gliding on (101) planes. The
curve called `equilibrium calculation' corresponds to the minimization
of the total energy ${\cal F}_{tot}$ (eq. \ref{eq:eqrelax}) with respect
to the distance $d$ between dislocations. 

It is clear that the calculated relaxation within the equilibrium model
is overestimated. The nonperiodic nature of the real dislocations arrays
may explain this discrepancy. Indeed, even if the nucleation is
homogeneous and the initial dislocation distribution is periodic, it is
difficult to maintain the periodicity as the thickness and the number of
dislocations increase. The periodicity can be maintained only if all
existing dislocations move continuously to smaller spacings to make room
for new dislocations. It has been shown \cite{Gosl92} that this process
requires more energy than the introduction of new dislocations in a
nonperiodic way. Numerical calculations have been performed by Jain et
al. 
\cite{Jain92} involving a uniform-random or a Gaussian dislocation
distribution showing that the energy of the dislocation arrays is
increased with respect total energy of periodic arrays having the same
average dislocation density (i.e. the same relaxation ratio). This leads
to a lower or slower relaxation rate. However, the real distribution of
dislocations depends on the mechanism of nucleation and propagation. The
nucleation may be the limiting process, when no threading dislocations
are expected from the substrate of high quality.  Nucleation plays also
an important role in selecting a particular glide system of dislocations. 
The propagation of
dislocations may also be the limiting process due to inhomogeneities of
the stress field present in the layer where the dislocations are moving.
The inhomogeneities of the stress field may result in a retaining force
which may block the dislocation propagation. 
Such a phenomenon, when the source of inhomogeneous stresses are 
the other misfit dislocations, is called  
'blocking mechanism'. It is the subject of the following paragraph.


\subsection{Blocking mechanism}
\label{chbloc}

Just above the critical thickness, dislocations propagate to form two
arrays of 
orthogonal dislocations.
 Freund~\cite{Freu90} has shown that the interaction between a moving 
dislocation ${\cal D}_1$ and an orthogonal dislocation  ${\cal D}_2$
impedes the motion of the
segment ${\cal D}_1$. Then this 
 results in a lower relaxation rate than predicted by previous
equilibrium  calculations. This is shown schematically on
Fig.~\ref{fig:blocking}.  A moving 
dislocation ${\cal D}_1$ encounters a
misfit dislocation  ${\cal D}_2$. The stress field 
generated by the blocking dislocation can completely suppress
the traction  force due to the residual stress in the epilayer. 
To bypass the blocking dislocation, 
${\cal D}_1$ may alter its path by moving 
in the same glide plane but closer to  the surface i.e. in a channel of
width $h^*<h$. This reduction of the thickness reduces
the driving force acting on ${\cal D}_1$.
Furthermore, the retaining line tension increases due to interaction
with the surface  (image forces). 
When ${\cal D}_1$ passes above ${\cal D}_2$,
three forces are acting on it : 

\noindent - the driving force due to the residual homogeneous strain
$\epsilon_{res}$

\noindent - the retaining force due to the line tension of  ${\cal D}_1$

\noindent - the interaction force associated with the stress field 
resulting from  ${\cal D}_2$.

Let us express the retaining forces acting on 
the element of  ${\cal D}_1$ in 
terms of the residual strain needed to just compensate these forces.

i) The residual strain $\epsilon_{line}$ which exactly compensates
the dislocation line tension can be found readily within the homogeneous
strain 
approximation of the relaxation by minimizing 
${\cal F}_{tot}$ in (\ref{eq:etotball}) 
with respect to $\epsilon_{res}$,  
and replacing $h$ and $R_{cut}$ by $h^*$
(which corresponds to neglecting interactions between dislocations) :  

\begin{equation}
{\epsilon_{line}} = 
  {\frac{\left( 1 - \nu \,\cos^2 \beta  \right) \,b\,\log ({\frac{\alpha
\,h^*}{b}})}
    {8\,\pi \,\left( 1 + \nu  \right) \,h^*\,\sin \beta\,\cos \phi}}
\label{eq:epsilonline}
\end{equation}

ii) The residual strain, $\epsilon_{res}$ compensating the 
stress field $\sigma_D(x,z)$ of  ${\cal D}_2$ 
can be expressed in terms of the forces resolved in the glide plane
acting on an elementary segment of the dislocation:
 
\begin{equation} 
\sum_{\alpha \gamma}\sigma _{D,\alpha \gamma}\,n_\gamma\,b_\alpha + 
\sum_{\alpha \gamma}\sigma_{res,\alpha \gamma}\,n_\gamma\,b_\alpha =0
\label{eq:sigmares}
\end{equation}

where $\vec{n}=(1/\sqrt{2})<1\bar{1}0>$ is the glide plane normal unit
vector  and $\vec{n}=(1/2)<110>$  the Burgers vector. The homogeneous
biaxial stress  tensor is related to the residual strain by :

\begin{equation}
{{\sigma }_{res}}= 
  -\frac{E }{1 - \nu }
\left(\begin{array}{clcr}
 \epsilon_{res} &  0 & 0 \\ 
 0 &  \epsilon_{res}  & 0 \\
 0 & 0 & 0
  \end{array}\right)
 \label{eq:sigmavseps}
\end{equation}

Now, the residual strain compensating the stress due to the blocking
dislocation ${\cal D}_2$ for 
an elementary segment of
 ${\cal D}_1$ passing at the position $(x,h^*)$ 
may be written for this glide system as :

\begin{equation}
\epsilon_{res}(x,h^*) = 
  \frac{\left( 1 - \nu  \right)}{E }
   \,\left[ \left( \nu  - 1 \right) \,
\sigma_{D,xx}(x,h^*) + \nu \,\sigma _{D,zz}(x,h^*) \right]
\label{eq:epsilond}
\end{equation}

To bypass the orthogonal dislocation stress field in the channel of
width $h^*$, the residual strain $\epsilon_D(h^*)$ has to be greater
than $\epsilon_{res}(x,h^*)$ for every 
position $x$.
 Then $\epsilon_D(h^*)$ is the maximum value of $\epsilon_{res}(x,h^*)$
when $x$ is varying, this condition may be written :

\begin{eqnarray}
\epsilon_D(h^*)=\epsilon_{res}(x_{max},h^*) 
\label{strumpf}\\ 
\frac{\partial \epsilon_{res}(x_{max},h^*)}{\partial x}= 0 
\label{eq:xmax}
\end{eqnarray}

Since linear elasticity theory has been used, 
the two residual strains required to balance 
both retaining 
forces should be added to obtain the total residual strain as a function of
the channel 
width $h^*$. Since the channel width is a free variable, the actual
residual strain 
below which the moving dislocations are stopped is :

\begin{equation}
\epsilon_{min}=\hbox{Min}_{h^*} 
\left[ \epsilon_{D}(h^*)+\epsilon_{line}(h^*) \right]
\label{eq:epsilonmin}
\end{equation}

Relations (\ref{strumpf}), (\ref{eq:xmax}) and (\ref{eq:epsilonmin})
constitute Freund's blocking criterion.
They  mean that the residual strain corresponds to
a saddle point of the function $\epsilon_{res}(x,h^*)+\epsilon_{line}(h^*)$
in  the two-dimensional space ($x$,$h^*$).
It has been found \cite{Gill94} that Freund's blocking criterion 
predicts a residual  strain which is in good agreement 
with the strains found experimentally in 
Si$_{1-x}$Ge$_x$ film on Si(001). We have applied this blocking criterion
to the relaxation of MgO grown on Fe(001) presented  
above~\cite{Vass96}. The expressions of the stresses due to the blocking
dislocation may be derived~\cite{Gosl92}
from the Airy stress function as done
by Freund~\cite{Freu90} or more directly obtained from Dynna
et al.~\cite{Dynn94}. 
Fig.~\ref{fig:relaxres} displays the residual strains 
$\epsilon_D(h^*)$, $\epsilon_{line}(h^*)$ and the sum of these, 
$\epsilon_{tot}(h^*)$ for a 100 
\AA $\;$ layer of MgO deposited on Fe(001).
Both residual strains behave as expected. 
The residual strain compensating the line tension is maximum near the
surface and  then decreases when the channel width, $h^*$ increases. 
On the contrary, the residual 
strain compensating the orthogonal dislocation stress field increases as
the channel 
width approaches the layer thickness $h$. The minimum of the total
residual strain occurs at about $h^*=70 
$ \AA $\;$ and $\epsilon_{tot}=1.5\%$. Then at this layer thickness, the
expected relaxation is 
estimated to $2.1\%$ (the misfit between MgO and Fe is $3.6\%$). 
This value is in good agreement with the experimental one. On the
Fig.~\ref{fig:relaxmgo}, the relaxation calculated from the blocking
criterion is  represented. 
It is in better agreement than the equilibrium one presented above at
least for high 
thicknesses. Although Freund's criterion predicts no relaxation when
$h<$ 27 to 30 \AA,  
relaxation is clearly observed experimentally when $h>20 $ \AA.
This relaxation is possible because the density of blocking misfit
dislocations is 
insufficient to block all of the threading segments present in the MgO
film. Above 
50 \AA $\;$ the agreement between theory and experiment improves
continuously and 
the agreement is excellent at about 100 \AA. 
The experimental results presented in Fig.~\ref{fig:relaxmgo}
correspond to room 
temperature growth. For layers grown at higher temperature, the
relaxation is 
faster~\cite{Vass96,Gosl92}. This can indicate that nucleation is enhanced
or that the blocking 
barrier is overcome by thermal activation. 
This example shows that other aspects
of relaxation must be taken into account, e.g. 
the frictional stresses~\cite{brad90}, 
the non-periodic nature of the dislocation array~\cite{Jain92} or the
multiplication mechanism \cite{bean95}. The prediction 
of the relaxation rate depends on the growth 
conditions and is  currently an active field of research.


\section{The Asaro-Tiller-Grinfeld instability and related phenomena}
\label{tersoff}

\subsection{Three-dimensional clusters as an alternative to
misfit dislocations}
\label{tsf1}

\subsubsection{Experimental facts}
\label{tsf1a}

If a crystalline material is coherently adsorbed on a substrate which
has a different 
atomic distance, it does not like it and tries to escape from this 
unpleasant situation which costs energy. There are several ways to
escape.
The most usual way is to make misfit dislocations (Fig. \ref{2instab}a)
as seen in the previous chapter. 
However,   mismatched  epitaxial films  can  relax their strain 
through a different mechanism, namely a deformation of the surface 
or `2d-3d transition'.  In  III-V semiconductors, for example,
this frequently  occurs for a lattice mismatch above typically 2\% in
layers
under compressive stress and for lower mismatch in layers under tensile
stress. If the  misfit is moderate, this deformation 
leads to alternating hills and valleys  (Fig. \ref{2instab}b
and \ref{ponch2}) while  for a
 stronger misfit, the adsorbed film splits into clusters (Fig.
\ref{ponch1}). These clusters  will
be studied in the next chapter.

The  base of each hill is still highly constrained by 
the substrate, and the elastic energy density is high. 
However, at the top and
on the sides of the hills, the adsorbate lattice constant is closer
to that of the free adsorbate: the adsorbate is said to be `relaxed'.
The adsorbate is still coherent although dislocations always form in the
later 
stages of the growth. The gain in elastic energy at the top of the
clusters is the reason of the deformation, as will be seen more
quantitatively in this chapter and in the following one.

For a given substrate and a given adsorbate, stress relaxation can occur
through misfit dislocations or, without loss of coherence, through
surface deformation  according to the
growth conditions.
For instance, in InAs/GaAs or
InAs/InP systems,  cluster formation precedes  dislocation formation on
the (100) substrate, and on several high index surfaces\cite{notzel,notzel',nishi}.
On the contrary it is not observed on the
other low index surfaces, e.g.\cite{joyce} (110), (111)A and (111)B. 
 A Ge deposit forms coherent clusters on Si(100), while it
directly forms large incoherent clusters on Si(111)\cite{Goldfarb}. The
substrate orientation obviously determines the surface energetics and
kinetics.
This hesitation of nature between the two kinds of instability 
has been analyzed for instance by Joyce et al.~\cite{joyce} and is not 
very well understood. 

Nevertheless, we will try to understand a part of these phenomena.


\subsubsection{Energetics}
\label{tsf1b}

It is of interest to wonder about the ground state of a coherent,
epitaxial film, even though it is never reached. Let us first
ignore the interface energy and assume that there is only a surface
energy and an elastic energy. For a large thickness, the elastic energy
is 
proportional to the volume and dominates the surface contribution,
so that the adsorbate tends to separate from the substrate and to form 
a single big cluster  almost detached from the substrate
(Fig. \ref{clus}, last picture).   This state obviously minimizes the
elastic 
energy, and therefore the total free energy.

In reality, the ideal ground state can of course not be reached since
the 
atoms would have to diffuse on very long distances, but, if the
deposit is thick enough, big clusters can form and lower the
energy.

The waves of figure \ref{2instab}b and  \ref{ponch2}
are presumably the incipient stage of the instability 
which leads to isolated clusters.

However, it turns out that big, coherent clusters
(Fig. \ref{clus}, last two pictures) are never observed. 
Dislocations appear
in the cluster when it reaches a critical 
size~\cite{madhux,chen,pon4,freund3} 
(Fig. \ref{ponch1}) but the above argument 
helps us to understand why there is an instability, and also
to predict the evolution of a population of 
clusters --a problem which will 
be addressed  in  chapter \ref{dots}. Moreover, even though
shapes corresponding to the  last two pictures of
Fig. \ref{clus} are never observed,
the shape of
coherent clusters does depend on the size, in contrast with 
incoherent clusters of chapter \ref{wet}. In particular,
it will be seen in the next chapter that small clusters, dominated
by the surface energy, are `platelets' (Fig. \ref{clus}, first picture).

To close these qualitative remarks on energetics, it is appropriate to 
consider the interface energy. When one wishes to make good 
epitaxial layers, one needs to have a strongly
negative interface energy $\sigma_{sa}$, so that (\ref{young}) is 
satisfied. 
This highly negative value is obtained by the choice of the materials,
of the temperature, the fluxes, the growth rate, etc.
If the interface energy is sufficiently negative, the first adsorbed
monolayer(s) is/are stable even if the misfit is fairly strong.
This stable layer is called a `wetting layer'~\cite{koba}.
It is sometimes related to complex chemical reactions, e.g. in
InAs on GaAs, where a (In,Ga)As alloy forms~\cite{joyce}.

In the simplest cases, 
the existence of the wetting layer implies only a small modification of
the above discussion. Cluster formation
and wavy deformations do lower the free energy,
but only after  the wetting layer has formed. This picture applies 
to Ge/Si. In more complicated cases, there is experimental evidence 
that a considerable transfer of matter takes place from the wetting
layer to the clusters. This seems to be the case in InAs/GaAs(001). 
This feature can be taken into account by simple theories~\cite{dobbs}
but, for the sake of simplicity, it will be ignored in the remainder of
this review.


\subsubsection{Kinetics}
\label{tsf1c}

The thermodynamical analysis is probably not sufficient to 
determine the structure of an epitaxial deposit.
Kinetic factors are presumably often essential,
for instance for the choice between misfit dislocations 
and a three-dimensional modulation. 
The crystal chooses the 
fastest relaxation path in the phase space. 
The formation of misfit dislocations 
requires, as seen in the previous chapter, 
a purely microscopic atom motion inside 
the material, on a distance of the order of the atomic distance.
In contrast,
the formation of a three-dimensional structure involves atomic diffusion
on the surface (which is much easier than in the bulk) but on much
larger distances,
typically hundreds of atomic distances.  
It is difficult to say which mechanism will be preferred.

A way to make the discussion more quantitative is to calculate the
activation energy. That of misfit
dislocations has been calculated in the previous chapter.
The activation energy for the formation of a coherent cluster
will be evaluated 
in the next one in the case of a singular surface.

\subsubsection{Singular and non-singular orientations}
\label{tsf1d}

In this chapter, the onset of the 2d-3d instability will be studied 
in the only case where a quantitative theory
exists,~\cite{Asaro,Grinfeld}
namely the case of a non-singular surface ($T>T_R$,
see chapter \ref{Intro}).

Unfortunately, most of the experiments and most of the technological
applications, are concerned with (001) 
surfaces which are believed to be singular 
at the growth temperature.  Nevertheless, experiments on non-singular
surfaces would be feasible, and this justifies the 
forthcoming study. Moreover, the wavy deformations observed in
weak misfits are in qualitative agreement with the theories of this chapter,
as will be seen.

It will be shown in the
present chapter that a plane surface of  non-singular
orientation is unstable with respect to 
spontaneous deformations of wavelength
larger than some value $\lambda_c$. This value will be evaluated.
This instability of a  non-singular surface
will be found to involve no activation energy, 
although the time of establishment of the 
instability will be seen to be very long for small misfits. The case of
singular
surfaces, e.g. the (001) surface of body centered cubic crystals, will
be 
investigated in the next chapter and their instability
will be seen to be subject to an activation energy. 


\subsection{The Asaro-Tiller-Grinfeld instability: thermodynamics}
\label{tsf2}

The stability of an initially planar system was investigated first by 
Asaro and Tiller~\cite{Asaro} and later in great detail by 
Grinfeld~\cite{Grinfeld} 
and other authors~\cite{Nozieres,Srolovitz,Yang}.

The simplest possible approach is the following: one calculates the
energy of 
the system (substrate + adsorbate) when its 
surface has a weak sinusoidal plastic 
modulation of amplitude $\delta h$ (Fig. \ref{grfld}a)
and  compares this energy to that of the planar surface ($\delta h=0$).
Here, the word `plastic'  means that the height modulation results from 
an irreversible modulation of the number of atoms, in contrast with an elastic
strain.\footnote{The word `plastic' is often understood as implying dislocation
formation. It is not the case here.}  
If the modulation produces an energy decrease, the planar surface is
unstable.
The method is in fact applicable to any  {\it weak modulation},
because such a modulation can be Fourier transformed, and the strains
produced
by the various Fourier components below the modulated region 
can be calculated by linear elasticity theory
and  can just be added as in any linear problem. This additivity rule is
not
applicable in the modulated region but it does not matter since this 
region is very thin. This approach, in which the problem becomes linear,
is called {\it linear stability analysis}. This method has already been
used in 
chapters \ref{PP1} and \ref{PP2}. 

In this approach, the height of the surface can be assumed sinusoidal,
namely
\begin{equation}
z(x,y) = \langle z \rangle + \delta h \cos(qx)
\label{sinus}
\end{equation} 
where $q$ is a real, positive number.

Before going further, one should mention that the assumption of a
sinusoidal,
and therefore {\it continuous}
modulation is acceptable only above $T_R$. 
Below $T_R$, continuous height changes are not possible as noticed in
chapter
\ref{Intro}. In the remainder of the present chapter, the temperature
$T$
will be assumed higher than $T_R$. The case $T<T_R$, which presumably
corresponds
to a defect-free, high symmetry crystal surface, will be addressed in
the next 
chapter \ref{dots}. 
The applicability of the various theories to real surfaces will 
be addressed in paragraph \ref{tsf2'}.

It will be seen that (if gravity is neglected) the modulation
(\ref{sinus}) 
lowers the energy if $q$ is small enough, so that the planar
surface is {\it always} unstable. 

This energy gain comes from the fact that, in  the
modulated part of the adsorbate, one can  strain the material, so that
its  lattice constant becomes closer to the one 
it would like to have (i.e. that of the free adsorbate). 

The free energy can be most conveniently written if the strain is
measured with respect to the completely  strained state, 
in which the lattice constant 
in the $x$ and $y$ directions is that of the substrate.\footnote{
If it were measured from the free adsorbate, it would be
discontinuous at the interface.  Thus,
the quantity denoted $\epsilon$  in this chapter and the following ones
was called $\delta a/a + \epsilon_{res}$ in chapter \ref{Marty1}.}
If it were measured from the free adsorbate, it would be
discontinuous at the interface. 
The free energy change
resulting from the strain  can be written as

\begin{equation}
\delta {\cal F}_1 = - \hbox{Const} \times \frac{\delta a}{a} \int_A d^3r 
[\epsilon_{xx}(\vec r) + \epsilon_{yy}(\vec r)] 
\label{2.a}
\end{equation}
where $\int_A$ denotes the integral over the adsorbate. 
The strain $\epsilon_{\alpha \gamma}(\vec r)$ is measured with respect to the 
completely  strained, coherent state, in which the lattice constant 
in the $x$ and $y$ directions is that of the substrate.
A positive misfit corresponds to an adsorbate bigger than the substrate,
so that the coherent adsorbate is compressed by the substrate 
(in the directions parallel to
the surface). Relaxing this compression
corresponds to
a positive strain  (from the completely  strained state)
 and to a negative $\delta {\cal F}_1$.
The constant depends on the elastic constants of the adsorbate. The
integral
on $x$ and $y$ vanishes for complete layers, so that (\ref{2.a}) is
proportional to  the thickness $\delta h$
of the modulated region. It is also proportional to 
the  strain modulation amplitude $\epsilon$, the misfit
and of course  to the area ${\cal A}$. Therefore

\begin{equation}
\delta {\cal F}_1 = -C_1 E_a \delta h \frac{\delta a}{a} \epsilon {\cal
A}
\label{2.1}
\end{equation}
where $E_a$ is a typical  elastic constant (e.g. the Young modulus) of
the 
adsorbate and $C_1$ is a positive constant of order unity. The
detailed calculation for an isotropic elastic medium\footnote{The
calculation might be done for a 
cubic crystal bounded by a (001) surface, using the
corresponding elastic Green's functions~\cite{por77}.} 
is available in original articles
and textbooks~\cite{Nozieres,Asaro,Grinfeld,Srolovitz,Yang}. 
The result is 
\begin{equation}
C_1 = 1/(1-\nu_a)
\label{C_1}
\end{equation}
where $\nu_a$ is the Poisson ratio of the adsorbate.

The energy gain (\ref{2.1}) is partially compensated by the quadratic
part of the
elastic energy, which can be written as

\begin{equation}
\delta {\cal F}_2 =  \frac{1}{4}\sum_{\alpha,\gamma,\xi, \zeta} \int
d^3r 
\Omega_{\alpha,\gamma}^{\xi, \zeta}(\vec r)
\epsilon_{\alpha,\gamma}(\vec r)  \epsilon_{\xi, \zeta}(\vec r) 
\label{2.1'}
\end{equation}
where $\alpha$, $\gamma$, $\xi$, $\zeta$ =$x,y,z$ and 
$\Omega_{\alpha,\gamma}^{\xi, \zeta}$, are the elastic constants,
which are (known) functions of $\vec{r}$ 
because they are not identical in the substrate and in the adsorbate. 

The exact minimization of the total elastic energy 
$\delta {\cal F}_1+\delta {\cal F}_2$
would be a hard task, on which mathematicians of the nineteenth century,
e.g. Green and Dirichlet, have concentrated their efforts. For a small
modulation 
$\delta h$, the problem is
easy~\cite{Nozieres,Asaro,Grinfeld,Srolovitz,Yang}.
The strain which minimizes $\delta {\cal F}_1 +\delta {\cal F}_2 $ when
the
surface is given by (\ref{sinus})
is expected to be the real part of an expression of the form
\begin{equation}
\epsilon_{\alpha \gamma}(x,y,z) = \epsilon^0_{\alpha \gamma}
\exp(iqx)f(z)
\label{sinuss}
\end{equation}
where $\epsilon^0_{\alpha \gamma}$ and $f(z)$ can be determined from 
the equations of elasticity which have the form~\cite{Landau} 
\begin{equation}
\sum_{\gamma \xi \zeta}\Omega_{\alpha \gamma}^{\xi \zeta}\partial_\gamma 
\partial_\xi u_\zeta =0
\label{eqel}
\end{equation} 
 
According to (\ref{sinuss}), the operator $\partial_x$
should be replaced by $iq$ and $\partial_y$ by 0. Therefore,
(\ref{sinuss})
can be solved by replacing $\partial_z$ by $\alpha_0 q$, where the
constant 
$\alpha_0 $ depends on the elastic constants, and should have a positive
real part 
since  the perturbation has to vanish for $z = -\infty$. In the case of
an isotropic 
elastic medium, $\alpha_0=1 $~\cite{Nozieres,JVAP}. In all cases, the
penetration depth 
of the perturbation is proportional to the wavelength $2\pi/q$.

The quadratic part of the elastic energy in the bulk is thus
proportional to $1/q$. In addition it is proportional to 
$\epsilon^2$ and to the area ${\cal A}$. Finally,

\begin{equation}
\delta {\cal F}_2 = C_2 E \epsilon^2  {\cal A}/q
\label{2.2}
\end{equation}
where $E$ can be chosen as the Young modulus of the substrate,
and the constant  $C_2$ is of order unity. If the wavelength is much
longer 
than the adsorbate thickness $h$, $C_2$ does not depend 
on the elastic constants of the adsorbate.

In the case of a isotropic solids~\cite{Nozieres,JVAP},

\begin{equation}
EC_2 = \frac{E}{1-\nu^2}
\label{C_2}
\end{equation}
where $\nu$ and $E$ are  the Poisson ratio of the adsorbate (if $qh\gg 1$)
or of the substrate (if $qh\ll 1$). If both materials have identical 
elastic constants $E$ and $\nu$, (\ref{C_2}) applies without restriction.

Minimization of the total elastic energy
$\delta {\cal F}_{el}=\delta {\cal F}_1+\delta {\cal F}_2$ 
with respect to $\epsilon$ yields (assuming $E_a=E$, which is sufficient 
for our qualitative analysis)

\begin{equation}
  \epsilon =  \frac{q C_1}{2C_2} \frac{\delta a}{a} \delta h
\label{2.3}
\end{equation}
and the total elastic energy
$\delta {\cal F}_{el}=\delta {\cal F}_1+\delta {\cal F}_2$ 
is
\begin{equation}
\delta {\cal F}_{el} = -\frac{q \delta h^2 C_1^2}{4C_2} 
 \left(\frac{\delta a}{a}\right)^2 E{\cal A}
\label{2.4}
\end{equation}

Thus, the elastic free energy change resulting from the sinusoidal 
deformation is negative.

However, there is another contribution to the free energy, which opposes
the sinusoidal modulation. Indeed, this modulation 
increases the surface area, 
and therefore the surface energy or `capillary' energy.  The
area increase is easily seen to be ${\cal A} \delta h^2q^2/4$,
and the capillary energy increment $\delta {\cal F}_{cap}$ is obtained by 
multiplying by a coefficient $ \tilde\sigma$, 
called {\it surface stiffness}. Hence,

\begin{equation}
\delta {\cal F}_{cap} =  {\cal A} \tilde\sigma \delta h^2q^2/4
\label{capi}
\end{equation}

 If the strain is small, its effect on 
$ \tilde\sigma$
can be neglected and the total energy is the sum 
\begin{equation}
{\cal F}_{tot}= {\cal F}_{cap}+{\cal F}_{el} 
\label{wtot}
\end{equation}

Thus, the total energy increment per unit area is 
obtained by adding (\ref{2.4}) and (\ref{capi}), and equal to

\begin{equation}
\delta {\cal F}_{tot}/{\cal A} = -\frac{q \delta h^2 C_1^2}{4C_2} 
 \left(\frac{\delta a}{a}\right)^2 E +  \tilde\sigma
 \delta h^2q^2/4
\label{2.5}
\end{equation}

We conclude that, if a material is adsorbed on a substrate which has a 
different lattice parameter, its surface cannot be planar. The planar
shape 
is unstable with respect to modulations of wavelength 
larger than

\begin{equation}
\lambda^\ast = \frac{2\pi C_2}{C_1^2} 
 \frac{ \tilde\sigma}{E}
 \left( \frac{a}{\delta a}\right)^2
\label{2.5A}
\end{equation}
since (\ref{2.5}) is negative is that case.

It follows from (\ref{2.5}) and  (\ref{2.5A}) that the free energy gain
per 
unit area is proportional to $(\delta a/a)^4$ for a given amplitude 
$\delta h$.

In the instability discussed in this paragraph, 
the effect of the substrate is merely to produce an anisotropic,
external stress.
Such a stress can also be produced mechanically. This is the situation 
addressed for instance by Nozi\`eres~\cite{noz'}, Grilh\'e~\cite{grilh}
and 
Kassner \& Misbah~\cite{kkm} in
their theoretical, nonlinear analysis, and by 
Thiel et al~\cite{thiel} in their experiments on He. An instability
arises in
both cases, but the long time
evolution is quite different. In a mechanically
stressed homogeneous material, the amplitude of the surface modulation
can be very large. In contrast, when the stress is due to a substrate, 
the modulation is not expected to penetrate into the substrate.
The former situation will not be addressed in the following, while the
consequence 
of the presence of the substrate-adsorbate interface 
below the free surface will be discussed in paragraph \ref{tsf3}.

If, as we have just seen in this paragraph, 
the initially planar surface is unstable
with respect to modulations of all wavelengths longer than (\ref{2.5A}), 
the actual structure which will appear is not yet clear.
For short times, one can expect that the
wavelength which will dominate the actual structure will be the one
which develops
more rapidly than the other ones. Thus, the preceding thermodynamic
study
is not sufficient and a kinetic treatment is necessary.


\subsection{The Asaro-Tiller-Grinfeld instability: kinetics}
\label{tsf2''}

\subsubsection{Classical theory near equilibrium}
\label{tsf2''a}

  Without calculation, 
we expect that fluctuations of longer wavelength require a longer time
to form, since 
the atoms should come from longer distances to form the modulation.
A more precise study of the evolution
in time of the modulation (\ref{sinus}) was initially done by 
Spencer et al.~\cite{spencer,spencer'} and is reproduced below. 

The translational invariance of the problem 
implies that, at short time, a small initial sinusoidal deformation
remains 
sinusoidal, namely

\begin{equation}
z(x,y,t) = \langle z \rangle + \delta h(t) \cos(qx)
\label{sinust}
\end{equation}

Insertion of  (\ref{Lan_eq}) into (\ref{sinust}) yields, neglecting the
noise,
\be
\frac{d \delta h}{d t} \cos(qx) = -\partial j(x,t) / \partial x
\label{spen1}
\ee
where the surface current density $j(x,t)$ is usually
assumed~\cite{spencer} to be related to the 
chemical potential $\mu$ by a linear relation as explained in paragraph 
\ref{Mul_cur}, 
\be
j(x,t) = - \Upsilon  \partial \mu(x,t) / \partial x
\label{spen3}
\ee

Relations (\ref{spen1}) and (\ref{spen3})  imply that the chemical
potential $\mu$ has the form

\be
\mu(x,t) =  \mu_1(t) \cos(qx)
\label{spen4}
\ee
where the function $\mu_1(t)$ is related to $\delta h(t)$
by the following  relation which follows from formulae 
(\ref{spen1}) to (\ref{spen4}).

\be
\frac{d \delta h}{d t} = -\Upsilon q^2 \mu_1(t) 
\label{spen11}
\ee

On the other hand, $\mu(x,t)$ is  the free energy per atom, so that
the free energy variation per unit time is,
if  $a^2c$ is the volume per atom
\be
 \frac {d {\cal F}}{d t} = (a^2c)^{-1}
\int dx dy \mu(x,t) \frac {\partial z(x,t)}{\partial t}
= (a^2c)^{-1} \frac {d h(t)}{d t} \mu_1(t) \int dx dy  \cos^2(qx) 
=  \frac{1}{2} (a^2c)^{-1} {\cal A}  \frac {d h(t)}{d t} \mu_1(t)  
\label{spen13}
\ee

Now, $d{\cal F}/dt$ can be related
to the free energy calculated in paragraph \ref{tsf2}
for a non-singular surface. In order to extend the theory to
singular surfaces, we shall assume, following 
textbooks~\cite{Landau'} that, as in any system, the free energy of weak
fluctuations is quadratic with respect to the variables which describe
the system, which are here the height variations. Because
of translational invariance, it is appropriate to use the Fourier 
components $z_k$, which diagonalize the free energy increment

\be
\delta{\cal F} = \sum_k B(k) |z_k|^2
\label{spen14}
\ee

If (\ref{sinust}) holds, only the components $z_q$ and $z_{-q}$ are
appreciable and

\be
\delta{\cal F}(t) =  B(q) {\cal A} \delta h^2(t)
\label{spen15}
\ee

Taking the time derivative of (\ref{spen15}) and comparing with
(\ref{spen13}), one finds

\be
 \mu_1(t)  =  4 a^2c B(q)  \delta h(t) 
\label{spen16}
\ee

For a rough (i.e. non-singular) surface, $B(q)$ is obtained if 
one identifies  (\ref{spen15}) with (\ref{2.5}).
This yields $B(q)= - B_1|q| + B_2q^2$, where 

\begin{equation}
B_1  =  \frac{  C_1^2}{4C_2} \left(\frac{\delta a}{a}\right)^2 E 
\hbox{~~~~~and~~~~~} B_2= \tilde\sigma/4
\label{spen17}
\end{equation}
We have written  $|q|$ instead of $q$ in
order to include 
the case $\vec{q}=(q_x,q_y)$ with $|q|=q_x^2+q_y^2$.

An appropriate treatment of  a singular surface requires renormalization group
methods \cite{Nozieres,tang}. If one wishes to preserve simplicity
at the expense of rigor, one can 
introduce into (\ref{spen14})
a positive coefficient $B_0$ of order 0.

\begin{equation}
B(q)= B_0-B_1|q| + B_2q^2
\label{spen20(t)}
\end{equation}

This heuristic form does not satisfy invariance under the translation
$z \; \rightarrow \; z+c$, but correctly describes the stability with
respect to deformations of any wavelength if $T$ is sufficiently
lower than $T_R$.  Moreover, the form (\ref{spen20(t)})
is justified in the presence of a substrate
since translational invariance is suppressed. We shall come back to
this point in  paragraph \ref{tsf3}.
 For an infinite adsorbate 
thickness, the coefficient $B_0$  vanishes at and above 
$T_R$.\footnote{Formula (\protect\ref{spen20(t)}) is only acceptable for
fluctuations
near equilibrium. The energy of a bump of thickness equal or larger 
than the atomic distance
on a singular surface is given by a quite different formula
as will be seen in the next chapter. }

Insertion of (\ref{spen16}) into (\ref{spen11}) yields

\be
\frac{d \delta h(t)}{d t} = -4\Upsilon a^2c B(q) q^2 \delta h(t) 
\label{spen21}
\ee
the solution of which is

\be
\delta h(t) = \delta h(0) \exp (\omega_q t )
\label{spen22}
\ee
with
\be
\omega_q = -4  a^2c \Upsilon B(q) q^2 = -4\Upsilon a^2c ( B_0-B_1|q| +
B_2q^2)q^2
\label{spen23}
\ee

The condition of stability of a plane surface is $\omega_q<0$
for any $q$. 
For a non-singular surface, $B_0=0$, $\omega_q$ is always positive 
for small $q<2\pi/\lambda^\ast$, given by (\ref{2.5A}), and the plane 
surface is unstable in agreement with paragraph \ref{tsf2}.

Moreover, as expected, the absolute value of
expression (\ref{spen23}) decreases rapidly with $q$
for $q<2\pi/\lambda^\ast$, so that long wavelength components $z_q(t)$
develop very slowly. The components which will actually appear are those of 
wavelength slightly larger than $\lambda^\ast$.
Those which are  shorter are stable, those which are much larger
are too slow.

Even in the case of a
singular surface, there may be a window 
$\lambda^\ast<\lambda<\lambda^{\ast \ast}$, in which (\ref{spen23})
is positive and the surface linearly unstable (i.e. without 
an activation barrier). In the absence of growth, this can probably 
occur only very near $T_R$. 
It will  be argued in the next chapter that MBE growth conditions can 
alter this conclusion.

Before doing that, it is of interest to give more details
on the instability. 
The  above formulae tell us how an initial perturbation evolves,
but say nothing about the initial perturbation. In a somewhat 
heuristic approach {\it \`a la} Langevin, one can assume that it is produced 
randomly by some `noise' and write an equation analogous to
(\ref{Lan_eq}). Eliminating the current as above and 
performing a Fourier transformation, one obtains

\begin{equation}
\frac{\partial z_q}{\partial t} = \omega_q z_q + \varphi_q(t)
\label{rumore}
\end{equation}
where the `noise' $\varphi_q(t)$ satisfies the following relations
analogous to (\ref{bruit})

\begin{equation}
\langle \varphi_q(t) \rangle =0\;\;\;\;,\;\;\;\;
\langle \varphi_q(t) \varphi_{q'}^\ast(t') \rangle =
\varphi_0^2 \delta(t-t') \delta_{qq'}
\label{rumoroso}
\end{equation}

In the absence of noise, $\varphi_0=0$, (\ref{rumore})
yields (\ref{spen22}). In the presence of noise, integration of
(\ref{rumore}) yields

\be
z_q(t) = \int_0^t e^{\omega_q(t-t')} \varphi_q(t') dt'
\ee
and use of (\ref{rumoroso}) gives


\begin{equation}
\langle z_q(t) z_{q'}^\ast(t) \rangle = \delta_{qq'} 
\frac{\varphi_0^2}{2\omega_q} \left[ e^{2\omega_qt} - 1 \right] 
\label{clamore}
\end{equation}
which describes the surface morphology at time $t$.


\subsubsection{High symmetry surfaces and MBE growth}
\label{Calvin}

The previous calculation applies near equilibrium. It is therefore not 
clear that it can be applied to a growing crystal. 

The theory of the previous  paragraphs applies to non-singular
surfaces near equilibrium. Instabilities are generally observed during growth, 
e.g. MBE, and the surfaces are often believed to be singular. In this
paragraph,  attention will
be focussed  on high symmetry orientations, (001) or (111).

Such surfaces are quite different in the equilibrium state and during growth.
The most obvious difference is the existence of long, closed steps which bound 
large terraces or valleys which appear and die at the frequency $F_0a^2$
of the RHEED oscillations. Another difference is that the average adatom density 
can be much larger than at equilibrium, and this can increase the
absolute value of the average current,
and even change its sign, as will be seen. 

The effect of  short-lived terraces is difficult to analyze but can presumably 
favour an instability since  these terraces create a kind of 
roughness, which actually increases with time~\cite{JVAP} and evolves toward 
true roughness, with diverging height fluctuations.  

The remainder of this paragraph is devoted to the effect of the current due to 
freshly deposited adatoms from the beam, which have not yet been
incorporated by a step. The surface density $\rho_{inc}(x,y,t)$ 
of these adatoms  can be much larger than the 
equilibrium adatom density $\rho_0(T)$, in particular in the growth of
an element (e.g. Si or Ge or a metal) when $\rho_0(T)$ is 
given by  (\ref{rho0}). 
In the case of metals,  values of the adatom energy $W_0$ are
given by Stoltze~\cite{Stolze}. The  density $\rho_{inc}(x,y,t)$ in MBE
is discussed in App.~\ref{ell_D}. At low temperature, it
depends only on the ratio $F_0/D$ of the flux rate
to the diffusion constant~\cite{JVAP,Ven,SK} and is much larger
than $\rho_0(T)$. This inequality is not so clear at the temperatures 
used in MBE growth. In the case of III-V semiconductors,
it has been argued~\cite{teretal} 
that the equilibrium adatom density, given by  (\ref{rho0'}), can be
large, of the same order as the  adatom density during growth. 

In this paragraph,  $\rho_0 $ will be assumed to be much smaller 
than $\rho_{inc}$, and the consequences of this situation will be investigated.
The current $j_{inc}(x,y,t)$ of freshly deposited adatoms on the surface
contains a part
$j_{inc}^{el}(x,y,t)$ which results from elasticity, and a part  
$j_{inc}^{(2)}(x,y,t)$ which results from other factors, in particular
terrace nucleation.
The latter has been discussed  in chapter \ref{PP1}, where its main
effect was found to be a modification of the coefficient $B_2$ in
(\ref{spen20(t)}).
In the absence of Ehrlich-Schwoebel effect, $B_2$ is increased with respect to
its equilibrium 
value. The  Ehrlich-Schwoebel effect  is generally weak at usual growth
temperatures, at least in the case of semiconductors, and will be neglected in 
this chapter and in the following ones.
The part which results from elasticity can be roughly written
as proportional
to the average $\bar{\rho}_{inc}$ of the incoming adatom 
density.\footnote{In reality, $\rho_{inc}(x,y,t)$ is strongly modulated
and
vanishes in the neighbourhood of a step. The elastic potential is also
modulated. 
In a correct treatment, the 
current should be proportional to the average value of 
$\rho_{inc}(x,t){\cal V}_{el}(x,t)$
on a distance large with respect to the distance between steps. Here, we
approximate 
the average value of this product by the product of the average values.
A more 
accurate treatment\protect\cite{DP'V} has been done in the case of 
a vicinal surface and leads to the 
prediction of step bunching if the coefficient $K_1$ of formula
(\protect\ref{huss12}) 
has the sign of $\delta a /a$. In the case of a high symmetry surface,
step bunching is likely to appear too, thus leading to higher harmonics
to be added
to (\protect\ref{sinust}). This effect does not show up in 
the linear stability analysis.}

\begin{equation}
j_{inc}^{el}(x,t) = -\beta D \bar{\rho}_{inc} \partial 
{\cal V}_{el}(x,t)/\partial x
\label{huss11}
\end{equation}
where ${\cal V}_{el}$ is the elastic potential energy of an adatom. It
has to be proportional to the strain,
  
\begin{equation}
{\cal V}_{el}(x,t) = - K_1 \epsilon(x,t) 
\label{huss12}
\end{equation}
and the strain associated to the modulation (\ref{sinust}) has the form

\begin{equation}
\epsilon(x,t) = \epsilon(t) \cos(qx)
\label{strain}
\end{equation}
where $\epsilon(t)$ is related to the amplitude $\delta h (t)$ by
(\ref{2.3}).

Now, there is a surprise \cite{DP'V}. The coefficient $K_1$ can be
positive or negative, independently of the sign of $\delta a /a$! 
The reason is the following. The  coupling of adatoms  to the strain,
expressed by (\ref{huss11}), is partly due to the size of the atom, but
also
has chemical reasons which, a priori, are hard to predict. Only the part
of $K_1$ which results from size effects has necessarily 
the sign of $\delta a /a$. 

Formulae (\ref{huss11}), (\ref{huss12}) and
(\ref{strain}) yield

\begin{equation}
j_{inc}^{el}(x,t) = -\beta D K_1 \bar{\rho}_{inc} q  \epsilon(t) \sin(qx)
\label{huss13}
\end{equation}
or, replacing  $\epsilon(t)$ by its expression (\ref{2.3}),
\begin{equation}
j_{inc}^{el}(x,t) = -\beta D K_1 \bar{\rho}_{inc} q^2  
\frac{C_1}{2C_2} \frac{\delta a}{a} \delta h(t) \sin(qx)
\label{huss14}
\end{equation}

When inserted into (\ref{spen1}), this expression yields a contribution 
to (\ref{spen23}) proportional to $q^3$, and therefore replaces $B_1$
by $B'_1$ given by

\begin{equation}
B'_1 / B_1 = 1 +
\frac{\bar{\rho}_{inc}}{\rho_0}\frac{K_1}{2a^2cEC_1}
 \frac{a}{\delta a}
\label{huss15}
\end{equation}

If $ K_1\delta a/ a<0$, the instability can disappear if $\rho_{inc}/\rho_0$
is large enough. 
If $ K_1\delta a/ a>0$, the instability should  develop  more
rapidly during growth than in the absence of flux. 
Formula  (\ref{2.5A}) which gives $\lambda^\ast$ should be multiplied by
(\ref{huss15}),
so that the wavelength of the surface modulation can be smaller than in
the absence of growth.

Another possible effect of MBE growth is a high concentration 
of surface vacancies instead of surface atoms. The extension
of the previous argument to this case would be straightforward.



\subsection{Experiments}
\label{tsf2'}

As will be seen, instabilities are observed in heteroepitaxial growth,
and can be explained by the mechanism of paragraphs 
\ref{tsf1} to \ref{tsf2''}.
However, a precise comparison is not possible, in particular 
because   thermodynamical
surface properties ($T_R$, $\tilde\sigma$~...) are generally not well known.

The instability of coherent, epitaxial films has been observed
for instance in the growth of Ge on Si(111)~\cite{voigt}, 
and of Ge$_{1-x}$Si$_x$ on Si(001)~\cite{pidd,duta,isabelle}.
The observed features depend on the Germanium concentration, presumably
because the misfit does. The misfit of pure Ge on Si is 0.0417.
The non-stoichiometric composition allows to obtain a tunable misfit,
but makes comparison with theory difficult since the concentration
can become inhomogeneous for kinetic~\cite{guyer} or 
thermodynamic~\cite{glas} reasons.

For small misfits (low Ge concentration), it is possible to grow rather
thick layers, and at a `critical' 
thickness\footnote{This critical thickness for the 2d-3d transition 
is generally different from the critical thickness for 
misfit dislocation formation, introduced in chapter \ref{Marty1}.}
a three-dimensional modulation appears, which is 
`quasi-periodical'\cite{duta} and reminiscent of formula (\ref{sinus}).

For large misfits, the adsorbate splits into clusters and this happens
at a much lower critical thickness. For instance, this thickness is
about 100 nm for Ge$_{0.83}$Si$_{0.17}$ on Si(001)~\cite{duta}, 
and only 3 monolayers~\cite{terso} for pure Ge on Si(001) and 2
bilayers~\cite{voigt} 
for Ge on Si(111), where the misfit is about 0.04. In the case of InAs on GaAs(001), 
the misfit is even larger (0.07) and
the adsorbate splits into clusters at an even lower thickness, about 1.6
monolayers~\cite{moison}.

Berb\'ezier et al.~\cite{isabelle} have found that the 
amplitude of the modulation increases proportionally to 
$(\delta a/a)^2$. According to previous paragraphs, 
this amplitude, at a 
given time,  should be given by (\ref{clamore}), where 
$\omega_q$ is given by (\ref{spen23}). From these formulae,
the amplitude would be expected to increase more rapidly than
$(\delta a/a)^2$ for long times.

The case  Ge$_{1-x}$Si$_x$ on Si corresponds to $\delta a/a>0$, i.e. 
the adsorbate is compressed by the substrate. A modulation has
also been observed (Fig. \ref{ponch2}) for $\delta a/a<0$~\cite{pon1}.

The case of large misfits will be addressed in the next chapter, but the
splitting into clusters is in agreement with the argument of paragraph
\ref{tsf1}. The cluster size is just limited by diffusion.

In the case of small misfits, the roughly sinusoidal
modulation which has been observed is presumably a transient state,
which {\it apparently} does not evolve because the dynamics is too
slow. Separation into clusters would still lower the 
free energy. 
The qualitative agreement between the theory of paragraph 
\ref{tsf2}, which predicts an instability, 
and the experiments, which provide an observation 
of this instability, does not imply that everything is well understood.
In particular, the theory was made for a non-singular  surface, 
i.e. a rough surface in the sense of chapter \ref{Intro}.
However, a
(001) surface is generally believed to be singular at the temperature at
which 
the   crystal is grown. The complete solution of this puzzle 
would require a quantitative theory which is not yet available,
but the following elements may contribute to the explanation.

i) The surface may contain steps or other defects 
to be  thermodynamically rough. 

ii) Even if the surface of the `free' material is smooth, 
the surface of the constrained material may be rough or closer to
roughness. It is especially interesting that waves appear in tensile
films 
($\delta a /a < 0$) at a lower misfit than in compressive films.
An increase of the atomic distance clearly favours surface melting
(as it favours ordinary, bulk melting) and therefore surface roughness
(since a
liquid surface is rough).

iii) As seen in paragraph \ref{Calvin}, 
a growing surface can be linearly unstable even 
if a surface which has a similar morphology, 
but is  at equilibrium, is stable.

In any case, it is not surprising that separated clusters have not been 
observed for a moderate misfit. Indeed, theories based on
an activated process predict a  huge activation energy proportional
to $(\delta a/ a)^{-4}$, as seen from  formula \ref{flouc}) below.
On the other hand, if the plane surface is linearly unstable,
the instability is established at a rate proportional to
$\lambda^{\ast-3}$
if the stabilizing terms of (\ref{spen23}) are ignored, and
$\lambda^\ast$
is very large in weak misfit as seen from  (\ref{2.5A}).

One can wonder whether the existence of a `critical' thickness below
which the plane surface is apparently stable is a thermodynamic or
kinetic property. In other words, is a three 
monolayer thick deposit of Ge on Si
thermodynamically stable or does it only lack time to 
deform?
This point is not yet clarified. The fact that the `critical'
thickness can be considerably increased by surfactants, as will be seen
in chapter \ref{surf},
suggests that  kinetic effects are important. The same conclusion 
is suggested by the argument of paragraph \ref{tsf1}, according to
which cluster formation always lowers the energy. However, it will be
argued
in the next paragraph that the stability of thin layers can also have 
thermodynamical grounds.


\subsection{Effect of the adsorbate thickness}
\label{tsf3}

If the stress is produced by a substrate, the previous treatment is only 
reliable if the adsorbate is sufficiently thick, as noticed  by 
Grinfeld~\cite{grin4,grin5}. A finite  thickness
of the adsorbate has in particular the following effects.

i) The elastic constants are usually not the same in the adsorbate and
in the 
substrate. This introduces computational complications~\cite{spencer',junqua}, 
apart from simple particular cases, e.g. if the adsorbate is very thin.
This case is treated in paragraph
\ref{dot1'''} below in the case $T<T_R$, and the case $T>T_R$ might be
treated as well. This effect modifies the instability threshold
$\lambda^\ast$~\cite{spencer',junqua} but  does not introduce a critical 
thickness,\footnote{However, a thickness threshold, below which the 
instability is not observable in practice, has been been introduced by
various 
authors~\cite{spencer',sny}.  } 
except  in the unphysical case of an infinitely hard, undeformable
substrate.

ii) A greater thickness implies a longer growth time, and the
instability is enhanced by growth as argued in paragraph~\ref{tsf2''}.

iii) Another effect of a finite adsorbate thickness 
$\langle z \rangle$ is the limitation of
height fluctuations in the direction $\delta z <0$, where 
$\delta z = z-\langle z \rangle$. Indeed a
local fluctuation always costs capillary energy, but if 
$\delta z < -\langle z \rangle$, it does not bring any elastic  energy.
It is thus tempting to assume that  such a fluctuation  is 
completely forbidden, i.e. the condition

\begin{equation}
z \geq 0
\label{z>0}
\end{equation}
is satisfied on the whole surface.
This is a reasonable approximation if
the substrate (but not necessarily the adsorbate) is below its
roughening
temperature. 
For instance, if the adsorbate  becomes thinner than 1 monolayer, 
the height is 
in most places equal to 0 or 1, as the spins of an Ising model.
Now, the free energy of an Ising paramagnet
has the form (\ref{spen14}) (where $z$ would be replaced by the
magnetization $M$ in a magnet) with a {\it non-vanishing} 
 coefficient $B_0$. More generally, a finite adsorbate thickness 
introduces a non-vanishing  coefficient $B_0$, which is now allowed by 
symmetry, even above $T_R$. 
Thus, a small adsorbate thickness
makes the instability more difficult. Moreover, if $\delta z$
is equal to 0 or 1, a continuous variation of the form (\ref{sinus})
is hardly acceptable and the instability should have different 
features. 

These special features are  well illustrated by recent
experiments by Ponchet et al.~\cite{pon2,pon3}.
Layers of InAs of different thickness were deposited on InP(001)
 by MBE during a short time at about 500$^{\circ}$C,
then maintained in an As flux (but without In flux) at 
the same temperature during 30 seconds, then buried under InP. The
result is the
following. Deposits thicker than 2 monolayers\footnote{In the case of
a binary semiconductor, a (001) monolayer designates a set
of an atomic layer of metal and an atomic layer of metalloid.} 
give rise to a fairly regular 
array of  small clusters. Deposits of thickness comprised between
1.5
and 2 monolayers give rise to an irregular array of clusters whose volume is 
about 5 times as large as when  resulting from a thick deposit. 

A possible  interpretation of the experiment is the following. 
For thick deposits, the 
calculation of paragraph \ref{tsf2} will be assumed to be valid.
Then  a modulation of well-defined 
wavelength $\lambda$ appears --more precisely, 
two modulations in two orthogonal 
directions. The amplitude $\delta h$ increases with time, 
and when it becomes of the order 
of the adsorbate thickness, the substrate has to be taken into account
and  the adsorbed film splits into clusters. 
The shape of these clusters is the subject of the next chapter.
The wavelength $\lambda$, as seen above,  is
of the order of magnitude of $\lambda^\ast$. 

For thin deposits,  the substrate cannot be ignored even at 
the beginning of  the instability. Since there is a wetting layer, 
its effect is summarized by a condition 
stronger than (\ref{z>0}),  namely
$z \geq c$, where $c$ is the thickness of a monolayer.

For thin deposits, but thicker than 1 monolayer, the theory of
paragraph \ref{tsf2} has to be modified.
The instability does not appear in the linearized treatment.
The system is not `linearly unstable'. However, it is not stable either
since, as
argued in paragraph~\ref{tsf1b}, the formation of {\it big} clusters
does lower the energy. Thus, thin adsorbates are not stable,
but metastable. They
can  only reach their equilibrium state if they overcome an
activation barrier. This activated process may be 
expected to give rise to a 
rather disordered array of clusters, as observed experimentally. Indeed,
it is a localized effect rather than a cooperative one. The nucleation 
barrier can be overcome at some place and not elsewhere, while the
linear instability described in paragraph 
\ref{tsf2} corresponds to the increase 
of a density wave in the whole surface.

Another point which has to be understood is that 
clusters which arise from  thin deposits are bigger 
than those which result from thick deposits.
This may seem paradoxical since, for the same cluster size, 
the atoms has to diffuse on a
longer distance in the case of a thinner film.
The following arguments can explain the experimental facts.

i) A theoretical value of the typical linear  size 
$V_{c}^{1/3}$ of clusters arising from activated 
processes will be evaluated in the next chapter
(see formula (\ref{flic}) below). 
It is the product of $(\delta a/a)^{-2}$ by the ratio of a surface
tension divided by an elastic constant, just as
wavelength $\lambda^\ast$ given by (\ref{2.5A}).
This suggests that both lengths $V_{c}^{1/3}$ and $\lambda^\ast$ 
have the same order of magnitude.
However, the volume of the clusters should be,
in the activated case, essentially the cube of the
linear size, i.e. proportional to $(\delta a/a)^{-6}$, while in the
linearly unstable case, it should be proportional to $\langle z \rangle$
times the square of the linear size,  i.e. proportional to 
$\langle z \rangle (\delta a/a)^{-4}$ as seen above.
In other words,
in the case of nucleation, the volume is that of the critical cluster.
If there is a smooth modulation of the surface, the wavelength
is the critical wavelength.

ii) The following argument is slightly different, but related.
In the Asaro-Tiller-Grinfeld scenario,
assumed to be valid for thick films, the
instability is collective, all mounds appear at the same time and cannot
go fishing atoms outside their own domain which has a well-defined radius 
$\lambda^\ast$. In contrast, if clusters are randomly nucleated, 
the nucleation of new clusters in their vicinity does not occur at once, 
and the first created clusters have some time to sweep remote 
atoms atoms. 

iii) In the case of thick deposits, the instability 
can begin during the growth, and the wavelength can be smaller
as seen in paragraph \ref{Calvin}.

The above interpretation  of the experiments of Ponchet et al. is only 
tentative and can be subject to controversy. 
Indeed, the average cluster size has been supposed to be selected at
the onset of the instability 
in a two-dimensional layer, which then splits into `islands' or clusters
which have this size. 
In reality, the evolution after cluster separation is not 
negligible~\cite{joyce,koba,gerard1,gerard2,PL} 
as will be seen in chapter \ref{popolo}. 
A support of our view may be seen in the observation that 
deposits of InGaAs on GaAs or AlGaAs give more regular structures 
for (311)B surfaces\footnote{In the (311)B GaAs surface, the steps are
As, while they are Ga in the (311)A surface. The (311)A surface
is quite different and will not be considered here. } 
than for the (100) orientation, 
in the case of MOVPE~\cite{notzel,notzel'} as well as 
MBE~\cite{nishi}. Indeed, the roughening transition 
temperature is expected to be lower for a (311) face than for (100),
so that the argument of paragraph \ref{tsf2} 
is more likely to be applicable.


\subsection{Solid-solid interfaces and other generalizations}
\label{tsf4}

A case of interest is that of  coherent, epitaxial solid-solid 
interfaces. Such interfaces are commonly encountered in semiconductor
technology (quantum wells, multilayers). One can for instance 
consider~\cite{marty1,marty2} a quantum well 
constituted by a thin slab of a material $A$ in a infinite crystal of a
material $S$. 
The elastic properties are simpler in this case then a
multilayer or a single interface between two solids because the lattice 
constant is that imposed by the material $S$.  

 Using the same arguments as for a free surface, one can show that a
coherent 
interface is not planar at equilibrium above its roughening transition, 
which can be defined as that of a free surface.
The planar shape is linearly unstable with respect to small deformations,
just as the surface of a coherent adsorbate~\cite{marty1,marty2}.

If one goes beyond the linear stability analysis, the 
two cases are different. Large 
modulations of a solid-solid interface are not so  favourable 
as those of a free surface.

On the other hand, volume dynamics is very slow and solid-solid 
interfaces can remain planar during a long time. We do not need to worry
about our TV, its transistors will not be damaged by interdiffusion. 

We conclude this chapter by a hint on  possible generalizations of the 
Asaro-Tiller-Grinfeld instability.
In (\ref{2.4}), $\delta h$ is the amplitude of a height modulation. As a
matter of fact,
it might be the amplitude of the modulation of any continuous `field', 
for instance 
the  surface impurity density. If this field is linearly coupled
to the elastic strain, then a  modulation of this field lowers the {\it
elastic} energy as shown in paragraph \ref{tsf2}.
However, this does not give rise to an instability if the 
coefficient $B_0$ of formula (\ref{spen20(t)}) is strongly positive.


\section{Thermodynamics of an  epitaxial, coherent cluster }
\label{dots}

\subsection{Energy of a cluster }
\label{dot1}

In the previous chapter we have seen that a plane adsorbate surface 
is unstable if  adsorption is coherent. 
The final stage of this instability is an array of clusters. 
If the cluster  size has a narrow distribution, these can be used as
`quantum
dots' in microelectronics after being `buried' into  the material
which constitutes the substrate. These clusters are the subject of the
present chapter. In contrast with the previous chapter, 
we now study the final or nearly final result of the instability,
rather than its onset.  In the first few paragraphs of this chapter,
simple theoretical considerations will be made, which then  will
be compared with experimental observations.

In contrast with the three-dimensional clusters addressed in chapter
\ref{wet}, 
those which result from coherent, epitaxial growth often form
a fairly regular array and have 
a narrow distribution of 
sizes~\cite{notzel,notzel',moison,pon2,pon3,gerard2,petroff,MGIBB},
so that they can be used as quantum dots. 
This phenomenon is called {\it self-organization}. Its
mechanism is not well understood, but a few conjectures will be 
presented in the next chapter. We shall first 
consider a single cluster and  determine its 
equilibrium  shape. The standard way to do  this is to minimize its
energy.

In the linear stability analysis of chapter \ref{tersoff}, 
the energy was supposed to be the sum 
of an elastic part resulting from the elastic strain in
the whole sample, and a surface (or `capillary') term independent of the
strain. This assumption is in principle not correct for strongly
deformed surfaces,
and the surface
energy does depend on the strain through the `surface stress'.
The surface stress can be viewed either as a strain-dependent part of
the surface energy,
or as an elastic stress which is concentrated on the surface and should
be added to
the part which varies continuously in the solid and has only
discontinuities at interfaces.
In other words, the  surface stress is a delta-function singularity of
the elastic
stress at the surface.  As will be seen in paragraph \ref{dot4b},
certain experimental
properties might be explained by taking the surface stress 
into account, as was done for instance  by Shchukin et
al.~\cite{schukin} and M\"uller \& Kern~\cite{muller}.
For the sake of simplicity, the surface stress will generally be ignored
in the present chapter. This is correct  in 
certain limiting cases, e.g. i) for weak strains, 
which do not affect very much the surface energy. 
ii) For big clusters, where the elastic energy  is not much affected by
the surface.

While in chapter \ref{tersoff} the temperature was assumed to be 
higher than the transition temperature $T_R$ which corresponds to
the average surface orientation, in the present chapter the condition
$T<T_R$ will be assumed to hold. 


\subsection{Capillary energy of a cluster }
\label{dot1'}

If a smooth film splits into clusters, this modifies the free energy 
density. As in the previous chapter, one can distinguish a `capillary'
energy, localized near the interfaces, and an elastic energy, which is
not.
The elastic energy is simpler because its variation in space obeys
linear equations which only depend on the known elastic constants of the
materials. The capillary energy is  difficult to evaluate
from first principles, and even from empirical potentials since it
depends,
for instance, on surface reconstruction, which is an essential feature 
of all semiconductors. We shall present here a qualitative view.

The capillary energy of a cluster contains in principle two parts. One
corresponds to the free surface, and the other to the
adsorbate-substrate interface. The main effect of the latter is to
produce, in many cases,
as seen above, a wetting layer. The energy of the cluster-wetting layer
interface will be ignored.

 The property $T<T_R$ implies that the capillary energy  
increment $\delta  {\cal F}_{cap}$ resulting from a modulation 
is no longer proportional to $\delta h^2$ as it was in (\ref{capi}).
Instead,
it is proportional to $\delta h$ for small values of $\delta h$. Indeed,
the 
perturbed surface
may be viewed as formed by flat parts separated by steps. The step free
energy 
per unit length  $\gamma$ is positive below $T_R$ and
the number of steps is proportional to $\delta h$. 

Similarly, a cluster of height $h$ whose basis has a linear size $R$
has a capillary free energy equal to
$$
{\cal F}_{cap} = \hbox{Const} \times \gamma h R/c
$$
where  the constant is positive since a negative value would imply that
the 
plane surface is unstable even in the absence of misfit. 
The constant  depends on the details of the shape of the cluster.
We shall assume that it is a truncated pyramid (Fig. \ref{pyr}) 
with a square basis of side
$R$, and that the sides make an angle $\theta$ with the substrate
surface.
Then the detailed expression for small  $\theta$ is
\begin{equation}
{\cal F}_{cap} = 4\gamma (h^2/c) \frac{1+x}{1-x} \cot \theta
\label{flac}
\end{equation}
where
\begin{equation}
x  = R'/R = 1 - \frac{2h }{R} \cot \theta
\label{xx}
\end{equation}
is the ratio of the sides $R$ and $R'$ of the two square faces.

Expression, (\ref{flac}) is correct for $|\tan \theta | \ll 1$
and  qualitatively valid  for $|\tan \theta|<1$. For larger values,
the interaction between steps are not taken into account, but 
(\ref{flac}) still gives the right order of magnitude.

In the following, it will be assumed that the  cluster is a truncated
pyramid.
The ambition of a theoretical treatment is thus restricted 
to find the three parameters $R$, $h$, $x$ which minimize the free
energy. In the absence of elastic effects, it is possible to do much
better and to find the equilibrium shape without any restriction by the 
{\it Wulff construction}~\cite{Nozieres,JVAP}. The existence of an
elastic energy 
makes things more difficult.  The experimentally observed shape is 
not always a truncated pyramid. Fig. \ref{exppyr} shows a more
complicated shape which results from the fact that
the surface of a semiconductor has a rectangular rather than square
symmetry. The cluster shape will be addressed in greater detail
in paragraph \ref{dotexp}.


\subsection{Scaling laws for the elastic free energy of a cluster}
\label{dot1''}

The explicit form of the elastic Green's functions 
for a semi--infinite isotropic medium~\cite{Landau} 
or cubic crystal~\cite{por77} bounded by a planar, 
stress--free (001) surface is known. 
However, these Green's functions are not directly applicable in the
presence of 
adsorbed clusters except if these clusters 
are very thin. This special case will be addressed in paragraph
\ref{dot1'''} 
for  an isotropic medium.

In the general case, 
the only simple statements which can be made about the elastic free
energy are
scaling laws, which hold for any cluster shape. 

As in the previous chapter, the energy
will be counted from a state where the 
substrate has  its `natural' lattice constant and where the adsorbate
has
the lattice constant of the free substrate. The energy of that state
(set equal to 0 by our choice of origin) is higher than that of the
equilibrium state,
since the elastic strain precisely tries to lower the elastic energy.
Therefore,
with our convention, the elastic energy at equilibrium is {\it
negative}.  
The strain will also be counted, as in the previous chapter, 
from the above defined state  where the substrate 
will be said to be `totally unconstrained' and the adsorbate `totally
constrained'.

A list of  scaling laws, {\it which hold within linear elasticity
theory and ignoring surface stress,} will first be given.

A. The elastic free energy of a coherent, epitaxial cluster of 
a given shape and size
is proportional to the square  $(\delta a/a)^2$ of the misfit.

B. The elastic free energy of a coherent, epitaxial cluster of 
a given shape with
a given  misfit is proportional to the cluster volume.

In other words, if all linear dimensions are multiplied by $\lambda$ 
the elastic free energy is multiplied by $\lambda^3$. The shape may be a
pyramid, a 
truncated pyramid, a half-sphere, etc.

C. The strain created by a coherent, epitaxial cluster of 
a given shape
 at a  long, given  distance $\vec r$ is proportional to the cluster
volume $V$.

D. The strain created by a coherent, epitaxial cluster of given shape
and size  at  a
  long  distance $\vec r$ is proportional to $1/r^3$ for any given
direction $\vec r/r$.
 
 The same law holds for an isolated adatom.
 
E.  The elastic interaction free energy between two  
coherent, epitaxial clusters of given shapes at distance $\vec r$ is 
repulsive and proportional to
$1/r^3$ for large $r$.

\begin{equation}
V_{el}(r)= B/r^{3}
\label{elasBB}
\end{equation}

The same law holds for two isolated adatoms.

These scaling laws are derived in App.~\ref{apx}. Property (A) is a
straightforward 
consequence of the linear and continuous nature of elasticity theory,
which as a matter of fact is an approximation.
The other scaling laws, especially (D) and (E), are well-known for an
isotropic elastic medium. The quantitative formulae which give the
response of an isotropic solid to superficial forces
have been obtained by J.V. Boussinesq~\cite{bouss} at the 
end of the nineteenth century and
can be found in textbooks~\cite{JVAP,Landau}.
 
It is of interest to compare the elastic interaction between adatoms and 
the phonon-mediated interaction between electrons in the BCS theory of 
superconductivity. 
Although electrons ore quantum particles, there is a strong analogy
between 
their interaction and the elastic interaction between two point
impurities in a {\it bulk}
solid. However, the interaction between electrons is generally assumed
to be short ranged,
in strong contrast with (\ref{elasBB}). As a matter of fact, the elastic 
{\it strain} created by a point impurity in a {\it bulk} solid is 
proportional to $1/r^3$ as in the surface problem~... but the term
in $1/r^3$ of the interaction {\it energy} between two point impurities
turns out to vanish~\cite{JVAP}.


\subsection{ Case of a thin cluster: Tersoff approximation}
 \label{dot1'''}

We can say a little more if the adsorbed cluster  is sufficiently thin.
For the pyramid of Fig. \ref{pyr}, this means that 
$h \ll R$ or/and $\tan \theta \ll 1$.
In this case, an approximate
calculation of the elastic energy, often
used by Tersoff and his coworkers~\cite{legoues,TT} and hereafter called 
`Tersoff approximation' is possible. It relies on two simplifications.

i) In (\ref{2.a}), the strain is assumed independent on $z$, so that
\begin{equation}
\delta {\cal F}_1 = - \hbox{Const} \times \frac{\delta a}{a} \int dx dy 
[\epsilon_{xx}(x,y) + \epsilon_{yy}(x,y)] z(x,y)
\label{2.aaa}
\end{equation}
where the integral is on the substrate-adsorbate interface 
and $z$ is the height of the adsorbate surface
with respect to this interface.  Thus, the integral
is over the substrate-adsorbate interface, which is assumed to be a plane.
If the adsorbate is an isotropic solid of Young modulus $E_a$ 
and Poisson ratio
$\nu_a$, the constant is equal to $(\delta a/a)E_a/(1-\nu_a)$.

ii) Since the adsorbate is thin, its contribution to the term 
${\cal F}_2$ defined by (\ref{2.1'}) is negligible.

With these approximations, the problem is reduced to finding 
the response of a semi-infinite crystal limited by a plane (the substrate)
to known forces acting on the surface, which are fully defined by
(\ref{2.aaa}).

In the case 
of a truncated pyramid with $h \ll R$, the strain is proportional to $h$ for
fixed $R$ and $\theta$, so that ${\cal F}_2$  is proportional to 
$h^2$. The linear part ${\cal F}_1$  is also proportional to 
$h^2$ since $z$ is proportional to $h$ in (\ref{2.aaa}). 
Thus, the total elastic energy is proportional to $h^2$.
On the other hand, according to paragraph \ref{dot1''}, 
the energy is proportional to $R^3$ for fixed $h/R$ and $\theta$. 
We conclude that the 
elastic energy of a low pyramid is proportional to $h^2R$,  
in agreement with the quantitative formula  (\ref{elasA}) below,
valid for a coherent epitaxial cluster of a
pyramidal form in the limit 
$\tan \theta \ll 1$ in the isotropic limit ($c_{11}-c_{12}=2c_{44}$).
The (tedious) derivation of this formula  
essentially relies on the fact that, in the limit $\tan \theta \ll 1$,  
one can use the elastic Green's function 
(i.e. the elastic response to a localized stress) of a solid
bounded by a plane surface, which  is known. The Green's function 
method yields the energy as a multiple integral which
can be calculated for an isotropic solid. If both $R$ and $h$ are much 
larger than the atomic distance, one obtains~\cite{DPV,these}

\begin{equation}
{\cal F}_{el}= -12 
\frac{1-\nu_s^2}{(1-\nu_a)^2}\frac{E_a^2}{2\pi E_s}
\left(\frac{\delta a}{a}\right)^2
V G(x) \tan \theta 
\label{elasA}
\end{equation}
where $E_s$ and $E_a$, $\nu_s$ and $\nu_a$ are the Young modulus
and the Poisson ratio of the substrate and the adsorbate respectively,
$V$ is the
volume of the cluster, $x$ is defined by (\ref{xx}) and 
$G(x)$ is a function which
decreases smoothly from a value  close to 0.5 (for $x=0$, i.e.
a full pyramid) to reach the value 0 with 
a finite slope for $x=1$, i.e. a slab (Fig. \ref{GGG}). 
For a given shape, $x$ is fixed and ${\cal F}_{el}$ is proportional to
$\tan \theta$
which is itself proportional to $h^2R$ as stated above.

The analytic form of $G(x)$ is impressive, namely
      
\begin{eqnarray}
G(x) = && \frac{1}{1-x^3}
\biggl\{  \frac{1}{3}  
\left[\sqrt{2}-\ln\left( \sqrt{2}+ 1\right)\right] (2+x)(1-x)^2
+  \frac{2}{3} \left[ \sqrt{2} \ln \left( 1 + \sqrt{2}\right) -1 \right] 
(1+x^3)
- \Xi(x)
\biggr\}
\nonumber
\\&&
\label{elasB}
\end{eqnarray}
with
\begin{eqnarray}
\Xi(x) =&&
- \frac{2x}{3}\sqrt{2+2x^2}   
+ \frac{2x^3\sqrt{2}}{3} \ln \frac{1 + \sqrt{1+x^2}}{x}
+ \frac{2\sqrt{2}}{3}\ln \left( x + \sqrt{1+x^2} \right)
\nonumber
\\&&
- \frac{1-x-x^2+x^3}{2} \ln\frac{1+x+\sqrt{2(1+x^2)}}{1-x}
+\frac{1+x-x^2-x^3}{2} \ln\frac{1-x+\sqrt{2(1+x^2)}}{1+x}
\label{elasC}
\end{eqnarray}

Numerical calculations~\cite{cardon}  are in qualitative agreement with 
(\ref{elasA}) in the case of interest, i.e. $h \ll R$ and 
$\tan \theta \ll 1$.

  The Tersoff approximation makes calculations much easier, 
but reveals consistency problems. Consider a
coherent, epitaxial cluster which is infinitely long 
in the $y$ direction, and whose section by any plane parallel to $xOz$
is a triangle of thickness $2R$ and height $h$, so that the height $h(x,y)=h(x)$
is $h(x)= h-h|x|/R$ if $|x|<R$, otherwise $h(x)=0$. The strain 
on the surface of the substrate, at a distance $a$
of the cluster edge, has the form
\begin{equation}
\epsilon(a) = \hbox{Const} \times \int_{-R}^R \frac{h(x)}{(x+R+a)^2}\,dx
\label{legume}
\end{equation}
where the response function $1/(x+R+a)^2$ results from the scaling law (D)
of paragraph \ref{dot1''}, after integration on $y$. For large $R$,
(\ref{legume}) diverges as $\ln(R/a)$. A similar divergence would arise
in the case of a pyramidal cluster as well.  This divergence
is unphysical, since a strain cannot diverge.
It may be  an artifact of the
Tersoff approximation. However, a failure of linear elasticity
theory is not excluded, although more elaborate versions of 
linear elasticity have apparently
been applied successfully to this problem by Freund 
et al.~\cite{freund3,freund4} and by Shchukin et al.~\cite{schukin}.
The Tersoff  approximation is erroneous in at least two respects: 
i) The stress in the $z$ direction is not correctly taken into account. 
ii) When replacing the cluster by a surface density of external stress, 
one has to take into account the fact 
that the resistance of the cluster to deformation is proportional to its
thickness, so that surface elastic constants, proportional to the
cluster thickness, should be introduced. Since the logarithmic divergence 
is weak, and the results are compatible with the scaling laws,
the Tersoff approximation will be assumed to give the right order of
magnitude.



\subsection{Equilibrium shape of a coherent cluster: 2d-3d transition }
\label{dot2}

When the energy  (\ref{wtot}) is known, it can be minimized 
at a fixed volume $V$ with respect to $\theta$ and $x$, and this yields
the equilibrium shape.

For small volumes $V$,  the elastic energy (proportional to $V$) is
small
and the capillary energy is mainly due to the sides of the pyramid. 
The area of these sides should therefore be as small as possible.
This area is proportional to $\sqrt{Vh}$ multiplied by a coefficient
of order unity which depends on the shape (see formula (\ref{flac})). It
follows that 
the cluster should be as thin as possible, i.e. 
it should  reduce to a single 
monolayer.

For larger values, the elastic energy 
dominates the capillary energy. It is given by  (\ref{elasA}) with an 
acceptable approximation
 for $\tan \theta<1$. For larger values, other approximations can be
used, and it can be seen~\cite{DPV}
that the elastic energy is minimized if the cluster 
is three-dimensional, with $h\approx R$ and $\tan \theta\approx 1$.
As  a 
matter of fact, experiments are 
consistent with a smaller value, $\tan \theta\approx 1/4$ to 1/2. 

We conclude that there is a `2d-3d' (two-dimensional to
three-dimensional)
transition when the volume increases. An analysis based on the
formulae of the previous chapter shows that this transition is 
discontinuous. Clusters initially have the shape of platelets, then
abruptly change their shape and become pyramids with  $\tan
\theta\approx 1/4$. 
The transition takes place for a volume $V_c$ which will be now roughly
evaluated.
A detailed calculation will be found elsewhere~\cite{legoues,DPV}.

The energy of a cluster is the sum of (\ref{flac}) and (\ref{elasA}) or,
dropping constants which depend of the sample shape, but are generally
of order 
unity,
 
\begin{equation}
{\cal F}_{clus} \approx \gamma V^{2/3}/c-E\left(\frac{\delta
a}{a}\right)^2 V 
\label{floc}
\end{equation}

This expression is positive for small volumes $V$, reaches a maximum 
when $V$ increases, and is negative for a linear size larger than
\begin{equation}
 V_{c}^{1/3}\approx \frac{\gamma}{cE} \left(\frac{\delta
a}{a}\right)^{-2}
\label{flic}
\end{equation}

The maximum of the energy is obtained for a value of order
$V_{ac} \approx 2V_c /3$ and the corresponding energy is the activation
energy 
 
 \begin{equation}
{\cal F}_{ac} \approx \frac{\gamma^3}{c^3E^2}\left(\frac{\delta
a}{a}\right)^{-4} 
\label{flouc}
\end{equation}

 Thus, the theory predicts a first order transition (`2d-3d
transition') from
flat, one monolayer thick  clusters to three-dimensional clusters when
the volume
increases. The transition occurs for a volume close to $V_c$.

As a matter of fact, the 2d-3d transition had been observed
experimentally 
before any theoretical prediction, as will be seen in paragraph
\ref{dotexp}. 

In spite of its success in `predicting' the 2d-3d transition, formula
(\ref{floc})
is only qualitatively correct. The elastic part  (\ref{elasA}), 
in particular, must be modified in the case of
clusters of thickness one monolayer. The correct formula contains 
a logarithmic correction, and 
the energy of such a platelet turns out to be~\cite{legoues,TT}  

\begin{equation}
{\cal F}= 4\gamma R -8 \frac{1-\nu_s^2}{(1-\nu_a)^2 }
\frac{E_a^2}{2\pi E_s}c^2 \left( \frac{\delta a}{a}  \right)^{2} R \ln
\frac{RÊ\eta}{a}
\label{tt}
\end{equation}
where 
$c$ is the monolayer thickness, $a$ is such that $a^2c$ is the volume of 
the unit cell,  and 
\begin{equation}
\ln \eta = 2\ln 2 -2 + \sqrt{2} - \ln ( \sqrt{2} +1 ) = -0.081= \ln
0.922 
\label{tt''}
\end{equation}

 The logarithm in the elastic part
(second term) of expression (\ref{tt})  violates 
the scaling law ${\cal F}_{el} \sim V$
of paragraph~\ref{dot1''}.\footnote{In
short, the logarithm arises from the following fact. Any  two 
atoms of the adsorbate at points $\vec r$ and $\vec r'$ 
produce a strain in the substrate, which in turn produces an effective,
elastic  interaction proportional 
to $1/|\vec r-\vec r'|^3$ between the two atoms
according to (\ref{elasBB}). 
The calculation of the energy of a platelet requires 4 integrations,
namely on the two components of $\vec r$ and $\vec r'$.
Three of these 4 integrations  
yield the logarithm of (\ref{tt}), while the last
integration yields the factor $R$ before this logarithm. It is
noteworthy
that, in (\ref{elasA}), the logarithm has been cancelled by further 
integration on the vertical coordinate.}

Relation (\ref{tt}) has been obtained by applying elasticity
theory to an object of atomic thickness. This is obviously 
not quantitatively correct. An idea of the correction to be made is
obtained by taking into account the surface stress~\cite{muller}.

It is of interest to use (\ref{tt}) to obtain an 
order of magnitude of the elastic energy per atom.
Typical values of the Young modulus 
are $E_sc_s^3=12$ eV for GaAs and $E_ac_a^3=9.3$ eV for InAs.
The energy per atom is proportional to $\ln(R/c)/(R/c)$, with
a multiplying factor of the order of 0.01 eV for InAs/GaAs, proportional
to the 
square of the misfit. Elastic forces are really weak at the atomic level
because the misfit cannot be very large. But they are important for a
big enough cluster because of their long range character.


\subsection{  More ambitious treatments}
 \label{dot1''''}

In this review we addressed   an isotropic material and approximated 
the energy 
by formula (\ref{wtot}),  ignoring the effect of the strain on 
the surface energy, i.e., the surface stress~\cite{Nozieres,JVAP}.

Shchukin et al.~\cite{schukin} have given expressions 
which are far more ambitious in three respects. Firstly, they are valid 
in the physical case of a cubic crystal. Secondly they take into account 
the surface stress (which has also been considered by Moll 
et al.~\cite{moll}).  Thirdly, they are not restricted to
thin adsorbates. However, the formulae of Shchukin et al. contain 
the elastic functions of a single material and therefore 
hold only if the adsorbate has the same elastic constants as the
substrate. 

The complications with respect to (\ref{elasA}) in the work of  Shchukin
et al. are  moderate.
The surface stress just introduces two additional terms, which are
negligible if
the cluster size is large enough.  Shchukin et al.~\cite{schukin} 
do not give explicit expressions of the various terms, but
write useful scaling laws in terms of functions which can be 
calculated numerically by means of numerical integration of the equations 
of elasticity (the so-called finite element method).

Using this method, Freund et al.~\cite{freund4} and  Johnson \& 
Freund~\cite{freund3} demonstrated
a spectacular deviation from the predictions of the Tersoff
approximation for thick clusters. 
They find that the strain perpendicular to the substrate surface
changes its sign beyond a height of about 0.22$\times 2R$,
where $2R$ is the basis diameter. Usually, cluster formation
or other  instabilities of the Asaro-Tiller-Grinfeld type are observed
when the 
substrate is bigger than the adsorbate (i.e. $\delta a>0$), 
otherwise other types of instabilities 
generally occur, e.g. dislocations or twins. Then, near the interface, 
the adsorbate is obviously compressed by the substrate in the $x$ and
$y$
directions. However, at a height larger
than 0.22 $a$, it is stretched! An intuitive explanation for this
surprising behaviour 
may be 
as follows. The compression in the $x$ and $y$ directions is accompanied
by an
extension in the $z$ direction, because the Poisson ratio is positive in
all materials.
Thus, the lower layers of a cluster push the upper layers upward. This
can produce
a stretching in the perpendicular direction, as happens if you press the
center of a 
rubber leaf. Freund et al. were able to find an analytic version of
their numerical 
results in a much older article by Ling~\cite{ling} who himself used
previous analytic
theories~\cite{weinel,isibasi} made during the second world war.
Elasticity is indeed 
a very old science, and an accurate bibliography requires a very good
library!

A quantitative treatment of  the capillary energy is even more difficult
than for the
elastic energy. One can represent the interaction between 
atoms by an empirical potential~\cite{AFLP} or make a first principle 
calculation~\cite{pehlke}. In both cases  many reconstructed 
structures  are {\it a priori} possible, and  are a source of 
serious difficulties, since it is in principle necessary to 
compare the energies of all of them, to take
into account the possibility of transitions
from a structure to another, etc.
 Moreover, the reconstructed structures
are experimentally known only for simple surfaces, and not for all the 
orientations which appear, for instance on the cluster sides.


\subsection{Experiments}
\label{dotexp}

An essential prediction of the simple theory presented in this chapter
is the discontinuous 2d-3d transition. 
This is consistent with the  experimental fact
that, for a large enough misfit, 
three-dimensional clusters  appear rather suddenly after the deposition of an
appreciable mass of material, which, as expected, decreases with
increasing misfit.
For InAs/GaAs(001), it is about half a monolayer~\cite{gerard1,gerard2}  
above the wetting layer. 

Another prediction of the simple theory is that big clusters are
somewhat 
sharper than smaller ones. This is also in agreement with observations.
As a matter of fact, the first clusters which appear in the growth of Ge
on Si(001) 
are `huts'~\cite{mo}
(Fig. \ref{exppyr}). These elongated islands are favoured by the
symmetry of the 
(001) faces of the diamond lattice, but seem to be
metastable~\cite{MRKOSW}.
The stable shapes, according to Medeiros-Ribeiro et al.~\cite{MRKOSW}
are pyramids
(square-based islands bounded, as huts, by \{105\} facets) and `domes' 
(``structures with a large number of facets that look rounded at lower 
resolution''~\cite{MRKOSW}). Domes seem to be the stable shape
of bigger clusters~\cite{MRKOSW}, and the fact that they are also
steeper~\cite{MRKOSW}
is in qualitative agreement with theory.\footnote{A transition from
huts to domes has also been observed with a weaker misfit
of 0.008, in  SiGe/Si(100)~\protect\cite{floro}.}  

However, this agreement is only qualitative and the experimental
situation
seems more complicated than the simple theoretical description given
above.
First of all, the size distribution is `bimodal', i.e. it has two maxima
which
respectively correspond to pyramids and domes. Second, the sides of the
pyramids 
have a low slope which would be expected to be unstable.

Intermediate stages between platelets and and three-dimensional
clusters are difficult to observe, in agreement with the fact that they should
be excluded in a discontinuous transition. 
An indirect
investigation is however possible using luminescence study. This method is
essentially a measurement of the energy of the electronic wave functions
localized in the adsorbate.
and is  more sensitive to the thickness $h$ of the
adsorbate than to its lateral size $R$. As it is applied to buried layers,
the 2d-3d transition process is stopped, and a comparison of
different steps of the adsorbate evolution is allowed.

Rudra et al.~\cite{rudra} have deposited a thin InAs layer on InP 
(2 ML for example) and
covered it by InP. When the InAs layer is immediately
covered, they observe a single luminescence line, corresponding to a thin,
two-dimensional quantum well. When the coverage is performed after a certain time, 
they
observe several luminescence  lines, which can be attributed to
different thicknesses of the layer. Calculation of the energy of transition
showed that each line corresponds to an entire number of ML, between 4 and
8 MLs\cite{houdr93}. When a longer growth interruption is performed
before coverage by InP, the luminescence exhibits a wide line,
characteristic of an array of InAs islands~\cite{rudra}. 
Similar results were obtained  by Lambert et al.\cite{lambert}.

An interesting study of the smallest possible three-dimensional clusters
has been made for InAs/GaAs  by  Colocci et al.~\cite{col} using luminescence. When
increasing the coverage, they
obtained an increasing number
(1, 2, 3...) of luminescence lines, which they attributed to an increasing
number of layers of the clusters.

This seems to contradict the theoretical statement that the cluster shape
should jump abruptly from a two-dimensional state (1 layer + the wetting
layer) to a three-dimensional shape with many layers. The reason of the
unexpected observation of two-monolayer-thick clusters may be attributed to
a lack of thermal equilibrium. The quasi-critical clusters of Colocci et al.
have  a size of several thousands of atoms, in qualitative agreement with
theoretical predictions~\cite {DPV}.


\section{Formation and evolution of a population of clusters}
\label{popolo}

\subsection{The three stages}
\label{3steps}

In the present chapter, cluster formation will be addressed, as well
as the evolution of the resulting population of clusters. In contrast
with 
chapter \ref{tersoff}, the misfit is assumed to be rather large, so that 
three-dimensional clusters form directly on the wetting layer, while a
smaller misfit
allows several adsorbate layers to be deposited  before the onset of the 
instability. The case of large misfit corresponds, for instance, 
to InAs/GaAs(001), where $\delta a/a \approx 0.07$.

The evolution proceeds as follows. i) Formation of two-dimensional
clusters on 
the wetting layer. Their number is determined by  kinematic factors,
mainly the 
diffusion constant $D$. At lower temperature, $D$ is small, adatoms 
cannot
diffuse very far, and there are many small islands. For a given
coverage, the
island size increases with temperature. ii) As more matter is deposited, 
clusters grow and become three-dimensional (the 2d-3d transition studied
in the previous chapter).  iii)  As growth proceeds, the  cluster size 
increases and the cluster distribution evolves. 

For lower misfits, e.g. in the case of Ge on Si, the beginning of the
growth will be different, but stage (iii) is fairly analogous.

During stage (i), addressed in paragraph \ref{dot4}, 
the effect of elasticity is presumably not very strong, because 
the elastic interactions between adatoms and very small clusters are
generally weak with respect to chemical, short range interactions
responsible for capillary effects. In contrast, stage (ii) is an elastic effect.
In phase (iii), the importance of elasticity is not so clear.

As will be seen in paragraph \ref{dottorexp}, the 
size distribution of three-dimensional clusters  is generally  
narrow. This fact, which is not observed for incoherent clusters
\cite{Goldfarb,zn} is often attributed to 
elasticity. Actually, as will be seen in paragraph \ref{dot4},
 a population of 
two-dimensional, coherent clusters at equilibrium, would have a narrow 
distribution, but the 
case of three-dimensional clusters is controversial.

The next three paragraphs  \ref{dot4}, \ref{dot3b}, \ref{dot3}
address simple, well understood
and commonly accepted theoretical concepts which, unfortunately, do not
give the complete solution of the essential problem, 
which is the evolution of a set of  three-dimensional clusters.  
The experimental data about this problem are 
treated in paragraph \ref{dottorexp} and compared with theoretical views
in paragraph \ref{dot4b}.


\subsection{When a monolayer starts growing}
\label{dot4}

As summarized in App.~\ref{ell_D},
the growth of an uncompleted monolayer (or `submonolayer') is rather
well understood~\cite{JVAP,Ven,SK,Zin,chrz,bartel} 
when elastic effects are  absent, and indeed, as argued in paragraph
\ref{3steps}, elasticity is presumably not essential at low coverage.

The usual theory~\cite{JVAP,Ven,SK,Zin} deals with the growth of an
element.  Atoms of a single species are deposited at the rate 
of $F_0$ monolayers per unit time (Fig. \ref{grow})
on an initially planar, high symmetry surface (e.g. (001)).
As  seen in chapter \ref{GG1}, application to binary
semiconductors is possible although some caution is necessary.  

The best physical insight is provided by rate
equations~\cite{Ven,SK,Zin}
which describe the evolution of the number of atoms and  islands.
In the simplest approximation (acceptable at low temperature)  
only the adatom density $\rho(t)$ and the cluster
density $\rho_>(t)$ are considered, and 
very simplified forms of the rate equations are, omitting numerical 
factors:
\begin{equation}
\frac{d \rho}{dt}  = F_0 - 2D\rho^2 - D \rho \rho_>
\label{rate1}
\end{equation}
and

\begin{equation}
\frac{d \rho_>}{dt}  = D\rho^2 
\label{rate2}
\end{equation}

The various terms correspond to deposition from the beam, formation of a
pair and sticking of an adatom on a cluster.
Since local effects are ignored, numerical check by
Monte-Carlo simulations is necessary\cite{bartel}.

The rate equations express the fact that freshly deposited atoms have
two possibilities to form chemical bonds. They can
either  go to already formed terraces or nucleate new terraces. 
The former process requires a long diffusion on the surface, which is
more likely to be possible (for a given coverage) 
if the diffusion constant $D$ is high and if the 
growth rate $F_0$ is slow.\footnote{Of course, nucleation dominates at the
very 
beginning of the growth process, while it becomes very unlikely when the
coverage is an appreciable fraction of a monolayer}
Therefore, the number of terraces per unit area after deposition of, 
say, 10\% of a monolayer, is an increasing function of $F_0$ and a {\it 
decreasing function} of $D$. More precise formulae are given in 
App.~\ref{ell_D}. 
In practice, $F_0$ is rarely much larger or much
smaller than 0.1 monolayer per second (within a factor 10) but $D$ is
very sensitive to temperature and can also be influenced by additives or
`surfactants' (see chapter \ref{surf}). 

If elastic effects and other causes of instability are not too
important, terraces begin to coalesce when the coverage becomes close to
1/2, and a complete layer forms when the coverage is equal to 1. 
This happens for the first (`wetting') layer
in the case of InAs/InP and  InAs/GaAs. The formation of the next
layers, in coherent epitaxy, is only possible if the misfit is very weak.


\subsection{Elasticity and the 2d-3d transition}
\label{dot3b}

 While elasticity can be neglected at the very beginning of the
growth, it should be taken into account when clusters are
big enough to become three-dimensional, and the rate equations
\ref{rate1} and
\ref{rate2} must be generalized  to incorporate
the 2d-3d transition. This has been done by Dobbs 
et al.~\cite{dobbs,DZV}. They write a set of three equations for 3
time-dependent
(but space-independent) quantities which are
the numbers of adatoms, of two-dimensional clusters and of
three-dimensional clusters. Since these quantities are
space-independent,
the rate equations are global and cannot describe the effect of elastic
forces 
on the current of matter. 
 Elasticity is implicitly contained in the theory since
 it is the cause of the 2d-3d transition. Moreover,
elasticity  determines the detachment rate of atoms from 
steps.\footnote{ The concept of detachment rate, in the frame of
global rate equations, is ambiguous. For instance, 
if an atom detaches from a cluster and attaches immediately after that
to the same cluster, should this
event be called `detachment'? Dobbs et al.~\cite{DZV} do not,
since, in their formula (23), they introduce a barrier to detachment 
from 2-D clusters, defined as the elastic energy necessary to detach
an atom and to bring it to an infinite distance.} 

This detachment rate is especially important for two-dimensional
clusters, because ``atoms detaching from the edge of a three-dimensional 
island are
more likely to find the energetically favourable sites in the higher
levels
than to completely detach from the island''~\cite{DZV}. In the case of
two-dimensional islands, the  detachment rate depends on the island size
and this effect favours a particular island size as will be seen 
below.

Even though we are mainly concerned with growth, it is of interest to 
investigate the equilibrium distribution of a cluster population. A
good reason for this interest is that it is often convenient to anneal
a deposit after it has been grown. 

The equilibrium cluster size $R_0$ at a given coverage 
can be determined by minimizing the
total free energy, or alternatively the free energy per unit adsorbate
volume, which is obtained by dividing the free energy per cluster
(\ref{tt})
by the cluster volume $V=cR^2$. Thus, one has to minimize

\begin{equation}
{\cal F}_{clus}/(R^2)= 4\gamma R^{-1} -8 \frac{1-\nu_s^2}{(1-\nu_a)^2 }
\frac{E_a^2}{2\pi E_s}c^2 \left( \frac{\delta a}{a}  \right)^{2}  R^{-1}
\ln \frac{RÊ\eta}{a}
\label{tt22}
\end{equation}
with respect to $R$. The result is a finite value, namely
\begin{equation}
R_0= \exp\left[ \frac{\pi\gamma(1-\nu_a)^2E_s}{(1-\nu_s^2)E_a^2c^2} 
\left( \frac{\delta a}{a}  \right)^{-2} - 1 \right]
\label{tt33}
\end{equation}

We conclude that the equilibrium state of a population of platelets 
corresponds to nearly identical platelets with a finite size close to
$R_0$,
with rather weak thermal fluctuations. 

The above argument neglects interactions between platelets,
but is probably qualitatively correct even if the coverage is not very
low.

Thus, if a population of epitaxial, coherent platelets is annealed, their
distribution should tend toward the equilibrium distribution and
therefore, 
toward a narrow size distribution... provided they do not become
three-dimensional, as they are expected to do. 
However, Priester and Lannoo~\cite{PL} have pointed out that the
existence of 
an optimal size  $R_0$ for two-dimensional clusters 
can contribute to the mechanism responsible for
the narrow size distribution of {\it three-dimensional}  
epitaxial, coherent clusters. 
The kinetics associated to this mechanism have been studied by 
Dobbs et al.~\cite{dobbs,DZV}.
The time evolution of the size and number of three-dimensional clusters
is well reproduced by Dobbs et al.
but they cannot say anything about the size distribution since it is not
included in the theory.


\subsection{Annealing of a population of incoherent 
clusters in the absence of growth: 
the Lifshitz-Slyozov mechanism}
\label{dot3}

Before investigating more in detail the evolution of a population of
clusters
on a growing surface, it is of interest to examine this problem in the
simpler 
case of a surface which does not grow (and does not evaporate either).
In the absence of elastic effects, when only the surface energy has to
be 
taken into account, the problem has been solved by Chakraverty \& 
Avignon~\cite{chakra,avignon1,avignon2}.
The evolution tends to reduce the 
surface energy, and therefore the total surface of the clusters.
Since the adsorbate volume is constant,  the ratio 
Surface/Volume decreases, and therefore the average  cluster size
increases.
The microscopic mechanism  is the following. Adatoms escape from smaller
clusters
and diffuse on the surface to bigger clusters where  
the free energy per atom (i.e. the
chemical potential) is lower on smaller clusters (Fig. \ref{LSW}). 
This mechanism, called `Ostwald ripening'
is rather universal and was first discovered in demixing solid 
alloys~\cite{oswald,lifshitz}. 

The theory predicts that the average cluster size $\bar R(t)$ is
proportional to 
$\bar R(t) \simeq t^{1/4}$ for three-dimensional, incoherently adsorbed
clusters~\cite{chakra,avignon1,avignon2} and 
$\bar R(t) \simeq t^{1/3}$ in three-dimensional
alloys~\cite{oswald,lifshitz}. 
The effect is also present for two-dimensional 
adsorbed clusters~\cite{europhys} 
in the absence of elasticity, and then $\bar R(t) \simeq t^{1/3}$.

The expected size distribution is broad.

Experiments on incoherent clusters \cite{Goldfarb,zn}
are in qualitative agreement with theory, although the exponents do not
always 
fit very well. The behaviour  $\bar R(t) \simeq t^{n}$ 
is presumably only valid at very large times~\cite{bray}. 

How are the above conclusions modified by elasticity? In particular, 
does the average size increase with time? A negative answer
might reasonably be expected since the elastic interaction between
adatoms is
repulsive and long-ranged. These features explain the fact that 
two-dimensional, coherent, epitaxial 
clusters have an optimal size. In the case of three-dimensional
clusters, the answer is not so clear, but qualitative considerations 
(App.~\ref{apx}) suggest that the total elastic energy of a
population of 
three-dimensional clusters is a decreasing function of the average
size and therefore favours an increase of this size. In other words,
atoms are still expected to detach preferably from small clusters and to
diffuse toward small ones. As seen in the next paragraph, this seems
to be in contradiction with experiment. This discrepancy suggests that 
new concepts should be introduced, as will be seen in paragraph
\ref{dot4b}.


\subsection{Experiments}
\label{dottorexp}

        In this paragraph are reported some experimental results on the
evolution and size distribution of coherent clusters where elasticity has a
role to play.

Many studies have been devoted to Ge/Si(001), and only a small part
 of them will be reported here. 

According to Goldfarb et al.~\cite{Goldfarb}, 
qualitative agreement with the Lifshitz-Slyozov theory has been observed
in  this system  at temperatures $T>$700 K. However,
the size distribution, as said in paragraph \ref{dotexp}, has two
maxima. 
The maximum which corresponds to small pyramids disappears at
sufficiently 
high coverage. Then, the micrographs and histograms
published by Ross et al.~\cite{ross} show a remarkably narrow 
size distribution. Moreover, there is no hint of
any increase of the average size of the coherent `domes' which
correspond to the 
higher maximum.  This seems hardly compatible with  the Lifshitz-Slyozov
theory.

 The tendency to ripening predicted in 
paragraph \ref{dot3}
(decreasing number of clusters, increasing size) has been observed in III-V
compounds under certain growth conditions. When the cluster size reaches a
certain value (`critical size'), dislocations appear in the cluster,
and the situation described in  paragraph \ref{dot3} becomes effective. 
However, a precise comparison with theory 
is often difficult because of the coexistence of coherent 
and incoherent clusters, where the latter are larger 
than the critical size and the former are
smaller. This has been reported for example~\cite{pon4} in 
InAs/InP(001)~\cite{pon4} 
and InAs/GaAs(001)~\cite{madhux}.
A general feature when misfit dislocations have formed is that the size
distribution is  wide, as expected from theory.
However, it should be pointed out that this 
evolution is possible only under certain favourable growth 
conditions, for instance a
long annealing after  deposition~\cite{pon4}.
Actually, very often, the whole
population of clusters remains coherent, which 
probably implies that their size does
not exceed the critical value. G\'erard et al.~\cite{gerard2} 
have observed in the
InAs/GaAs system that after the initial nucleation of numerous small
islands, the average cluster size tends to increase when an annealing is
performed after the deposit. However, the size does no longer increase
after a certain annealing time.

These results suggest that there is a competition between two effects
and that there is a barrier which opposes
usual Ostwald ripening. The different
experimental results reported above show a strong dependence on the growth
conditions and on the lattice mismatch. This implies that the evolution
towards cluster ripening, although it is energetically  
favourable, can be kinetically delayed.

        What happens when the clusters are still coherent (and strained) is
rather complex. An interesting set of results concerns the evolution of the
average size as a function of the quantity of strained material. When this
quantity is large, the cluster density is large too, and the average size
increases with the deposited quantity~\cite{moison}
as it is usually
expected. This is contrast to what has been observed in the 
InAs/GaAs~\cite{koba,gerard2}
and InAs/InP~\cite{pon2} systems 
when the deposited quantity is
very small, i.e. just enough to induce the 2d-3d transition. When the
deposited amount increases, the cluster density increases too, but the
average size decreases.

A remarkable experimental feature is the
relatively narrow size distribution of such population of coherent
clusters. The size distribution is particularly narrow when the cluster
density is high. This fact suggests a possible role of elastic interactions
between islands~\cite{koba,gerard2}.
According to Kobayashi et al.~\cite{koba}, a
sufficiently dense population of adsorbed clusters evolves, through
inter-cluster exchange of matter, towards a more uniform size distribution.
This is exactly opposite to what is predicted for incoherent adsorption, as
seen in paragraph XII-D.


\subsection{Open questions}
\label{dot4b}

Various explanations have been proposed for the narrow cluster size  
distribution and for the absence of Ostwald 
ripening, at least in certain cases.  
Our feeling is that the situation is unclear, and we shall mainly 
formulate questions,  which are subject to controversy at the moment, 
but  will presumably be answered in the next future, since research in this
field  is particularly active.

\vskip5mm

{\it Question} 1. What is the potential energy of an adatom outside 
clusters? Is there an energy barrier near the cluster surface?

As seen in paragraph \ref{dot1''}, the strain produced by a cluster
increases with decreasing distance. The elastic potential felt by an
adatom should also increase in absolute value.
In principle, although the elastic interaction between two adatoms or
between two large, coherent clusters is repulsive, the cluster-adatom
interaction can be attractive~\protect\cite{dupo}. Such an attraction
would enhance Ostwald's ripening and therefore would not help to 
explain the experimental facts.
Therefore, we will accept the general belief that an  energy 
barrier~\cite{barab} exists near a cluster. It opposes attachment of
adatoms to clusters but, according 
to the detailed balance principle, it should also oppose adatom
detachment from
islands\footnote{So far as we understand, we disagree on this point
with Barab\`asi~\protect\cite{barab}.} 
and therefore slow down the evolution during annealing. 
This suggests a second question.

\vskip5mm

{\it Question} 2. Are there big differences between the evolution 
when the beam is on and when it is off?

As reported in paragraph \ref{tsf3}, this difference 
is clear when the 2d-3d transition takes place 
during annealing~\cite{pon2,rudra,carlin}.

In the case of three-dimensional clusters, one could imagine that there 
is no evolution during annealing 
without beam, and a rapid change of the size distribution during growth.
This would mean that atom detachment from islands is negligible and that 
the evolution is fully due to freshly deposited adatoms. 
Actually, the evolution during annealing can be slower than 
under irradiation~\cite{gerard2}, but is yet present.

\vskip5mm

{\it Question} 3. How high can be an energy barrier to adatom
attachment,
and how does it depend on the cluster size? 

The  barrier height is related to the local strain.
In the limit of  very large  clusters, the strain should reach a finite 
limit. The barrier height is also expected to have a limit 
$W_{bar}$ which is reached for  some value $R_{lim}$ of the cluster radius. 
According to Barab\`asi~\cite{barab}, this limit can be comparable to 
the chemical bonding energy, and thus  the  cluster size
cannot grow beyond  $R_{lim}$. This saturation is a non-linear effect
which suggests a breakdown of linear elasticity theory~...

\vskip5mm

{\it Question} 4. Is linear elasticity theory applicable in the presence of 
coherent, epitaxial clusters? 

In paragraph \ref{dot1'''}, the strain in the vicinity of a  cluster 
has been evaluated by the Tersoff approximation and found to diverge 
with the cluster size. This may be an artifact of the approximation, 
but, in the present state of our knowledge, a  divergence of the strain
calculated by linear, continuous elasticity theory is not excluded. 
This would mean that this theory breaks down for big enough clusters, and 
a plausible criterion is that it breaks down when the elastic energy per bond
is comparable to the  chemical energy, in agreement with Barab\`asi's
statement~\cite{barab}. For reasonable cluster sizes, 
this breakdown takes place only 
for large misfits, since the logarithmic divergence of (\ref{legume}) is 
very weak.

\vskip5mm

{\it Question} 5. What is the effect of the 2d-3d transition?

The 2d-3d transition can favour a
narrow distribution of clusters~\cite{PL}. The following mechanisms
are possible.
First, platelets have a preferred size and therefore
should have a rather uniform
distribution before they transform. The distribution  may remain
rather uniform after transformation.
Second, the barrier for incoming adatoms is somewhat higher for
three-dimensional clusters than for platelets. Thus, 
platelets grow faster than three-dimensional clusters.
Third,  there is a purely geometric effect, 
when the biggest clusters are already three-dimensional while some 
smaller clusters are still flat, the incoming atoms can be expected to
go preferably to the latter, just because they occupy more space on the
surface. Thus, in the neighbourhood of the 2d-3d transition, 
small platelets grow faster than bigger, three-dimensional clusters.
However, as seen in the previous paragraph, 
there is experimental evidence that equalization still operates 
a long time after the 2d-3d transition.

\vskip5mm

{\it Question} 6. Can adatom detachment from big
islands be made easier by elasticity, in comparison with 
smaller islands? 

This has been suggested by Kobayashi~\cite{koba} on experimental grounds,
and by various theorists~\cite{barab,ratsch}. If it is so, this may
produce a current of atoms from big to small islands and favour a finite
cluster size. The difficulty is to explain this easy detachment.
An energy barrier at the surface does not provide an explanation 
since it opposes both attachment of adatoms
to clusters and detachment from clusters, as said above. 
The explanation can result from a failure of linear elasticity
theory, since the elastic
energy gain will be lower than predicted by  linear elasticity,
when the cluster size becomes larger than the threshold $R_{lim}$
defined above. 

Other possible solutions have been proposed, which suggests a last question.

\vskip5mm

{\it Question} 7. Is the picture given in chapter \ref{dots} correct?

The postulates of this description are the following.

i) Microscopic details, e.g. the strain dependence of the diffusion
constant, have been disregarded.

ii) The only contributions to the energy are the elastic energy (the
volume
integral of a continuous energy density) and a strain-independent
surface energy. The surface stress, which generates a
strain-independent surface energy, is neglected.

iii) There is no three-dimensional diffusion, no mixing of the substrate
and  adsorbate materials.

iv) The clusters have their equilibrium shape and the equilibrium
energy.

The modulation of the diffusion constant has been addressed in 
various papers reviewed by Schroeder \& Wolf~\cite{Kanzler}. In
our opinion, such effects should not be essential for a qualitative 
understanding of phenomena. For instance, they can modify the speed of 
Ostwald's ripening or the critical thickness, 
but cannot explain the narrow cluster size distribution. 
The effect of surface stress has been already addressed in paragraph
\ref{dot1''''}. According to Shchukin et al.~\cite{schukin} it can 
stabilize an intermediate cluster size.\footnote{This size is 
rather metastable than stable,
since the most stable state is constituted by big bubbles (strictly
speaking, a single bubble) according to the argument given at the
beginning of chapter \ref{tersoff}.} As in the case of two-dimensional 
clusters  addressed in paragraph \ref{dot3b}, this can explain a 
narrow size distribution  around this size.
A more complete investigation seems necessary to
check whether this interpretation is compatible with the
relatively large cluster sizes which are observed. 

Chemical mixing depends on the materials and on temperature. In 
the case of InAs/GaAs, Joyce et al.~\cite{joyce} have given good
arguments 
in favour of a strong mixing, both in the wetting layer and within
clusters.  Indeed, they succeeded in measuring the 
total volume of clusters with a reasonable 
accuracy, and found that it is larger than the deposited volume.
Mixing can strongly influence the evolution of the clusters~\cite{Ter}. 
In case of 3d cluster, the alloy segregation is probably enhanced
by  strain-driven atomic diffusion~\cite{PL'}.
For instance, one might imagine 
that penetration of the substrate material
into the weakly relaxed cluster core lowers the elastic energy, and that
this penetration 
becomes more difficult as  clusters grow, thus making the growth of
big clusters more difficult, in agreement with experiments. 
As mentioned in chapter \ref{GG1}, chemical inhomogeneities can also result 
from segregation or demixing
within the adsorbate. An example is provided by the
clusters resulting 
from the deposit of GaInAs alloy on GaAs(100). Indeed,
an enrichment in Indium on the top of clusters was experimentally evidenced
by different methods~\cite{tilmann,grandj} 

The fact that clusters need an appreciable time to reach equilibrium  
is testified by experiments~\cite{MRKOSW,MRKOSW'} which demonstrate,
for example,  the occurrence of metastable states. Jesson 
et al.~\cite{jesson1,jesson2} have described mechanisms which slow down 
the establishment of equilibrium. They consider a pyramidal cluster with 
faceted sides, and argue that adding a new layer on
a side requires
to overcome an increasingly high activation barrier when the size 
increases. This implies that the sides of the clusters are faceted, 
with a roughening temperature higher than the growth temperature, a 
property which is not necessarily general, even though faceting
at room temperature is rather well established (see Fig. \ref{exppyr}).


\section{Stacking faults and twins}
\label{poly}

In this short chapter we come back to kinetic instabilities, 
those which are present at low temperature but not at high temperatures.
A particular type of kinetic instabilities was studied in great detail 
in chapters \ref{PP1} and \ref{PP2}, namely those which arise from the 
Ehrlich-Schwoebel  effect. The emphasis   given to 
Ehrlich-Schwoebel instabilities 
in the present review is justified by the large amount of 
theoretical work which has cast a good physical insight on these
phenomena.
There are, however, other types of instabilities, and it might be
misleading to omit them completely  in this review, since one might
be tempted  to attribute to the 
Ehrlich-Schwoebel  effect instabilities which are due to other causes. 

Another possible and perhaps more frequent source of growth instability
is constituted by stacking faults. These defects can arise during the
growth of a crystal of a  material which has another crystal form with a
comparable energy.

A well known example is provided by the face centered cubic Bravais
lattice,
which is the most common  crystal among metallic elements. It is well
known that a different stacking of the (111) planes leads to another
Bravais lattice which is the hexagonal close packed lattice. Each
atom has the same number of 
neighbours in both lattices, the atoms form triangles
in both cases, and the energy difference can be expected to be 
small.  Similarly, random stacking faults which 
destroy the periodicity are expected
to have a low energy and to form rather easily during growth in the
(111) direction. A growth model in which atoms are represented by hard
spheres should reproduce this feature.

Stacking faults are also possible in semiconducting
elements, such 
as Si or Ge, although the interactions are quite different.
Indeed, atoms cannot be  represented by hard spheres. The rigid unit is
not the atom,
but tetrahedra of 5 atoms (a central one and its 4 neighbours).
Nevertheless, as in the case of metals, stacking faults 
in the (111) direction have a fairly low energy. Indeed they can be
obtained by rotating the tetrahedra without distorting them. 

Gallas et al.~\cite{gallas} have observed growth instabilities
resulting from stacking faults  during the growth of Si(111) in
the (111) direction at low 
enough temperature  (about 300$^{\circ}$C) although a perfect crystal
easily grows at higher temperature.
These stacking faults observed in the growth of Si(111) are certainly of 
kinetic origin since they do not appear at higher temperature.
We have seen in chapter \ref{Marty1} that  stacking faults
of thermodynamical origin can also appear in heteroepitaxy.
The example of Si is interesting because the Ehrlich-Schwoebel effect  
has also be suggested~\cite{EagGil} to explain instabilities observed 
at low temperature is the growth of Si in the (001) direction.
The controversy about existence or non-existence
of an Ehrlich-Schwoebel effect is reported in paragraph \ref{ES_int}.

Twins (Fig. \ref{fig:tem_twin})  are more severe crystal defects
than stacking faults, in the sense that they do not
break only the periodicity, but also the orientational long range order.
They are directly responsible for the formation of polycrystals, 
as seen from Fig. \ref{fig:afm_twin}.
As  stacking faults, twins can be interpreted as resulting from 
dislocations, as seen from Fig.~\ref{fig:twin_schem}.



\section{Effect of additives (surfactants)}
\label{surf}

In soft matter physics, `surfactants' are materials which lower the
surface  free energy. A typical example is soap in soap bubbles. 

Applied to crystal growth, the word  `surfactant' has taken a completely 
different meaning. It designates an additive which i) stays on the
surface during 
growth, instead of being incorporated as impurities, and 
ii) modifies the kinetic and/or thermodynamic properties of the surface.
Those surfactants which facilitate layer by layer growth with
respect to three-dimensional growth in heteroepitaxy 
are of particular interest. Examples for Ge/Si
are As, Sb, Bi, Te~\cite{voigt'}. The coverage in these instances is of
the order of 1/3 of a monolayer.
Without any surfactant, as seen in paragraph \ref{tsf2'},
it is possible to grow 3 Ge monolayers on Si(001) at 700 K before forming 
three-dimensional clusters. With As, one can grow 50 layers 
and probably more~\cite{copel1,copel2}.
At such high thicknesses, the stress is relaxed, 
and this implies that the
adsorbate is no longer coherent, although it is
epitaxial~\cite{copel1,copel2}.

Factors which may explain the effect of surfactants in crystal
growth are the following.

\begin{itemize}

\item
 
Reduction of the surface energy. It is the essential property of
surfactants in
soft matter physics. In the case of interest here, it may have the
important effect of  favouring the floating of
surfactants during growth~\cite{voigt'} but is not especially
appropriate for 
growth of good crystals~\cite{gallas}.
Most of the current theories of surfactant-favoured growth invoke
modifications of the kinetic behaviour rather thermodynamic effects.

\item
Modification of the surface diffusion
coefficient~\cite{gallas,voigt',copel1,copel2,unter,reuter,rosen}. 

Slow surface diffusion makes the equilibrium state harder to reach 
and therefore should help to avoid thermodynamical instabilities in
heteroepitaxy. A way to reduce surface diffusion is to reduce
temperature,
but this can favour kinetic instabilities. Surfactants can therefore be
a
good alternative method. Of course, diffusion should not be {\it too} 
low.
Another caveat is that a low diffusion coefficient increases the adatom
density,
and this effect tends to accelerate the evolution, so that the overall
result is not easy to evaluate quantitatively, except by Monte Carlo
simulations. 

Voigtlaender et al.~\cite{voigt'} have observed  the effect of several
surfactants 
during the growth in the (111) direction i) of  pure Si(001) and ii) 
epitaxial Ge/Si(001). They argued that  surfactants which favour layer
by layer 
growth in heteroepitaxy are precisely those which reduce 
surface diffusion in the pure material.
Conversely, `surfactants' which accelerate diffusion (Ga, Sn, In, Pb)
favour 
three-dimensional growth. 

The paper of 
Voigtlaender et al.~\cite{voigt'} has the merit to propose a simple
picture. It also suggests an explanation of the different effect of 
different surfactants: those which accelerate diffusion are 
elements which just provide 
the number of electrons necessary to passivate dangling bonds, while 
those which provide more electrons can also retain adatoms and therefore
slow down diffusion.

Even if not fully correct, a simple picture has always the merit to
provide a starting point for further progress and discussion. 
A point which  has been criticized~\cite{gallas,KaKa} 
is  the method used by Voigtlaender et al. to estimate the 
surface diffusion velocity. Their criterion was the distance between
islands after deposit of a submonolayer (see App.~\ref{ell_D}). 
In the simplest cases, 
a small surface diffusion favours a short  distance between islands.
However, the
case of surfactant-covered Si(001) at high temperatures may be more
complicated
as suggested by first principle calculations and Monte Carlo simulations
of Kandel \& Kaxiras~\cite{KaKa}.

\item

Modification of the detachment rate from steps. 
This is the other way to slow down the evolution toward equilibrium.
Kandel \& Kaxiras~\cite{KaKa} have performed Monte Carlo simulations
where 
a barrier to detachment from steps was taken into account in a
phenomenological way
and obtained a good agreement with experiment. 
In spite of the controversy between Kandel \& Kaxiras and  Voigtlaender
et al.,
both points of view are to some extent complementary. The 
remarkably synthetic picture given 
by  Voigtlaender et al. provides much insight, but it involves 
oversimplifications which are not always confirmed
by first-principle calculations~\cite{schroed}.

\item

Modification of the Ehrlich-Schwoebel effect. 
For example, Markov~\cite{markov} has suggested that a surfactant can 
induce an {\it inverse} Ehrlich-Schwoebel effect, 
which favours downward motion of the atoms.
Roughly speaking, the reason would be that the atoms willing to stick to
an upper 
terrace ($k_-$ process of Fig. \ref{tbv}) are prevented to do so because 
the surfactant is already there, while an atom coming from above 
($k_+$ process) can easily push the surfactant. 

\item

Segregation of the surfactant on the lower terraces~\cite{rosenf}. The
result would  be that the adatom mobility would be different 
on the lower and upper terraces.

\end{itemize}

This enumeration shows that the action of surfactants is complex and
controversial.

Note that, when surfactants do not prevent cluster
formation, they can affect their size and shape~\cite{mass1,mass2}.


\section{Conclusion }
\label{concl}

This review is the common work of theorists and experimentalists, and
has been an opportunity for many discussions between them.

Such discussions are not always easy. Theorists tend to introduce
oversimplified  models, which help to understand some basic  
mechanisms, but may conceal other mechanisms which are not less basic.
For instance, they have devoted more attention to phenomena which 
can be described by the simple S.O.S. model~\cite{JVAP} than to
twinning.

This review is, in this sense, rather theoretical. For instance, 
the distinction between kinetic and thermodynamic instabilities is 
an oversimplification because 
both kinetics and thermodynamics may contribute to the same process.
Such mixed effects have been taken into account in the present
review, but more attention  should be paid to this point in the future.

Another source of misunderstanding between theorists and
experimentalist, and 
technologists even more, is that when a difficulty appears, the
theorist tries to understand it, while the technologist tries to avoid
it, usually by using a more complicated method.
For instance, instead of wondering how one can have a better
self-organization
of quantum dots, one can organize them artificially. Artificial
organization
of quantum dots is clearly outside the scope of the present review, but 
can be a better technology. 
Even then, it is of interest, both for fundamental and
technological purposes, to understand the self-organization
of quantum dots. This is still an open problem.

Another point  which is not completely understood is why the 
Asaro-Tiller-Grinfeld theory seems to be applicable to surfaces
which are believed to be singular.

In both cases, qualitative hints have been suggested in  the present
review,
but a quantitative  treatment is still to be done. It will
presumably be  numerical to a large extent, and we hope
that the qualitative considerations presented here will help
in the choice of the models to use in the simulations.

The comparison between theory and experiment is probably easier in
heteroepitaxy
because most of the basic mechanisms are well identified. But the growth
of a simple material can also be difficult. A typical example is
diamond, of which it is impossible to produce big, artificial crystals.
We would have liked to include this problem in this review, 
but it looks difficult to isolate simple concepts in a complicated
chemical game, in spite of interesting theoretical attempts \cite{diam1}.
The above considerations suggest directions for future research.
We would recommend

-A deeper fundamental understanding of basic mechanisms in experimental
facts.

-A more detailed theoretical description, to compare with a more 
systematic experimental study of the various factors which affect
the various physical or chemical systems. 

This difficult program, is now being carried out by many research groups
throughout the world, and a rapid progress can be expected.


\acknowledgements

It is a pleasure to acknowledge illuminating
discussions with Isabelle Berb\'ezier, Jean Grilh\'e, Reinhold Koch,
Max Lagally, Martin Rost, Vitali Shchukin,
Martin Siegert, Pavel \v{S}milauer, Bert Voigtlaender,
Ir\`ene Xueref and many
other colleagues. We also thank Joachim Krug
for his critical reading of a part of the manuscript. We are grateful to our
colleagues Erwan Adam and Alessandro Verciani for their friendly
and efficient help in our
 conflicting relation with computing systems.
P.P.~gratefully acknowledges financial support from Alexander von
Humboldt Stiftung.

\vfill\newpage
\appendix

\section{List of Symbols}

\begin{tabular}{l|l}

$a$ & atomic distance, or a related length\\

$d$ & distance, or space dimension\\

$D$ & surface diffusion coefficient of adatoms\\

2d, 3d & jargon for two-dimensional and three-dimensional\\

$c$ & monolayer thickness\\

$E$ &  Young modulus\\

ES & Ehrlich-Schwoebel (effect, length, ...)\\

$F$, $F_0$  & Flux of atoms in MBE \\

$h$  &  typical linear size of bumps, mounds, clusters, etc.
perpendicular to the terrace direction (`height')\\

$j$, $\vec j$  &  one and two-dimensional current density 
(in atoms per second and per length unit)\\

$k_B$  &  Boltzmann constant\\

$m$, $\vec m$  &  slope (has the dimension of a reciprocal length in chapter
\protect\ref{PP1}, but is dimensionless in chapter \ref{PP2})\\

$n$ & coarsening exponent, see paragraph \ref{PPexpp} \\
		     
$q$ &  a wave vector\\

$R$   &  typical linear size of bumps, mounds, clusters, etc. in the
terrace direction\\

$T$  &  temperature\\

$T_R$  &  roughening transition temperature (depends on the surface
orientation)\\

$v$ & velocity \\

$V$  &  volume\\

$x$  &  a coordinate parallel to the terraces\\

	 &   In chapter \ref{dots},  quantity defined by  (\ref{xx})\\

	& concentration of an alloy\\
	
$y$  &  a coordinate parallel to the terraces\\
	  
$z$  &  the coordinate perpendicular to terraces\\
 
  & in chapters \ref{PP1} and \ref{PP2}, 
this coordinate divided by the monolayer thickness $c$)\\

$\alpha$  &  kinetic coefficient, ratio of the current density to the
surface slope (see paragraph \ref{ch4D1})\\
		
		  & coordinate $x,y,z$ (e.g. in a summand)\\

$\beta$   &  1/$(k_BT)$\\

$\delta a/a$ & lattice misfit (positive if the  adsorbate is bigger than 
the substrate)  \\
$\gamma$   &  step free energy per unit length, 
step stiffness (paragraph \ref{Mul_cur})\\

		 &   dimensionless exponent (paragraph \ref{ch4F})\\
		  
		 &   coordinate $x,y,z$ (e.g. in a summand)\\

$\tilde\Gamma$  &  $\tilde\Gamma=\Omega\beta\gamma$  \\

$\epsilon$   &  strain (chapters \ref{Marty1} and following)\\

		 &   dimensionless  distance from the instability
threshold (chapters \ref{PP1} and \ref{PP2})\\

$\kappa$   &  reciprocal width of a domain wall (paragraph
\ref{ch4F})\\

		&    curvature  of a step (paragraph \ref{ch5B})\\
		    
$\mu$  &  chemical potential\\

		
$\nu$  &  Poisson ratio  (chapters \ref{Marty1} \ref{tersoff} and
\ref{dots})\\

$\rho$  &  number of adatoms per unit area\\

 $ \tilde\sigma$ & surface stiffness, surface energy per unit area \\
 
 $ \sigma$ & stress \\

$\Omega$   &  area per surface atom (chapter \ref{PP2})\\

$\Omega_{\alpha,\gamma}^{\xi, \zeta} $  &  Elastic constants\\

${\cal A}$  &  area of the growth front\\
 
${\cal F}$  &  free energy\\

$\ell$  &  distance between steps\\

$\ell_D$  &  diffusion length\\

\end{tabular}
 
\section{The diffusion length and the nucleation lengths}
\label{ell_D}

The diffusion length $\ell_D$ is the typical distance travelled by an
adatom on a flat, high-symmetry surface before being trapped. The
requirement
on the surface orientation ensures that the adatom is not trapped by a
preexisting step, but by another adatom (nucleation process) or by the
closed step of a growing island. The length $\ell_D$ can also be defined
as the
average distance between islands, in the stationary regime when 
the island density does not change~(see Ref.~\cite{VPW}). 

The simplest model (a nucleus of two adatoms is stable and it does not
diffuse) requires solely the knowledge of the 
diffusion constant $D$ and the flux
intensity $F_0$. Dimensional analysis suggests the expression (the
lattice
constant is put equal to one) $\ell_D=(D/F_0)^\delta$. A very simple
argument allows to determine the exponent $\delta$.

If $P(\ell)$ is the probability of nucleation per unit time 
in a region of size $\ell$, in the stationary regime it 
is given by the number $N$ of incoming adatoms times the probability
$p$ that an adatom encounters another one. $N$ is simply given by
$F_0\ell^d$ ($d=1,2$ being the dimension of the substrate)
and $p$ is the product of the number of distinct sites
visited by the adatom $(\approx\ell^d)$ times the probability a site is
occupied, i.e. the (average) density of adatoms. Finally, we have
\be
P(\ell) = F_0 \ell^{2d}\langle\rho\rangle
\ee

The diffusion length is given by the condition that the probability
of nucleation in a region of size $\ell_D$ during the deposition of one
layer of material is of order unity:
\be
P(\ell_D)\cdot {1/F_0} \approx 1 ~~~~\Rightarrow~~~~
\ell^{2d}_D\langle\rho\rangle \approx 1
\ee

Since there are no preexisting steps, adatom density is not influenced
by
ES barriers: $\langle\rho\rangle$ may be simply evaluated as the
solution
of the diffusion equation $F_0 +D\nabla^2\rho=0$ in a region of size
$\ell_D$, with $\rho$ vanishing on the boundary:
$\langle\rho\rangle\approx (F_0/D)\ell_D^2$. 
The final result is:
\be
\ell_D\approx\left({D\over F_0}\right)^{1\over 2(d+1)}
\ee

So $\delta=1/4$ and $\delta=1/6$, respectively in 1+1 and 2+1
dimensions.
The hypothesis that a dimer is stable and does not diffuse is valid only
at low temperatures: when the size of the largest cluster which is mobile
and `unstable' is larger than one, the exponent $\delta$ is 
modified~\cite{VPW,KKW}.

In the main text the `region of size $\ell$' must be meant as a terrace
of size $\ell$ which therefore can be a top, a bottom, or a vicinal
terrace (see Fig.~\ref{tbv}). This distinction is necessary in the
presence
of step-edge barriers, because the average density $\langle\rho\rangle$
depends on the type of the terrace.
In this context it is preferable to
speak of `nucleation lengths' for a top $(\ell_n^T)$, a bottom
$(\ell_n^B)$, and a vicinal $(\ell_n^V)$ terrace, rather than of
diffusion lengths. Since ES barriers are irrelevant for a bottom
terrace,
we simply have $\ell_n^B=\ell_D$. Conversely, for top and vicinal
terraces  $\langle\rho\rangle$ is 
obviously an increasing function of $\es$. For a vicinal one it has
a very weak dependence on $\es$ and goes to a finite value for
$\es\to\infty$;
in contrast, $\langle\rho\rangle$ diverges for a top terrace. 
In 1+1 dimensions the nucleation lengths are found 
via the implicit equations~\cite{EV,Psolo}:
\bea
&&(\ell_n^T)^4 +6\es(\ell_n^T)^3 -\ell_D^4 = 0
\; ~~~~~~~~~~\hbox{1+1~dim}\label{lnt}\\
&&(\ell_n^V)^5 +4\es(\ell_n^V)^4 -\ell_D^4(\ell_n^V +\es) =0
\eea
whose solutions are plotted in Fig.~\ref{ln-fig}.

In 2+1 dimensions we must assume that terraces have circular edges,
in order to obtain an analytical expression for the adatom density.
In fact, if $\rho$ depends solely on the radius $r$ the diffusion
equation reads $\rho''(r) +(1/r)\rho'(r) = -F_0/D$, whose general
solution is $\rho(r) = \rho_0 + \rho_1 \ln~r - (F_0/4D)r^2$.
For bottom and top terraces, analyticity in $r=0$ requires
$\rho_1=0$. The diffusion length is found to be $\ell_D=
(8D/F_0)^{1/6}$, while the nucleation length for a top terrace is
determined
by the equation:
\be
(\ell_n^T)^6 +4\es(\ell_n^T)^5 -\ell_D^6 = 0
\;~~~~~~~~~~\hbox{2+1~dim}
\ee
whose limiting solution for $\es/\ell_D\to\infty$ differs from the
one dimensional case. In units of $\ell_D$, $\ell_n^T\simeq
(1/6\es)^{1/3}$
in 1+1 dimensions and $\ell_n^T\simeq (1/4\es)^{1/5}$ in 2+1 dimensions.

The case of a vicinal terrace is not so easy: in fact it is
characterized by
two different quantities, the inner ($r_1$) and outer ($r_2$) radii.
If $\ell=r_2-r_1$, it is no more true that the number $N$ of incoming 
atoms per unit time is simply $N\approx F_0\ell^2$. The average adatom
density itself is a function of both $r_1$ and $r_2$, but in the limit 
$r_1\gg\ell$ it depends solely on $\ell$: $\langle\rho\rangle\approx
(F_0/2D)\ell^2$. In the opposite limit ($r_1\ll\ell$) the
attention must be paid to the logarithmic term, which diverges for
$r_1=0$.
This limit is relevant for the vicinal terrace 
neighbouring a top terrace, when $\es\to\infty$
(because in this limit $\ell_n^T\to 0$).

\section{Qualitative considerations on the coarsening exponent}
\label{app-coarsening}

Let us develop here some qualitative or semiquantitative 
considerations on the coarsening exponent $n$. 
Krug~\cite{Privman} and Tang et al.~\cite{Tang}
lay stress on noise, with the following argument:
the `coarsening time' is set by the condition that the
fluctuations of the height of a mound of size $L$ and average height 
$\bar h$, induced by shot noise, are of the same order of the height
itself.
In chapter~\ref{c_d} we wrote that, because of shot noise:
$\Delta h\sim\sqrt{t/L^d}$. Now we use the extra condition that
$\Delta h\sim\bar h\sim m^* L$, $m^*$ being the largest slope in the
profile. As a general rule: $m^*\sim L^\psi$, where $\psi=0$ for
a current with zeros at some finite $m_0$ and $\psi=1$ by following
the Burton-Cabrera-Frank model [it is sufficient to insert the
current~(\ref{j_II}) with $\gamma=1$ into Eq.~(\ref{L_M})].
Therefore we will have:
\be
L^{\psi+1} \sim \sqrt{t/L^d} ~~~~\Longrightarrow~~~~
L(t) \sim t^n~~~\hbox{with}~~~n={1\over d+2(\psi +1)}
\ee

If the slope of the
mounds is constant, $\psi=0$ and we obtain
$n=1/(d+2)$, i.e. $n=1/3$ in 1+1 dimensions and $n=1/4$ in 2+1
dimensions.
In one dimension, the value $n=1/3$ agrees with exact 
theories~\cite{KO,KM}; in two dimensions, the value $n=1/4$ seems to
agree with the coarsening exponent found on (100) oriented
substrates (quartic symmetry) or on isotropic 
substrates~\cite{SP,S97,S98}, but there is no agreement if the substrate
has 
a triangular symmetry~\cite{Tsui,S98}, in which case $n=1/3$.

A different, nonrigorous approach to coarsening has also been developed
by Golubovi\'c~\cite{Golub} and by Rost and Krug~\cite{RK}. 
We report it here (in a slightly modified way)
because it may clarify some differences
between 1+1 and 2+1 dimensions, and $-$conversely to the previous
argument$-$ it neglects noise and stresses the deterministic character 
of coarsening. We will concentrate on the case of an asymptotic constant
slope ($\psi=0$ in the previous notation), with in-plane symmetry. 

The evolution of the surface is governed by the current:
\be
\vec j = \vec j_{ES} + \vec j_M = 
\alpha\vec m(1-m^2/m_0^2) + K\nabla^2\vec m
\ee
As coarsening proceeds the two terms vanish separately, because 
$|m|\to m_0$ almost everywhere. So a first question is:
Do they remain of the same order, or does one of the two become
negligible?
The answer is: They {\it must} keep of the same order, otherwise the
surface
would not have the displayed `regular' profile and coarsening
(a deterministic process) would not be possible.

This condition implies a relation between the coarsening exponent $n$
and the exponent $\vartheta$ which governs the asymptotic behaviour of
$m$:
$(m_0-|m|)/m_0\sim t^{-\vartheta}$. In fact a qualitative evaluation
of the two terms give:
\be
|\vec j_{ES}|\sim |\vec j_M|~~~\Longrightarrow~~~ 
t^{-\vartheta}\sim {1\over L^2}~~~\Longrightarrow~~~ \vartheta = 2n
\ee

Once established that $|\vec j_{ES}|\sim|\vec j_M|$, two possibilities
are
left: $|\vec j_{ES}+\vec j_M|\sim |\vec j_{ES}|\sim |\vec j_M|$ or 
$|\vec j_{ES}+\vec j_M|$ is much smaller than each single term in the
current. To be more definite let us write the evolution equation for the
interface thickness $w^2(t)=\langle z^2(\vec x,t)\rangle$, starting from
the evolution equation for $z(\vec x,t)$~\cite{RK}:
\be
{1\over 2}\partial_t w^2(t) = \langle\vec m\cdot\vec j_{ES}\rangle
-K\langle (\nabla^2 z)^2\rangle
\ee

We have already argued that the two terms on the right hand side are of 
the same order of magnitude. Now their difference may be i)~of the same
order
too [and therefore $\partial_t w^2(t)\sim \langle\vec m\cdot\vec 
j_{ES}\rangle$] or ii)~it may be smaller. 
In the two cases we can write~\cite{RK}:
\be
{1\over 2}\partial_t w^2(t) \le \langle\vec m\cdot\vec j_{ES}\rangle
\ee

In the same spirit as before, its qualitative evaluation gives:
\be
{m_0^2 L^2\over t} \le t^{-\vartheta}~~~\Longrightarrow~~~
2n \le 1 - \vartheta ~~~\Longrightarrow~~~ n \le {1\over 4}
\label{Rost_ine}
\ee
where the equal sign corresponds to the case (i) mentioned just above.
The inequality $n \le {1\over 4}$ applies to 2+1 as well as 1+1
dimensions: since our model gives $L(t)\sim\ln t$ in 1+1 dimensions
(analytical
result) and $L(t)\sim t^{1/4}$ in 2+1 dimensions (numerical result), the
conclusion (\ref{Rost_ine}) on one side supports these results, and on
the
other side emphasizes an important difference between one and two
dimensions,
due to topology: in 1+1 dimension the current $\vec j$ which contributes
to the
evolution of the surface is vanishing small with respect to  each term 
of the current itself.


\section{Nonlinear analysis of the dynamics of a single step, in the
presence of step-edge barriers}
\label{app_mis}

We have to solve the differential equation: 
\be
D\nabla^2 \rho - \rho /\tau + F_0 =0
\label{de}
\ee
in the $(x,y)$ plane, with $y>\zeta(x,t)$ (one-sided model). 
Boundary conditions, if the step is locally in equilibrium, are:
\bea
\rho(x,\zeta(x,t),t) &=& \rho_{eq}^0 (1+\tilde\Gamma\kappa)\label{bc0}\\
\rho(x,\infty,t) &=& \tau F_0\label{bci}
\eea
where $\kappa$ is the curvature: 
$\kappa = -\zeta_{xx}/[1+\zeta_x^2]^{3/2}$.

The dynamics of the step is determined by the following relation:
\be
v_n = \Omega D \partial_n \rho|_\zeta = \Omega D \vec n\cdot\nabla \rho|_\zeta
\label{bcv}
\ee
where $v_n$ is the velocity normal to the step profile $\zeta(x,t)$:
\be
v_n = (0,v_0+\dot\zeta)\cdot\vec n = {v_0 +
\dot\zeta\over\sqrt{1+\zeta_x^2}}
\ee

In the previous equations, $v_0$ is the velocity of the straight step
and
$\vec n$ (the normal vector to the step) is:
\be
\vec n = {(-\zeta_x,1)\over\sqrt{1+\zeta_x^2}}
\ee

\subsection{Dimensionless equation}
\label{app_ae}

By introducing a new variable $u=\rho-F_0\tau$ and by rescaling $(x,y)$
with
respect to $x_s$ and $t$ with respect to $\tau$, 
Eqs.~(\ref{de},\ref{bc0},\ref{bci},\ref{bcv}) read:
\bea
&& \nabla^2 u - u =0\label{deA}\\
&& u(x,\zeta,t) = - {\Gamma\over\xi} - {\Gamma\zeta_{xx}\over
[1+\zeta_x^2]^{3/2}}\label{bc0A}\\
&& u(x,\infty,t) = 0\nonumber\\
&& {\Omega\Gamma\over\xi} + \dot\zeta = \Omega (u_y -\zeta_x u_x)
\label{bcvA}
\eea
where the quantities $\Gamma=\rho_{eq}^0\tilde\Gamma/x_s$ 
and $\xi=\Gamma/\tau\Delta F$
had already been defined in the body of the article.

As explained in the main text, we introduce the small dimensionless
parameter
$\epsilon\equiv (1/2)-\xi$ and 
the new variables:
\bea
X &=& x\sqrt{\epsilon}\\
T &=& t\epsilon^2\\
H(X,T) &=& \zeta(x,t)/\epsilon 
\eea
through which Eqs.~(\ref{deA},\ref{bc0A},\ref{bcvA}) read:
\bea
&&\epsilon u_{XX} + u_{YY} - u =0\label{deF}\\
&& u(X,\epsilon H,T) = -{\Gamma\over 1/2-\epsilon} 
-\epsilon^2{\Gamma H_{XX}\over
[1+\epsilon^3 H_X^2]^{3/2}}\label{bc0F}\\
&& {\Omega\Gamma\over 1/2-\epsilon} +\epsilon^3 H_T =
\Omega(u_Y -\epsilon^2 H_X u_X)\label{bcvF}
\eea

Now, $u$ and $H$ are expanded in powers of $\epsilon$:
\bea
u &=& u_0 + \epsilon u_1 +\epsilon^2 u_2 +\epsilon^3 u_3 + \dots \\
H &=& H_0 + \epsilon H_1 +\epsilon^2 H_2 +\epsilon^3 H_3 + \dots 
\label{Hex}
\eea

\subsection{Differential equation and boundary conditions at different
orders}
\label{app_adif}

Now the task is laborious, but straightforward: replace the expansions
for
$u$ and $H$ in Eqs.~(\ref{deF},\ref{bc0F},\ref{bcvF}), 
taking in mind that
we must keep all the terms till order $\epsilon^3$. For the differential
equation it is easy, and we obtain:
\bea
u_{0YY} - u_0 &=& 0 \nonumber\\
u_{0XX} + (u_{1YY} - u_1) &=& 0 \nonumber \\
u_{1XX} + (u_{2YY} - u_2) &=& 0 \label{expde}\\
u_{2XX} + (u_{3YY} - u_3) &=& 0 \nonumber 
\eea

For Eq.~(\ref{bc0F}), we start by rewriting the right-hand-side:
\be
u(X,\epsilon H,T) = -2\Gamma(1+2\epsilon +4\epsilon^2 +8\epsilon^3) 
-\Gamma\epsilon^2 (H_{0XX} +\epsilon H_{1XX})
\label{0right}
\ee
The left-hand-side must me handled with care:
\be
u(X,\epsilon H,T) = \sum_{n=0}^3 \epsilon^n u_n(X,\epsilon H,T)
\label{uex}
\ee
where:
\be
u_n(X,\epsilon H,T) = u_n + u_{nY}\epsilon H + u_{nYY} {\epsilon^2
H^2\over 2}
+ u_{nYYY} {\epsilon^3 H^3\over 6}
\ee
and $H^n$ itself must be expanded, using Eq.~(\ref{Hex}).
So, at third order in $\epsilon$, we obtain:
\be
u_n(X,\epsilon H,T) = u_n + u_{nY}(\epsilon H_0 +\epsilon^2 H_1
+\epsilon^3 H_2)
+ {u_{nYY}\over 2} (\epsilon^2 H_0^2 +2\epsilon^3 H_0 H_1) +
{u_{nYYY}\over 6}\epsilon^3 H_0^3 
\ee
where, here and in the future, an expression of the form
$u_{nY..Y}$ must be evaluated in $Y=0$.

By coming back to the expression of $u(X,\epsilon H,T)$ (see
Eq.~(\ref{uex})),
we have:
\bea
u(X,\epsilon H,T) &=& u_0 + \epsilon(u_{0Y}H_0 +u_1) +\epsilon^2
(u_{0Y}H_1 + u_{0YY}{H_0^2\over 2} + u_{1Y}H_0 + u_2)\\
&& + \epsilon^3 (u_{0Y}H_2 + u_{0YY}H_0H_1 + u_{0YYY}{H_0^3\over 6}
+ u_{1Y}H_1 + u_{1YY} {H_0^2\over 2} + u_{2Y}H_0 + u_3)
\label{uex2}
\eea

Comparing this expansion with the right-hand-side of Eq.~(\ref{0right}),
the boundary condition (\ref{bc0F}), at different orders in $\epsilon$
writes:
\bea
u_0 &=& -2\Gamma \nonumber\\
u_{0Y}H_0 +u_1 &=& -4\Gamma \nonumber \\
u_{0Y}H_1 + u_{0YY}{H_0^2\over 2} + u_{1Y}H_0 + u_2 &=& -8\Gamma -
H_{0XX}\Gamma
\label{expbc0}\\
u_{0Y}H_2 + u_{0YY}H_0H_1 + u_{0YYY}{H_0^3\over 6} 
+ u_{1Y}H_1 + u_{1YY} {H_0^2\over 2} + u_{2Y}H_0 + u_3 &=&
-16\Gamma -\Gamma H_{1XX}\nonumber
\eea

Finally, we have to evaluate Eq.~(\ref{bcvF}) up to $\epsilon^3$. This
requires 
to calculate $u_Y$ at the order $\epsilon^3$ and $u_X$ at the order
$\epsilon$.
Concerning $u_Y(X,\epsilon H,T)$, the result is identical to
Eq.~(\ref{uex2}),
once that $u_n$ is everywhere replaced by $u_{nY}$ (and $u_{nY}$ by
$u_{nYY}$,~\dots). Concerning $u_X(X,\epsilon H,T)$ at first order,
since
$u_0$ does not depend on $X$, the result is easily obtained:
\be
u_X(X,\epsilon H,T) = \epsilon u_{1X}(X,0,T)
\ee

By using this expression, but not yet expanding $u_Y$, Eq.~(\ref{bc0F})
reads:
\be
2\Gamma(1+2\epsilon +4\epsilon^2 +8\epsilon^3) + \Omega^{-1}\epsilon^3
H_{0T} =
u_Y -\epsilon^3 H_{0X}u_{1X}
\ee

If $u_Y$ is expanded according to the criterion given above, 
the velocity of the step (Eq.~(\ref{bcvF})) is determined at the various
order in $\epsilon$ by the following relations:
\bea
2\Gamma &=& u_{0Y} \nonumber \\
4\Gamma &=& u_{0YY} H_0 + u_{1Y} \nonumber \\
8\Gamma &=& u_{0YY} H_1 + u_{0YYY} {H_0^2\over 2} + u_{1YY} H_0 + u_{2Y}
\label{expbcv} \\
16\Gamma + \Omega^{-1} H_{0T} &=& u_{0YY} H_2 + u_{0YYY} H_0 H_1 
+ u_{0YYYY} {H_0^3\over 6} + u_{1YY} H_1 + u_{1YYY} {H_0^2\over 2} +
u_{2YY} H_0 + u_{3Y} - u_{1X}H_{0X}\nonumber
\eea

The next and final step is to solve, order by order, Eqs.~(\ref{expde})
with
boundary conditions (\ref{expbc0}) and (\ref{expbcv}).

\noindent {\it Zero order -} We have to solve: $u_{0ZZ} - u_0=0$, with
$u_0 = -2\Gamma$ and $u_{0Z}=2\Gamma$ as boundary conditions. The
solution is:
\be
u_0(Z) = -2\Gamma \exp(-Z)
\ee
We remark that the two boundary conditions are indeed equal, and this
will
be true also at the first and second order in $\epsilon$\,! This is not
surprising, because $H_{0T}$ $-$and therefore the motion of the step$-$
enters at the third order of the boundary conditions (\ref{expbcv})
only.

\noindent {\it First order -} The solution of the equation
$u_{1ZZ} - u_1=0$ is:
\be
u_1(X,Z,T) = A_1(X,T)\exp(-Z)
\ee
and the boundary condition $u_{0Z}H_0 +u_1= -4\Gamma$ implies:
\be
2\Gamma H_0(X,T) +A_1(X,T) = -4\Gamma
\ee

\noindent {\it Second order -} The differential equation
$u_{1XX} + (u_{2ZZ} - u_2)=0$ gives the solution:
\be
u_2(X,Z,T) = A_2\exp(-Z) + {A_{1XX}\over 2} Z\exp(-Z)
\ee
where $A_2$ (as well $A_1$) is a function of $X$ and $T$.
The boundary condition $u_{0Z}H_1 + u_{0ZZ}{H_0^2\over 2} + u_{1Z}H_0 +
u_2=
-8\Gamma - H_{0XX}$ gives:
\be
\Gamma^{-1} A_2 = -8 -2H_1 -H_0^2 -4H_0 -H_{0XX}
\ee

\noindent {\it Third order -} The differential equation
$u_{2XX} + (u_{3ZZ} - u_3)=0$ has the solution:
\be
u_3(X,Z,T) = \left[ A_3 +\left( {A_{1XXXX}\over 8} + {A_{2XX}\over
2}\right)Z
+ {A_{1XXXX}\over 8}Z^2\right]\exp(-Z)
\ee

This time the two boundary conditions (\ref{expbc0}) and (\ref{expbcv}) 
are not equivalent and this implies a condition on $H_0$ under the
form of a differential equation which governs the dynamics of the step.
In fact, Eq.~(\ref{expbc0}) gives:
\be
\Gamma^{-1} A_3 + 16 + H_{1XX} + 2H_2 +8H_0 + {H_0^3\over 3} +2H_0^2
+2H_0H_1 
+4H_1 =0
\ee
and Eq.~(\ref{expbcv}) gives:
\bea
&&\Gamma^{-1} A_3 + 16 + H_{1XX} + 
2H_2 +8H_0 + {H_0^3\over 3} +2H_0^2 +2H_0H_1\nonumber\\ 
&&+4H_1 + (\Omega\Gamma)^{-1} H_{0T} + (3/4)H_{0XXXX} - H_{0X}^2 +
2H_{0XX}=0
\eea

Their comparison provides the equation for $H_0$ we were looking for:
\be
(\Omega\Gamma)^{-1} \partial_T H_0 = -2\partial_X^2 H_0
-(3/4)\partial^4_X H_0 + (\partial_X H_0)^2
\ee


\section{Scaling laws for coherent epitaxial  clusters}
\label{apx}

\subsection{Scaling of the elastic free energy of a cluster with respect
to the misfit}
\label{scal1}

The first scaling law concerns the influence of the misfit for
an adsorbate of fixed shape and volume and fixed elastic constants. The
misfit
can be represented by an external  or force dipole density (or `stress')
acting inside the adsorbate. This stress will always be assumed weak
enough, so 
that the response is linear, i.e. the 
strain at a given point is proportional to the misfit $\delta a/a$, so
that the 
elastic energy density is proportional to $(\delta a/a)^2$. The total
elastic
energy is therefore also proportional to $(\delta a/a)^2$. The
determination of
the proportionality coefficient would be difficult. It is an integral
which
involves the elastic linear response functions (or `Green's functions')
of the system,
which are very complicated and depend on the adsorbate shape and elastic
constants.


\subsection{Scaling of the elastic free energy of a cluster with respect
to its volume $V$}
\label{scal2}

The second scaling law concerns the size dependence of the elastic free
energy 
for an adsorbate of a particular shape, whose linear dimensions can be 
varied but keep fixed ratios. 
The elastic free energy  is the sum of (\ref{2.a}) and (\ref{2.1'}) 
and can be written as 

\begin{equation}
\delta {\cal F}_{el}^{(a)} =  \int  d^3r 
\Phi\left(\vec r, \epsilon_{xx}(\vec r), \epsilon_{xy}(\vec r),... 
\epsilon_{zz}(\vec r) \right)
\label{risotto}
\end{equation}
where the function $\Phi$ is defined by  (\ref{2.a}) and (\ref{2.1'}). 

Expression (\ref{risotto}) is very general, and general expressions are
sometimes 
brain-twisting. It may therefore be a helpful exercise to consider the
example of
a cluster which is a half-sphere of radius $R$. 
If the origin of the coordinates are
chosen at the center of the sphere, the linear part of  $\Phi$ can be
written, 
according to (\ref{2.a}), as

\begin{equation}
\Phi^{(lin)}\left(\vec r, \epsilon_{xx}, \epsilon_{xy},... \epsilon_{zz}
\right) = 
\hbox{Const} \times \frac{\delta a}{a} 
[\epsilon_{xx} + \epsilon_{yy}] \theta(z) \theta (R-r)
\label{2.aa}
\end{equation}
where $\theta(z)=0$ for $z<0$ and $\theta(z)=1$ for $z \geq 0$. The
quadratic part 
$\Phi^{(quad)}$ can easily be written using (\ref{2.1'}).

Consider now  (Fig. \ref{olala}b)  
another structure of the adsorbate which is deduced from the 
first one (Fig. \ref{olala}a) by the similarity
transformation\footnote{The notation $\vec{r}$
designates the points of the totally constrained adsorbate, so that the
actual
locations of the atoms are $\vec{r}+\vec{u}(\vec{r}) $, where $\vec{u} $
is the elastic displacement}

\begin{equation}
\vec{r} \rightarrow \lambda\vec{r} 
\label{simil}
\end{equation}

The elastic free energy  of the new structure has the form

\begin{equation}
\delta {\cal F}_{el}^{(b)} =  \int  d^3r 
\Phi\left( \vec r/\lambda, \epsilon_{xx}(\vec r), 
\epsilon_{xy}(\vec r),...\epsilon_{zz}(\vec r) \right)
\label{lasagne}
\end{equation}
where the function $\Phi$ is just the same as before! The proof is easy 
and left to the reader, who can first treat the example of a half
sphere. 
This property is true {\it even if anharmonicity is taken into account}.
However, it would not be completely true if the atomic, discrete nature
of matter were not disregarded.

Now, if one makes the change of variables

$$
\epsilon_{\alpha, \gamma}(\vec r) = \eta_{\alpha, \gamma}(\vec
r/\lambda)
 \;\;\;\;,\;\;\;
\vec r = \lambda \vec \rho 
$$
formula (\ref{lasagne}) reads
\begin{equation}
\delta {\cal F}_{el}^{(b)}= \lambda^3 \int  d^3\rho 
\Phi\left( \rho, \eta_{xx}(\vec \rho), 
\eta_{xy}(\vec \rho),...\eta_{zz}(\vec \rho) \right)
\label{lasagne'}
\end{equation}
which is identical to (\ref{2.aa}) apart from the factor $\lambda^3$ and
the 
different names of the variables. If  (\ref{2.aa}) is minimized by the
strain $\epsilon_{\alpha, \gamma }(\vec r)=\epsilon^0_{\alpha, \gamma
}(\vec r)$,
then  (\ref{lasagne'}) is minimized by 
$\eta_{\alpha, \gamma }(\vec \rho)=\epsilon^0_{\alpha, \gamma }(\vec
\rho)$,
and therefore (\ref{lasagne}) is minimized by 
$\epsilon_{\alpha, \gamma }(\vec r)=\epsilon^0_{\alpha, \gamma }(\vec
r/\lambda)$.
And the minimum of (\ref{lasagne}) or (\ref{lasagne'}) is equal to 
the minimum of (\ref{2.aa}) multiplied by $\lambda^3$. Since the volume
of the transformed cluster is equal to the volume
of the original cluster multiplied by $\lambda^3$, 
{\it the elastic energy of a coherent,
epitaxial cluster of 
a
given shape is  proportional to its volume $V$}.


\subsection{Scaling of the strain at long distance $\vec r$ of a cluster 
as a function of its volume $V$}
\label{scal3}

The problem is analogous to the previous one but now we consider
the elastic displacement $\vec u(\vec r)$, at a fixed, large 
distance $\vec r$ of the center of the cluster. In the special case of 
an isotropic elastic medium, the complete calculation has been done by 
Boussinesq~\cite{bouss}.

The displacement at distance $r$ turns out to be proportional
to $1/r^2$, so that the strain is proportional
to $1/r^3$. We shall argue here that this scaling is also true for an
anisotropic 
elastic solid. As a lemma, we need to show first that, for fixed $\vec
r$, 
the displacement and  the strain are proportional to $V$.

Let us assume that the atomic displacements 
in a slab $(S)$ just below the adsorbate are known.
Its knowledge determines the displacement and
strain in the remainder of the crystal through the
equations of elasticity (\ref{eqel}). Outside the slab $(S)$, these
equations are 
independent of the existence or non existence of the cluster. On the
other
hand, in the slab, the displacements $\vec u_i$ are 
the same as those which would arise from 
external forces $\vec f_i$ acting on the $n$ 
atoms inside the slab in the absence of 
the
adsorbed cluster.
The forces $\vec f_i$ can in principle be determined by solving the
equations
of elasticity which usually yield the $\vec u_i$'s when the forces $\vec
f_i$
are known, but can be used also to determine the 
$\vec f_i$'s if the $\vec u_i$'s are known, since there are $3n$ linear
equations 
for $3n$ unknowns $f_{i\alpha}$. Since the total force must vanish, it
may be 
preferable to introduce a force dipole density  $\mu_{\gamma \xi}(\vec
r')$

The displacements $\vec u_i=\vec u(\vec r)$ at distance $\vec r$ are 
related to the  forces 
by linear equations of the form

\begin{equation}
u_\alpha(\vec r)=\int_{(S)}d^3r'
\sum_{\gamma \xi}\Gamma_{\alpha \gamma \xi}(\vec r-\vec r') \mu_{\gamma
\xi}(\vec r')
\label{uf}
\end{equation}
where the response function $\Gamma_{\alpha \gamma \xi}(\vec r-\vec r')$ 
is independent of the 
size and shape of the cluster, since the cluster has been removed.
It is also independent of the size and shape  of the 
slab. Actually, $\Gamma_{\alpha \gamma \xi}(\vec r-\vec r')$ 
is a Green's function of an elastic solid limited by a plane,
which can be found in textbooks... at least in the case of an isotropic
solid
or a cubic crystal limited by a (001) surface.

We now make the similarity transformation (\ref{simil}). It can easily 
be shown that the term linear with respect to the fictitious 
forces satisfies (\ref{lasagne}), so that 
the force or force dipole densities are the same in two corresponding
points.
Now, at a long distance, the dependence of $\Gamma_{\alpha \gamma
\xi}(\vec r-\vec r')$
with respect to $\vec r'$ can be ignored and, if the cluster is centered
at the
origin 0, (\ref{uf}) reads, before the similarity transformation,
 
$$
u^{(1)}_\alpha(\vec r)=
\sum_{\gamma \xi}\Gamma_{\alpha \gamma \xi}(\vec r) 
\int_{(S)}d^3r' \mu_{\gamma \xi}(\vec r\,')
$$
After  the similarity transformation, if the transformed slab is called
$(S')$, 
the displacement at the same point $\vec r$ is 
$$
u^{(2)}_\alpha(\vec r)=
\sum_{\gamma \xi}\Gamma_{\alpha \gamma \xi}(\vec r) 
\int_{(S')}d^3r' \mu_{\gamma \xi}(\vec r\,'/\lambda)
= \lambda^3 \sum_{\gamma \xi}\Gamma_{\alpha \gamma \xi}(\vec r) 
\int_{(S)} d^3r'' \mu_{\gamma \xi}(\vec r\,'')
$$
where $\vec r\,''=\vec r\,'/\lambda$. Comparison of the last two
formulae shows that
$u^{(2)}(\vec r)=\lambda^3u^{(1)}(\vec r)$. {\it The elastic
displacement
and stress at a long, given distance $r$ of a coherent, epitaxial
cluster of a
given shape is proportional to its volume.}

\subsection{Scaling of the strain at long distance $\vec r$ a cluster 
as a function of $r$}
\label{scal4}

A consequence of the previous paragraph  is that the strain on 
the adsorbate
surface at 
a 
long distance $r$ from the center of a cluster decays as $r^{-3}$.
Indeed, the strain  should be proportional to the volume, so that the
strain
$\epsilon_0(\vec{r})$ produced by the clusters (C)  and (C') satisfies 
$\epsilon_1(\lambda \vec{r})=\lambda^3\epsilon_0(\lambda \vec{r})$. 
But we have seen that 
$\epsilon_1(\lambda \vec{r})=\epsilon_0( \vec{r})$. Therefore 
$\epsilon_0(\lambda \vec{r})=\lambda^{-3}\epsilon_0(\vec{r})$ 
in agreement with our statement. In particular, the strain produced by a
single 
adsorbed
atom at a distance $r$ on the  substrate surface is proportional to
$r^{-3}$.

\subsection{Elastic interactions at long distance $\vec r$ between
clusters or adatoms}
\label{scal5}

One can deduce how the elastic interaction between two clusters at a
long distance $r$
behaves. Indeed, the energy change when depositing an adatom can be
shown to be 
a linear function of the local strain. Therefore, there is an elastic
interaction
between adsorbed atoms

\begin{equation}
V_{el}(r)= \hbox{Const}/r^{3}
\label{elasBB'}
\end{equation}

The constant  can be shown to be positive, so that the elastic interaction 
is repulsive~\cite{Nozieres,JVAP}. There is no contradiction with the
above statement that the {\it total} elastic energy is negative. 

The elastic interaction between two adsorbed atoms at a
long distance $r$ is given by a formula analogous to (\ref{elasBB'}),
which 
is  (\ref{elasBB}).

The result (\ref{elasBB'}) can be used to determine the scaling of the 
interaction energy of a set of coherent clusters as a function of their
average distance $r$ or their average diameter $R$ at constant coverage
$N/{\cal A}$. 
The distance $r$  will be assumed 
to be much longer than the average cluster diameter $R$. According to
(\ref{elasBB}),
the  elastic interaction must be proportional to $1/r^3$ multiplied by
the square of 
the average number $n$ of adatom per clusters (proportional to $R^3$) 
and by the 
number of clusters ${\cal N}$ which is proportional to $1/r^2$. Since
the total
adsorbed mass is proportional to $N=n{\cal N}$ and fixed, $R^3$ should
be proportional to
$r^2$ and the  interaction  energy between clusters is 
$$
{\cal F}_{inter}\approx \hbox{Const}\times {\cal N}n^2/r^3 \sim 
\hbox{Const}\times R^6/r^5 \sim 
\hbox{Const}/r\;\;\;.
$$
The constant is certainly positive
since it is positive for a single atom. Thus, the  `intercluster'
elastic energy
${\cal F}_{inter}$ 
is a decreasing function of $r$ and favours large clusters. 
In addition there is an `intracluster' energy discussed earlier in this
chapter.  Since it is proportional to the cluster volume, 
i.e. to the number of atoms,  
the total intracluster energy is not affected by atom exchange, at least
if the modification  of the shape is ignored. A more detailed analysis
taking shape modification into account does not modify the conclusion
that the average cluster size should increase
steadily with time, even if elasticity is taken into account. The 
possible reasons of the experimental facts which seem to contradict 
this conclusion are reviewed in paragraph \ref{surf}.


\vfill\newpage


\begin{tabular}{|c|c|c|}
\hline
 	& $T<T_R$ & $T>T_R$  \\  
\hline
\hline
Step free energy per unit length $\gamma$ & finite &  $\gamma=0$ \\
\hline
Typical radius of thermal fluctuations (bumps and valleys) & 
$R \approx k_BT/\gamma$ finite & $R=\infty$ \\
\hline 
Typical height $\delta h$ of thermal fluctuations & finite &  infinite
\\ 
\hline
 Excess free energy density of a height modulation &
$\delta{\cal F}/{\cal A}=2\gamma  |q|\delta h /(\pi c) $ &  
$\delta{\cal F}/{\cal A}=\tilde\sigma q^2 \delta h^2/4 $  \\
of small amplitude $\delta h$ and wave vector $q$  & +... & +... \\
\hline
Variation of the  free energy density resulting from a  rotation of &
$\delta{\cal F}/{\cal A}=\gamma |\delta \theta|/c$ &  
$\delta{\cal F}/{\cal A}= \tilde\sigma  \delta \theta^2/2 $  \\
the surface of small amplitude $\delta \theta$ around a high
symmetry orientation
 &  $+  ...$&  +... \\
\hline 
Continuous changes of surface height 
 &  impossible &  possible  \\
\hline
\end{tabular}

\begin{table}
\caption{Some typical properties of an infinite, plane crystal surface 
above and below $T_R$. The step free energy per unit length $\gamma$ is
assumed to be isotropic, and therefore equal to the step stiffness.
The infinite value of $\delta h$  above $T_R$ is an intuitive
consequence 
of the property $R=\infty$, which in turn results from the formula
$R \approx k_BT/\gamma$ since $\gamma=0$ above $T_R$. 
$c$ is the monolayer thickness. 
}
\label{rough_table}
\end{table}

\begin{tabular}{|c|c|c|c|c|c|c|c|c|}
\hline
Exp. system	& $T$(K) & $n$ & $L$ (nm) & Slope$^a$ & $N$ (ML) & Flux  
& Exp. technique $^f$ & Ref. \\  
 & & & & & & (ML/min) & & \\
\hline
\hline
GaAs(001) & 830/900 & & $< 10^3$ & 1$^\circ/2^\circ$ & 540$\;^e$ & 10 
& AFM,STM & \protect\cite{Orme,Joh} \\
\hline
Ge(001) & 330/500 & 0.4 &$<$ 200 & 2$^\circ/3^\circ\,^b$ & $1\mu$m & 40 
& STM & \protect\cite{Nostrand,ES-Ge}\\
\hline 
Cu(100) & 160 & 0.26 &  & (113) & 100 & 0.5 & TEAS &
\protect\cite{Ernst}\\ 
\cline{2-3}\cline{5-5}
         & 200 & 0.56 &  & (115) &        &          &      & \\ 
\hline
 Cu(100) & $< 180$ & & & (113) & &  & & \protect\cite{Jorri} \\
\cline{2-2}\cline{5-5}
         & $180/280$ & & & (115) & 150  &   0.6       & TEAS,SPALEED  &
\\
\cline{2-2}\cline{5-5}
 & $280/300$ & & & (117) &        &           &           &   \\
\hline
Cu(100) & 300 & 0.25(1) & $12.5\div 25^d$ & 5.6$^\circ\,^c$ & $1\div 20$ 
& 1.2 & LEED & \protect\cite{Zuo} \\
\cline{3-4}\cline{6-6}\cline{8-8}
         &     & 0.23(1) & $10\div 35$ &             &  $1\div 150$   
&           & STM  & \\
\hline
Fe(001) & 290 & 0.16(4) & $4\div 9$ & $13^\circ\pm 3^\circ$ & $20\div
600$ 
& 1.5/30 & STM,RHEED & \protect\cite{Stro}\\
\hline
Fe/Mg(001) & 400/450 & 0.23(2) & $20\div 60$ & $30^\circ$ (012) 
& $30\div 2000$ & 3 & STM,LEED & \protect\cite{thurmer}\\
\hline
Rh(111)/mica & 725 & 0.33(3) & $10\div 100$ & 1$^\circ$ & $1\div 300$ &
2 
& STM,RHEED & \protect\cite{Tsui}\\
\hline 
\end{tabular}

\begin{table}
\caption{Details on the table are given in the main text.
When a box is empty, it means that the datum is not
available.
\hfill\protect\newline
$^a\,$The slope is given in degrees and/or through the orientation of
the
corresponding facet. The notation $1^\circ/2^\circ$ means that the slope
is in between one and two degrees.
$^b\,$Authors study the distribution 
of the local slope and find that it is peaked at $2^\circ$ after 
$0.5~\mu$m have been deposited, and at $3^\circ$ after $1~\mu$m.
$^c\,$The experimental finding of a slope corresponding to $2.4^\circ$
is modified to $5.6^\circ$, 
because the former value is considered an `average' one.
$^d\,$The values of $L$ correspond to $2\pi/L^*$, $L^*$ being `the
diameter of the diffraction ring' in the SPA-LEED angular profiles.
$^e\,$In the case of GaAs, 1 ML means one ML of gallium and
one ML of arsenic.
$^f\,$AFM\,=\,Atomic force microscopy;
STM\,=\,Scanning tunneling microscopy;
TEAS\,=\,Thermal energy atom beam scattering;
LEED\,=\,Low energy electron diffraction;
SPALEED\,=\,Spot profile analysis LEED;
RHEED\,=\,Reflection high energy electron diffraction.}
\label{exp_table}
\end{table}


\begin{figure}
\caption{Cross-section of a ZnTe/CdTe multilayer seen by low resolution
(a)
and high resolution (b) electron microscopy.
Narrow white stripes in (a) are two monolayer thick ZnTe inclusions.
Since the lattice parameter of
both components is quite different, phase separation into ZnTe and CdTe
would presumably lower the free energy. Thus, the system is in principle
unstable or metastable. Courtesy of N. Magn\'ea and
J.L. Pautrat, C.E.A. Grenoble.}
\label{multi}
\end{figure}

\begin{figure}
\caption{Four mechanisms which can produce a growth instability.
(a) Diffusion instability. (b) Kinetic instability due to step-edge
barriers.
(c) Thermodynamic instability due to  different lattice constants,
$a$ and $a+\delta a$, of the
    substrate and the adsorbate.
In the present case $\delta a>0$.
(d) Geometric, shadowing instability.}
\label{eheh}
\end{figure}

\begin{figure}
\caption{(a) Principle of a diffusion instability. If a small bump
tends to form on a solid nucleus growing in a supersaturated solution,
diffusing atoms (circles) tend to go to the bump which thus grows more
rapidly
than the rest of the nucleus. Matter diffusion on the solid (dashed
arrows) has a stabilizing effect,
i.e. tends to destroy the bump.
(b) A snowflake.
(c) The Mullins-Sekerka instability in directional solidification.
The initially plane solid-liquid interface takes a wavy shape
which can reach a
stable shape if the velocity $v$ of the interface is low enough.}
\label{diff}
\end{figure}

\begin{figure}
\caption{A two-dimensional DLA pattern from the computer (a)
and its experimental realization
by electrodeposition of Zn in the electrolysis of a
quasi-two-dimensional solution of ZnSO$_4$ (b).  Courtesy of S. Bodea
and P. Molho, Laboratoire Louis N\'eel, CNRS, Grenoble.}
\label{DLA}
\end{figure}

\begin{figure}
\caption{Schematic description of  MBE growth in the case of a III-V
semiconductor, e.g., GaAs. Each cube represents a cubic unit cell.
Note that a large proportion of As atoms evaporate, while Ga atoms do
not.}
\label{fig1gg}
\end{figure}

\begin{figure}
\caption{Energy gap and lattice constant for
a few III-V
compound semiconductors. }
\label{fig2gg}
\end{figure}

\begin{figure}
\caption{Transmission electron microscopy image
of Al$_{0.48}$In$_{0.52}$As lattice matched grown  on
InP(001) by MBE (V/III BEP ratio equal 20,
 growth rate equal to 1 $\mu$m/h and
growth temperature equal to
$450^\circ$C ) revealing a fine
quasi-periodic contrast due to the
presence of In-rich and Al-rich clusters : (a) plane view, (b) cross
section .
Courtesy of O. Marty, Univ. Lyon I.}
\label{fig3gg}
\end{figure}

\begin{figure}
\caption{TEM image of Ga$_{0.47}$In$_{0.53}$As grown by MOCVD showing
CuPt ordering
along (111) and (11-1). Courtesy of O. Marty and C. Pautet, Univ.
Lyon I.
}
\label{fig4gg}
\end{figure}

\begin{figure}
\caption{Two examples of carrier confinement
(a) in a Ga$_{0.43}$In$_{0.57}$As quantum well between
Al$_{0.48}$In$_{0.52}$As barriers;
(b) in a heterojunction between highly n-type AlInAs and lightly n-type
doped GaAs.}
\label{fig5gg}
\end{figure}

\begin{figure}
\caption{Surface profile with top, bottom and vicinal terraces.
$D$ is the diffusion constant and $k_\pm$ are the sticking coefficients
from above and below (see the main text).}
\label{tbv}
\end{figure}

\begin{figure}
\caption{The adatom uses its condensation heat (a)~to look for a
high-coordination site or (b)~to knock-out and replace an edge atom.
Except for simple cubic structures, the solid-on-solid constraint
implies
(c)~that the capture area of a higher coordination site is larger.
Dotted circles are newly incorporated atoms, directly from the
flux~(a,c)
or after a knock-out process~(b).}
\label{non_the_pro}
\end{figure}

\begin{figure}
\caption{Profiles for the different potentials $V(m)$.
Full line: model~I, Eq.~(\protect\ref{I}).
Short-dashed: model~II with $\gamma=1$, Eq.~(\protect\ref{II1}).
Long-dashed: model~II with $\gamma=2$, Eq.~(\protect\ref{II2}).
The behaviour at small $m$ is the same in all the cases:
$V(m)\simeq\alpha m^2/2$.}
\label{V(m)}
\end{figure}

\begin{figure}
\caption{($a$)~Numerical solutions of the four-state clock
model~(\protect\ref{MS_clock}) and the surface growth model
with ($b$)~four-fold
in-plane symmetry~(\protect\ref{MS_growth}) and with ($c$)~in-plane
invariance~(\protect\ref{MR_growth}). Figures~($a$) and ($b$) are taken
from
Ref.~\protect\cite{S97} and are courtesy of Martin Siegert.
White, light and dark grey, and black domains correspond respectively to
the
orientations (1,1), (1,-1), (-1,1) and (-1,-1).
Figure ($c$) is unpublished and is courtesy of Martin Rost.}
\label{2d_growth}
\end{figure}

\begin{figure}
\caption{Example of a mound structure obtained with the Zeno model.}
\label{Zeno_mound}
\end{figure}

\begin{figure}
\caption{Surface morphology in kinetic Monte Carlo simulations of a vicinal
surface in presence of ES barriers. The steps become wavy~(a) due to the
Bales and Zangwill instability~\protect\cite{BZ}.
Subsequently, the ripples break down starting at the defects of the
ripple
pattern, and three-dimensional features appear on the surface~(b).
The figure is taken from Ref.~\protect\cite{Smil98} and it is
courtesy of Pavel \v{S}milauer.}
\label{Fig-Smil}
\end{figure}

\begin{figure}
\caption{Columnar growth.}
\label{colon}
\end{figure}

\begin{figure}
\caption{Droplets of the Volmer-Weber type (above) and of the
Stranski-Krastanov type
(below). The shape, and particularly the angle $\theta$, is independent
of the size.}
\label{drop}
\end{figure}

\begin{figure}
\caption{$[1\bar{1}0]_{Au}$ zone axis Fourier Filtered HREM image
showing a misfit
dislocation at an Au/AuNi(001) interface with a Burgers vector
$(1/2)<10\bar{1}>$.
The arrow shows only the projection of the Burgers vector in the
$(1\bar{1}0)$
observation plane. The dashed line is the Burgers circuit around the
dislocation core.
 The excess plane is also shown in black dashed line. Courtesy
of C. Dressler and J. Thibault, DRFMC/SP2M, CEA-Grenoble.}
\label{fig:HREMdisloc}
\end{figure}

\begin{figure}
\caption{Schematic cross-sectional view of a Shockley partial
dislocation of Burgers vector
(1/6)[112] which has glided in  a (111) plane from the surface to the
interface. A
stacking fault appears in the glide plane between the dislocation and
the surface, so that lattice planes have a discontinuity when crossing
the glide plane.}
\label{fig:disloc_partial}
\end{figure}

\begin{figure}
\caption{Schematic cross-sectional view of a dislocation
of Burgers vector $(1/2)[110]$
which has climbed from the surface to the interface with the
introduction of two supplementary planes (atoms in grey). The arrows
represent the displacement field in the epilayer.}
\label{fig:disloc_climb}
\end{figure}

\begin{figure}
\caption{Schematic cross-sectional view of a perfect dislocation
$(1/2)[101]$ which has glided in a (111) plane from the surface to the
interface. The arrows represent the displacement field in the epilayer.
There is no lattice discontinuity at the glide plane.}
\label{fig:disloc_parf}
\end{figure}

\begin{figure}
\caption{A dislocation of Burgers vector $\vec{b}$ moving in its glide
plane.}
\label{fig:disloc3d}
\end{figure}

\begin{figure}
\caption{Critical thickness versus misfit for the glide system
of Burgers vector  (1/2)[101]
and glide plane (111) in the FCC lattice of lattice parameter $a_0$
for two values of the
core parameter $\alpha$ (Dashed curve: $\alpha$=2, continuous curve:
$\alpha$=4).}
\label{fig:hcrit}
\end{figure}

\begin{figure}
\caption{A dislocation loop nucleating at the surface. An atomic step
appears at the intersection of the glide plane with the surface.}
 \label{fig:loop}
\end{figure}

\begin{figure}
\caption{Total energy of a half-loop
vs radius for 90$^{\circ}$ partial and
60$^{\circ}$ perfect dislocation in Au$_{80}$Ni$_{20}$/Au(001). }
\label{fig:loop_energy}
\end{figure}

\begin{figure}
\caption{Schematic representation of 2 arrays of orthogonal
alternating dislocations.}
\label{fig:dislocarray}
\end{figure}

\begin{figure}
\caption{Curves showing the total energy density of dislocations
versus the
residual elastic strain in the epilayer (lower axis) or versus the
dislocation
density (upper axis).  }
\label{fig:energy_strain}
\end{figure}

\begin{figure}
\caption{Blocking mechanism for the relaxation of MgO grown on
Fe(001). Circles:
experimental data from RHEED measurements. Continuous lines :
equilibrium and
blocking models.}
\label{fig:relaxmgo}
\end{figure}

\begin{figure}
\caption{Blocking mechanism : the progression of a threading
dislocation
${\cal D}_1$ may be disturbed by the stress field of a perpendicular
dislocation ${\cal D}_2$
lying at the interface, so that it bypasses this obstacle at a depth $h^*$.}
\label{fig:blocking}
\end{figure}

\begin{figure}
\caption{Residual strain needed to overcome   the line tension ($\epsilon_{line}$),
the orthogonal dislocation stress field ($\epsilon_D$) and
both retaining forces ($\epsilon_{tot}$)
as  functions of the channel width $h^*$. The value
$\epsilon_{min}$ corresponds to the possible residual strain before
blocking of further moving dislocations.}
\label{fig:relaxres}
\end{figure}

\begin{figure}
\caption{Two ways of relaxing the stress in an adsorbate. (a) Creation
of a misfit dislocation $D$ by a microscopic motion of atoms in a `glide
plane' inside the material.
(b) Creation of waves at the surface by atom diffusion on the surface on
long distances.}
\label{2instab}
\end{figure}

\begin{figure}
\caption{Experimental image of a modulation arising from the misfit
in a Ga$_{1-x}$In$_x$As multilayer on an InP(001) substrate
($x \approx 0.28$),
from Ref.~\protect\cite{pon1}.  The dark layers
 are under tensile stress ($\delta a/ a \approx -1.7$\%).
The bright layers are
lattice matched to InP  and are used as growth  markers.
Although the quantitative theory can be different
in a multilayer, the mechanism is probably the same as described for a
single adsorbed layer in paragraph \protect\ref{tsf2}. }
\label{ponch2}
\end{figure}

\begin{figure}
\caption{Experimental images of clusters of InAs on an InP(001)
substrate: (a)~plane view and (b)~cross-sectional observations.
The adsorbate (InAs) has a lattice constant
3.2 \% greater than the substrate (InP). Small clusters are
dislocation-free
while bigger ones are not, as most clearly seen from the lower picture }
\label{ponch1}
\end{figure}

\begin{figure}
\caption{ Schematic representation of clusters of increasing size
which arise from elasticity in coherent,
epitaxial growth. The shape and the contact angle depend on the size.
Small clusters are one monolayer thick while bigger ones are
`three-dimensional' with an increasing contact angle when the size
increases.
The spherical shape assumed for three-dimensional clusters is not
realistic. In
practice,  almost detached clusters such as the right hand one have
never been observed
in the dislocation-free state. }
\label{clus}
\end{figure}

\begin{figure}
\caption{Sinusoidal modulation of a surface (a) under the influence of an
anisotropic, mechanical stress and (b) under the influence of an
adsorbate.}
\label{grfld}
\end{figure}

\begin{figure}
\caption{A slab and a truncated pyramidal cluster.}
\label{pyr}
\end{figure}

\begin{figure}
\caption{A Ge cluster arising from the deposition of Ge on Si(001).
STM image kindly communicated by Max Lagally.}
\label{exppyr}
\end{figure}

\begin{figure}
\caption{The function $G(x)$.}
\label{GGG}
\end{figure}

\begin{figure}
\caption{ Growth of a submonolayer.}
\label{grow}
\end{figure}

\begin{figure}
\caption{ The Lifshitz-Slyozov mechanism. Atoms leave the small clusters
to go to the big ones where the free energy per atom (i.e. the chemical
potential) is lower.}
\label{LSW}
\end{figure}

\begin{figure}
\caption{Microscopic
twins in Au$_{0.5}$Ni$_{0.5}$ layers grown epitaxially
on Au(001). High resolution electron microscopy cross-section image, the
beam direction is [110].}
\label{fig:tem_twin}
\end{figure}

\begin{figure}
\caption{AFM image of the surface of a FePd epilayer deposited on
Pd(001). The steps oriented at about 45$^{\circ}$ with respect to the
edges of the figure correspond to the emergence of the microscopic
twins at the surface.}
\label{fig:afm_twin}
\end{figure}

\begin{figure}
\caption{Schematic representation of a microscopic
twin due to the gliding
of several Shockley partial dislocations on successive (111) atomic planes.}
\label{fig:twin_schem}
\end{figure}

\begin{figure}
\caption{Nucleation lengths $(\ell_n)$ for the different
kinds of terraces in a 1+1 dimensional model,
as a function of the ES length $(\es)$. All the lengths
are expressed in units of the diffusion length $\ell_D$. Starting from
the top, we have $\ell_n^B$ (which does not depend on $\es$),
$\ell_n^V$ (which goes to the constant value
$1/\protect\sqrt{2}$) and
$\ell_n^T$ (which slowly goes to zero). The behaviour of $\ell_n^T$
at large $\es$ can be directly found from Eq.~(\protect\ref{lnt}):
$\ell_n^T\simeq (1/6\es)^{1/3}$. The small exponent (1/3) explains
the slow decrease of $\ell_n^T$.}
\label{ln-fig}
\end{figure}

\begin{figure}
\caption{(a) and (b) represent two similar adsorbed clusters ($C$), ($C'$).
The strain in two corresponding
points is identical. (c) Appropriate external forces within a
thin slab ($L$) just below the surface produce the same elastic
displacements and strains
in the substrate as the cluster (a). (d) The slab ($L'$) deduced from ($L$)
by the similarity
transformation produces the same effect as the cluster ($C'$). On the
other hand, the
strain at long distance $\vec r$  is multiplied by $\lambda^3$.  }
\label{olala}
\end{figure}

\end{document}